\documentclass[aps,10pt,prd,groupedaddress,nofootinbib,notitlepage,eqsecnum,preprintnumbers]{revtex4-1}

\usepackage[utf8]{inputenc}
\usepackage{hyperref}
\usepackage{xcolor}
\usepackage{graphicx}
\usepackage{amsmath,amssymb}
\usepackage{bm}
\usepackage{floatrow}
\usepackage{capt-of}
\usepackage[font=small,labelfont=bf,tableposition=top]{caption}
\usepackage{booktabs}
\usepackage{varwidth}
\usepackage[export]{adjustbox}
\usepackage{slashed}
\usepackage{autobreak}

\newfloatcommand{capbtabbox}{table}[][\FBwidth]

\newcommand{\Deltab}{\mathbf{\Delta}}
\newcommand{\tr}{\mathrm{Tr}}
\newcommand{\trans}{\mathrm{T}}

\begin{document}

\title{Neutrino masses, vacuum stability and \texorpdfstring{\\q} quantum gravity prediction for the mass of the top quark}

\author{\textsc{Guillem Dom\`enech$^{a}$}}
		\email{{domenech}@{thphys.uni-heidelberg.de}}
\author{\textsc{Mark Goodsell$^{b}$}}
		\email{{goodsell}@{lpthe.jussieu.fr}}
\author{\textsc{Christof Wetterich$^{a}$}}
		\email{{c.wetterich}@{thphys.uni-heidelberg.de}}

\affiliation{$^{a}$\small{Institut f\"ur Theoretische Physik, Ruprecht-Karls-Universit\"at Heidelberg, Philosophenweg 16, 69120 Heidelberg, Germany}\\
          $^{b}$\small{Laboratoire de Physique Th{\'e}orique et Hautes Energies (LPTHE), UMR 7589, Sorbonne Universit{\'e} et CNRS, 4 place Jussieu, 75252 Paris Cedex 05, France.}
		}

\begin{abstract}
A general prediction from asymptotically safe quantum gravity is the approximate vanishing of all quartic scalar couplings at the UV fixed point beyond the Planck scale. A vanishing Higgs doublet quartic coupling near the Planck scale translates into a prediction for the ratio between the mass of the Higgs boson $M_H$ and the top quark $M_t$. If only the standard model particles contribute to the running of couplings below the Planck mass, the observed $M_H\sim125\,{\rm GeV}$ results in the prediction for the top quark mass $M_t\sim 171\,{\rm GeV}$, in agreement with recent measurements. In this work, we study how the asymptotic safety prediction for the top quark mass is affected by possible physics at an intermediate scale. We investigate the effect of an $SU(2)$ triplet scalar and right-handed neutrinos, needed to explain the tiny mass of left-handed neutrinos. For pure seesaw II, with no or very heavy right handed neutrinos, the top mass can increase to $M_t\sim 172.5\,{\rm GeV}$ for a triplet mass of $M_\Delta\sim 10^8{\rm GeV}$. Right handed neutrino masses at an intermediate scale increase the uncertainty of the predictions of $M_t$ due to unknown Yukawa couplings of the right-handed neutrinos and a cubic interaction in the scalar potential. For an appropriate range of Yukawa couplings there is no longer an issue of vacuum stability.
\end{abstract}
 \maketitle

\section{Introduction \label{sec:introduction}}

The standard model (SM) of particle physics has to be extended at least to accommodate the mass of the left-handed neutrinos \cite{Fukuda:1998mi,Ahn:2002up,Eguchi:2002dm,Tanabashi:2018oca}. Simple extensions involve adding three right-handed neutrinos, a SU(2) triplet scalar or triplet fermions (or a combination of them). These new fields give rise to the tiny mass of left-handed neutrinos by means of the so-called seesaw mechanism, respectively referred to as type I, II and III (e.g. see Refs.~\cite{Xing:2011zza,Xing:2014wwa} for a review and references therein). In this work, we will be interested in the high scale seesaw type I+II \cite{Magg:1980ut,Schechter:1980gr,Lazarides:1980nt} with masses $\gtrsim 10^8{\rm GeV}$. Despite being inaccessible at particle colliders, a high scale seesaw has very interesting applications in particle physics and cosmology. For instance, it could account for the baryon asymmetry of the universe through leptogenesis \cite{Ma:1998dx,Chakraborty:2019uxk,Parida:2020sng} (for a review see Refs.~\cite{Davidson:2008bu,Fong:2013wr}), constitute a dark matter candidate \cite{Anisimov:2008gg,Arina:2011cu,Arina:2012jp,Dev:2016qbd} and play a role for inflation \cite{Chen:2010uc,Arina:2012fb,Rodrigues:2020dod} or dark energy \cite{Wetterich:2007kr}.

There is also the issue of the SM vacuum stability \cite{Sher:1988mj,Schrempp:1996fb} which may require beyond SM physics (see Ref.~\cite{Markkanen:2018pdo} for a recent review). The running of the SM couplings is perturbatively valid up to the Planck scale, where most likely quantum gravity effects start to become relevant. However, the discovery of the Higgs boson \cite{Higgs:1964pj,Englert:1964et} placed the electroweak (EW) vacuum in a (or close to) a metastable region \cite{ATLAS:2014wva,Aad:2015zhl}. In other words, given previous measurements of the Higgs and top quark mass, with mean values respectively around $M_H\sim 125\,{\rm GeV}$ and $M_t\sim 173\,{\rm GeV}$ \cite{Aaboud:2018zbu,Sirunyan:2018gqx}, the Higgs quartic coupling becomes negative before reaching the Planck scale, roughly around a critical scale of $10^{10}\,{\rm GeV}$ \cite{Casas:1994qy,Espinosa:2007qp,EliasMiro:2011aa,Degrassi:2012ry,Bezrukov:2012sa,Buttazzo:2013uya,Branchina:2014usa,DiLuzio:2014bua,Bezrukov:2014ina,Bezrukov:2014ipa,Zoller:2014cka,George:2015nza,Espinosa:2016nld}. If there is no new physics before the critical scale, the SM is in a potential conflict with inflation \cite{Brout:1977ix,Starobinsky:1980te,Guth:1980zm,Sato:1980yn}, where vacuum fluctuations are of the order of the horizon scale during inflation and the probability of ending at the SM vacuum would be very low. Furthermore, even if the scale of inflation is low enough to circumvent this issue, the presence of a primordial black hole would catalyze the decay to the stable vacuum \cite{Burda:2015isa,Burda:2016mou,Cuspinera:2018woe,Cuspinera:2019jwt}. One possible interesting solution is that the vacuum can be stabilised by the inclusion of a non-minimal coupling of the Higgs field to gravity \cite{Barvinsky:2009ii,Barvinsky:2009fy,Gong:2015gxf,DiVita:2015bha,Herranen:2015ima,Espinosa:2015qea,Enqvist:2016mqj,Kohri:2016wof,Kohri:2017iyl,Espinosa:2018mfn,Markkanen:2018bfx,Han:2018yrk,Branchina:2019tyy,Cai:2020ndh} (for a review within Higgs inflation see Refs.~\cite{Rubio:2018ogq,Rubio:2020zht}), well motivated by quantum corrections of the SM in curved backgrounds \cite{Birrell:1982ix}.

The above discussion points already to an important role of quantum gravity for the stability of the SM vacuum. Still, there remains the question of the UV completion of gravity. In this respect, an interesting possibility is that gravity might be asymptotically safe non-perturbatively \cite{Weinberg:1976xy,Irvine_1980,Reuter:1996cp}. In other words, the non-perturbative renormalisation group equations (RGEs) including gravity might present an interacting fixed point beyond the Planck scale. While not demonstrated in general,\footnote{One usually needs to truncate the expansion of the effective action in terms of curvature invariants. The presence of a fixed point might depend on the matter content. Consequences of using a regulator in the averaged effective action that breaks diffeomorphism invariance as well as analytic continuation to Minkowski space and the shape of a consistent graviton propagator remain debated \cite{Niedermaier:2006wt,Wetterich:2019qzx,Eichhorn:2019tcj,Eichhorn:2020mte,Reichert:2020mja,Bonanno:2020bil}.} there is mounting evidence for the presence of such an interacting fixed point \cite{Reuter:1996cp,Dou:1997fg,Souma:1999at,Reuter:2001ag,Lauscher:2001ya,Machado:2007ea,Codello:2007bd,Codello:2008vh}, derived using the functional renormalisation group approach \cite{Wetterich:1992yh,Reuter:1993kw} (for reviews see Refs.~\cite{Niedermaier:2006wt,Wetterich:2019qzx,Eichhorn:2019tcj,Eichhorn:2020mte,Reichert:2020mja,Bonanno:2020bil} and references therein). A direct consequence of a fixed point is quantum scale invariance, which can explain the near scale invariance of the primordial fluctuations spectrum generated during inflation \cite{Wetterich:2019qzx} (e.g. see Refs.~\cite{GarciaBellido:2011de,Ferreira:2016vsc,Rubio:2017gty} for implementations of scale symmetry to inflation).

A rather robust prediction of asymptotic safe gravity is that all quartic couplings are almost zero at the fixed point.
This is the basis of the asymptotic safety prediction \cite{Shaposhnikov:2009pv} of the mass of the Higgs boson to be $126\,{\rm GeV}$ with a few GeV uncertainty, in agreement with later observations. More generally, irrelevant parameters at the quantum gravity fixed point translate into predictions for renormalisable parameters near the Planck scale. For a given particle physics model below the Planck scale the perturbative renormalisation group running translates this to observable quantities at low energy scales \cite{Shaposhnikov:2009pv,Eichhorn:2017ylw,Eichhorn:2017lry,Eichhorn:2019dhg}. The asymptotic safety scenario predicts that the Higgs boson is found at the limit where metastability of the vacuum sets in. There is therefore no problem of ``vacuum stability'' in this scenario. This is in line with new measurements of the top quark mass. With a precise measurement of the mass of the Higgs boson \cite{Aad:2012tfa,Chatrchyan:2012ufa,Tanabashi:2018oca} to be 
\begin{align}
M_H=125.18\pm0.16\,{\rm GeV}\,,
\end{align}
the result of Ref.~\cite{Shaposhnikov:2009pv} turns into a prediction of the top quark mass around $M_t\sim 171\,{\rm GeV}$ \cite{Wetterich:2019qzx}. This result is in very good agreement with the latest measurements of the top pole mass \cite{Sirunyan:2019zvx,Aad:2019mkw}, where we respectively have that
\begin{align}
M_t=170.5\pm0.8\,{\rm GeV}\quad{\rm and}\quad M_t=171.1^{+2.0}_{-1.6}\,{\rm GeV}\,.
\end{align}

An essential ingredient of asymptotic safety predictions for observable parameters of the standard model is the ``great desert''. This assumes that the running of couplings for scales sufficiently below the Planck scale follows the perturbative running with the particle content of the Standard Model. There may exist, however, some ``oasis in the desert''--some intermediate scale above which particles beyond the standard model play a role. These particles influence the running of the couplings between the Planck scale and the intermediate scale, and therefore modify the predictions at the Fermi scale. The size of such modifications is the topic of this paper. We focus on the ``neutrino oasis''. The generation of neutrino masses indeed suggests an intermediate scale where the symmetry of baryon-lepton number (B-L) conservation is broken \cite{Wetterich:1981bx}. Early estimates within a SO(10) GUT situate this scale around $10^{12}\,{\rm GeV}$.

In this work, we use the prediction from asymptotically safe quantum gravity that all quartic scalar couplings vanish near the Planck scale as a boundary condition for the perturbative RGE's below the Planck scale. This is a central ingredient for our results since it constrains not only the quartic coupling of the Higgs doublet, but also all quartic couplings of the Higgs triplet in the seesaw II mechanism. The contribution of the gravitational fluctuations to the flow equations for quartic couplings $\lambda$ is an universal anomalous dimension, 
\begin{align}\label{eq:lambdaRGERGE}
\partial_t\lambda=A\lambda+(...)\,.
\end{align}
The anomalous dimension $A$ does not depend on properties of scalar fields, it only involves couplings in the gravitational sector, as the effective dimensionless Planck mass or cosmological constant. For $A>0$ and $A$ not too small, every quartic coupling is driven towards a fixed point at a small value, $\lambda_*\approx 0$. This holds provided that the contributions from other couplings, that are denoted by the dots in Eq.~\eqref{eq:lambdaRGERGE}, are small enough. This is indeed the case for the contributions of gauge and Yukawa couplings. A possible non-minimal coupling, say $\xi$, of scalars to the curvature tensor, e.g. $\xi\varphi^2 R$, yields a contribution proportional to $\xi^2$ which is also small for small $\xi$ \cite{Wetterich:2019qzx}. For the Higgs doublet a first estimate yields for $\xi$ a few times $10^{-3}$ or smaller \cite{Pawlowski:2018ixd}. We assume here that similar bounds on $\xi$ hold for all scalar fields.

All investigations find indeed a substantial positive value of $A$  for the momentum region above the Planck scale where the gravitational fluctuations matter. For recent investigations see Refs.~\cite{Pawlowski:2018ixd,Wetterich:2019qzx,Wetterich:2019rsn} and references therein. On this basis we assume in this paper that all quartic scalar couplings at the Planck scale are very small and can be approximated by zero. The quartic couplings will flow away from the fixed point only for momenta below the Planck scale where the gravitational fluctuations decouple effectively. For the flow below the Planck scale we can take for all quartic scalar couplings the fixed point value $\lambda_*=0$ as a good approximation for an ``initial value'' at the (reduced) Planck mass $M_{\rm pl}$.

Given this asymptotic safety prediction for the ``initial values'' of quartic couplings we investigate seesaw, and whether asymptotic safety is compatible with vacuum stability. At first glance and only focusing on the sign of the contribution to the flow of the Higgs quartic coupling, the answer to the latter is most likely positive. Indeed, the presence of an extra scalar, the SU(2) triplet, gives a positive contribution to the $\beta$-function of the Higgs quartic coupling and stabilises the vacuum if the triplet mass is close to or below the SM critical scale \cite{Gogoladze:2008gf,Arina:2012fb,Dev:2013ff,Kobakhidze:2013pya,Haba:2016zbu,Chakraborty:2019uxk,Parida:2020sng}. However, the question becomes non-trivial when we add the input from asymptotic safety. The joint doublet-triplet potential has 5 quartic couplings, 1 cubic coupling and 2 mass parameters. It is not obvious that the requirement that all 5 quartic couplings vanish at the Planck scale yields a stable potential. In fact, we find that the condition of vacuum stability requires the presence of the right-handed neutrinos and places lower and upper bounds on the value of the neutrino Yukawa couplings.

\begin{figure}
\begin{center}
\includegraphics[width=0.49\columnwidth]{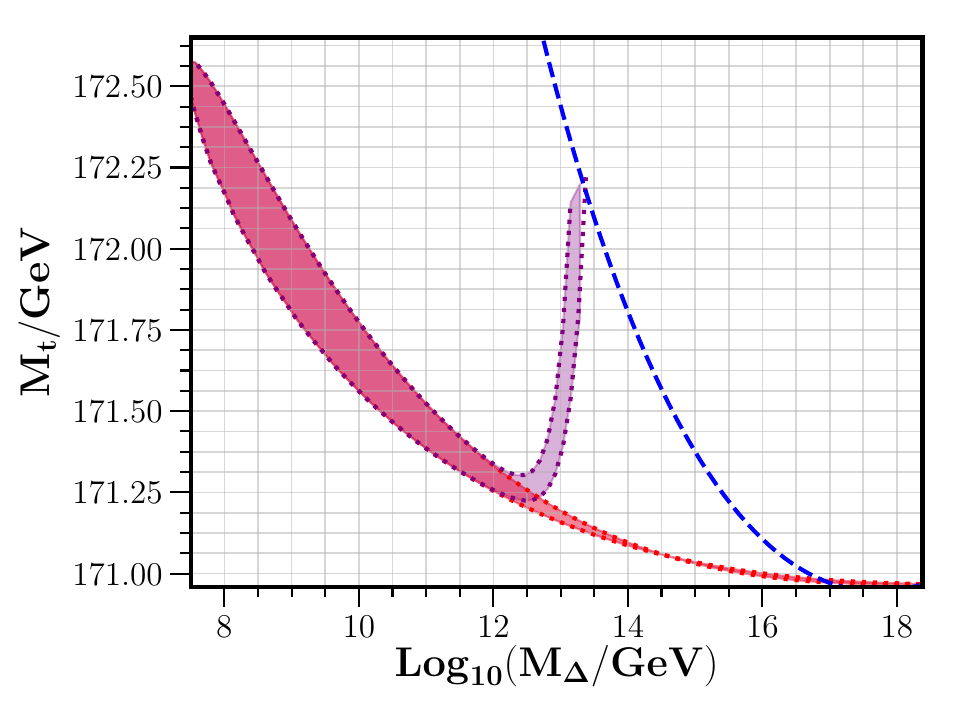}
\includegraphics[width=0.49\columnwidth]{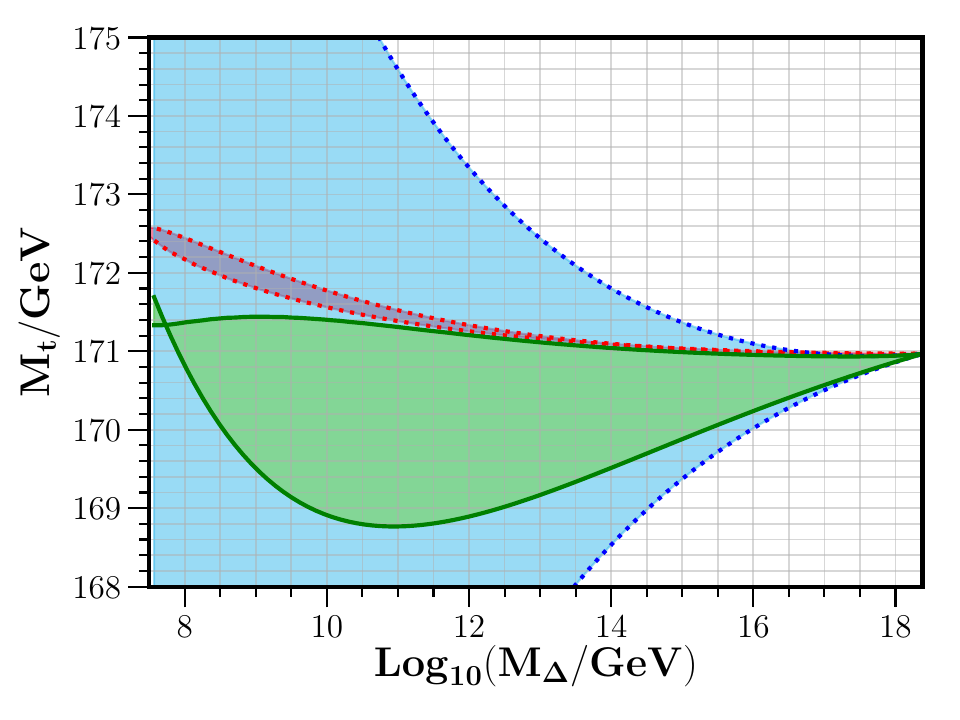}
\caption{Predicted range in the top quark mass from asymptotically safe quantum gravity assuming a Higgs mass of $M_H=125.15\,{\rm GeV}$. The predicted value of the top quark mass in the SM, $M_t\sim 171\,{\rm GeV}$, corresponds to the limiting case $M_\Delta=M_{\rm pl}$. Left: In red we see the predicted bounds derived considering the $SU(2)$ triplet alone and vanishing cubic coupling. In purple we present the bounds with the requirement that the mass of the light neutrinos mainly comes from the type II seesaw. The latter case implies bounds $M_t\gtrsim 171.3\,{\rm GeV}$ and $M_\Delta\lesssim 5\times 10^{13}\,{\rm GeV}$. In blue we show the maximum value for the top mass using the maximum value of the cubic coupling allowed by vacuum stability. Right: The vacuum stability of the full Higgs double-triplet potential requires the presence of right-handed singlet neutrinos. In red we show again the bounds using the triplet alone and vanishing cubic coupling. In blue we illustrate how the bounds change if the neutrino-Higgs Yukawa coupling $Y_\Phi$ is equal to the top-Higgs Yukawa coupling at the intermediate scale. The blue lower bound corresponds to a vanishing cubic coupling at the triplet scale. This bound depends on the value of $Y_\Phi$, with smaller $Y_\Phi$ pushing the bound upwards. A shift of the bound in the same direction occurs if the right-handed neutrino masses are larger than the triplet mass. The blue upper bound corresponds to the maximum value that the cubic coupling can attain, for a vanishing quartic doublet coupling at the triplet scale $M_\Delta$. The solid green curves obtain for a vanishing cubic coupling at the Planck scale, for two values of the doublet-neutrino coupling $Y_\Phi$ and right handed neutrino mass $M_\nu=M_\Delta$. For larger $M_\nu$ the allowed interval in between the curves shrinks.
\label{fig:mt}}
\end{center}
\end{figure}

Our main results are summarised in Fig.~\ref{fig:mt}. First, we find that the values of the Yukawa couplings of the light left-handed neutrinos to the scalar triplet are bounded by requiring vacuum stability of the doublet-triplet potential. Using these bounds, we obtain that the effect of the $SU(2)$ triplet is to raise the prediction from asymptotic safe gravity up to $1\%$ compared to the SM. This means that the top mass predicted in the SM, which corresponds to $M_t\sim 171\,{\rm GeV}$ and the limiting case $M_\Delta=M_{\rm pl}$ in Fig.~\ref{fig:mt}, can be as large as $M_t\sim172.5\,{\rm GeV}$ for $M_{\Delta}\sim 10^8\,{\rm GeV}$ assuming a negligible cubic interaction with the Higgs. Furthermore, if we require that the mass of the light neutrinos is solely due to a type-II seesaw mechanism, we find that the top mass prediction is within $172.5\,{\rm GeV}\gtrsim M_t\gtrsim 171.3\,{\rm GeV}$ and the triplet mass is bounded from above as $M_{\Delta}<5\times 10^{13}\,{\rm GeV}$. Including right-handed neutrinos and allowing for the maximum value of the cubic coupling the uncertainty in the prediction is $O(10\,{\rm GeV})$. In that case, experimental data constrains the right-handed neutrino sector. Further theoretical input on the size of the cubic coupling and the Yukawa coupling, for example from left-right symmetry at the intermediate scale, can narrow down the allowed range for the top quark mass.

A rather likely value for the cubic coupling at the Planck scale is $\gamma(M_{\rm pl})=0$. This obtains in case of left-right symmetry where $\gamma$ arises from a quartic coupling multiplied with the expectation value breaking this symmetry. Also in the general case a large enough gravity induced anomalous dimension will imply a fixed point value of $\gamma$ close to zero. We show in Fig.~\ref{fig:mt} two curves obtained with the boundary value $\gamma(M_{\rm pl})=0$, one for a doublet-neutrino coupling $Y_\Phi$ equal to the top Yukawa coupling $y_t$, the other for the minimal value of $Y_\Phi$ which ensures vacuum stability. The boundary value $\gamma(M_{\rm pl})=0$ narrows down the uncertainty considerably. The two curves are obtained for right-handed neutrino masses $M_\nu$ equal to the triplet mass $M_\Delta$. The predicted range narrows down further if $M_\nu$ is larger than $M_\Delta$, which seems rather likely.

This paper is organised as follows. In Sec.~\ref{sec:neutrinomasses} we review the type I+II seesaw mechanism. In Sec.~\ref{sec:vacuumstability} we derive upper and lower bounds on the neutrino Yukawa couplings by requiring vacuum stability within asymptotic safety. In Sec.~\ref{sec:predictions} we study the quantum gravity predictions for the top quark mass in terms of the triplet mass. We conclude our work in Sec.~\ref{sec:conclusions}. We also derive the general two loop RGEs for the type I+II seesaw. They are presented in App.~\ref{app:RGE2loop} which generalises the results of Ref.~\cite{Schmidt:2007nq}.

\section{Neutrino masses and the SU(2) Triplet scalar\label{sec:neutrinomasses}}

Let us start by reviewing the type I+II seesaw mechanism, which in addition to the SM fields includes a SU(2) triplet scalar with hypercharge Y=1 and a Yukawa interaction with two left-handed doublet leptons, as well as three (right-handed) neutrino singlets of SU(2) with Majorana mass. The Lagrangian may be written\footnote{We follow the notation of Ref.~\cite{Schmidt:2007nq} except for slight changes: $\lambda\to 2\lambda_3$, $\Lambda_6\to \sqrt{2}\gamma$, $Y_\Delta\to Y_\Delta/\sqrt{2}$ and $\Lambda_i\to \lambda_i$ with $i=\{1,2,4,5\}$.} as \cite{Magg:1980ut,Schmidt:2007nq}
\begin{align}\label{eq:LSM}
{\cal L}={\cal L}_{\rm SM}+{\cal L}_{\Delta}+{\cal L}_{\nu_R}+{\cal L}_{\rm Yukawa}\,,
\end{align}
where ${\cal L}_{\rm SM}$ is the SM part including the top quark and the standard potential for Higgs doublet, explicitly given by
\begin{align}\label{eq:VSM}
{\cal L}_{\rm SM}=-\left(D^\mu\Phi\right)^\dag D_\mu\Phi-V_{\rm SM}-y_t \overline{t_R}\Phi t_L+{\rm h.c.}\quad,\quad V_{\rm SM}(\Phi)=-m^2\Phi^\dag\Phi+\frac{\lambda_3}{2}\left(\Phi^\dag\Phi\right)^2\,,
\end{align}
and ${\cal L}_{\Delta}$, ${\cal L}_{\nu}$ and ${\cal L}_{\rm Yukawa}$ contain the new physics of the type I+II seesaw model. The Lagrangian describing the triplet, represented by a traceless $2\times2$ complex matrix, and its interaction with the doublet is given by 
\begin{eqnarray}\label{eq:LD}
{\cal L}_{\Delta}=-{\rm Tr}\left[\left(D_\mu\Deltab\right)^\dag D_\mu\Deltab\right]-V\left(\Phi,\Deltab\right)\,.
\end{eqnarray}
Her $D_\mu$ is the covariant derivative,\footnote{One could also work in the adjoint representation of SU(2) where the triplet is a three component complex vector transforming according to the generators of SO(3). In that case ${\rm Tr}[\left(D_\mu\Deltab\right)^\dag D^\mu\Deltab]=2\delta_{ij}\left(D_\mu\Delta^i\right)^\dag D^\mu\Delta^j$, where we have used that $\Deltab\equiv \Delta^i\sigma_i$ and ${\rm Tr}\left[\sigma_i\sigma_j\right]=2\delta_{ij}$. In the adjoint representation we have that $D_\mu\equiv\partial_\mu-i\frac{g}{2}t^aW_\mu^a-i\frac{g'}{2}B_\mu$, where $t^a$ are the generators of SO(3).} namely
\begin{align}
D_\mu\Deltab=\partial_\mu\Deltab-i{g_2}\left[\sigma^aW_\mu^a,\Deltab\right]-i\sqrt{\frac{3}{5}}g_1B_\mu\Deltab\,,
\end{align}
and we use the GUT normalisation for $g_1$ for easier comparison with the literature. The potential is given by
\begin{align}\label{eq:VD}
V\left(\Phi,\Delta\right)=M^2_\Delta{\rm Tr}\left(\Deltab^\dag\Deltab\right)&+\frac{\lambda_1}{2}{\rm Tr}^2\left(\Deltab^\dag\Deltab\right)
+\frac{\lambda_2}{2}\left\{{\rm Tr}^2\left(\Deltab^\dag\Deltab\right)-{\rm Tr}\left(\Deltab^\dag\Deltab\Deltab^\dag\Deltab\right)\right\}\nonumber\\&
+\lambda_4\Phi^\dag\Phi {\rm Tr}\left(\Deltab^\dag\Deltab\right)+\lambda_5\Phi^\dag \left[\Deltab^\dag,\Deltab\right]\Phi+{\gamma}\tilde\Phi^\dag\Deltab^\dag\Phi+{\rm h.c.}\,,
\end{align}
where $[,]$ refers to the commutator and ${\rm h.c.}$ to the hermitian conjugate. The part for the right-handed neutrino consists of the kinetic term and a majorana mass $M_{\nu}$,
\begin{align}\label{eq:NUR}
{\cal L}_{\nu_R}=-i\,\overline{\nu_R}\slashed{\partial}\nu_R-\frac{1}{2}M_\nu{\nu_R}C\nu_R\,.
\end{align}
In this work we will assume that the mass of the heavy neutrinos $M_\nu$ is either of the order of the Planck mass (pure seesaw II), or close to $M_\Delta$, reflecting a single intermediate scale of $B-L$ violation. The Yukawa interactions between the doublet, triplet and neutrinos are described by
\begin{align}\label{eq:Yukawa}
{\cal L}_{\rm Yukawa}=-Y_{\Phi}\bar{L}_L\tilde\Phi \nu_R -\frac{Y_{\Delta}}{\sqrt{2}} \overline{L^c_L} {\Deltab}L_L+{\rm h.c.}\,,
\end{align}
where $L_L$ is the lepton doublet and $L^c_L=i\sigma_2C\bar L^T_L$ and $\tilde\Phi=i\sigma_2\Phi^*$ are respectively the charge conjugates of $L_L$ and $\Phi$. Note that $C=i\gamma^0\gamma^2$ is the charge conjugation operator of the Lorentz group and $i\sigma_2$ is the charge conjugation operator of the SU(2) gauge group. To be more precise, there should be a sum over the three lepton generations and the Yukawa couplings would be a $3\times 3$ matrix. For simplicity, we consider in the main text only three generations of singlet neutrinos with equal mass and assume that only one has a non-vanishing Yukawa coupling. Nevertheless, the RGEs given in App.~\ref{app:RGE2loop} are completely general. The derived bounds for the top quark mass are indicative for the general case, but the detailed quantitative values will be influenced by assumptions on the generation structure.

After a SU(2) spontaneous symmetry breaking (SSB), where the neutral components of both the doublet and the triplet develop a non-zero vacuum expectation value (vev), the light neutrinos acquire a mass. In the unitary gauge  we  express the doublet and triplet expectation values as
\begin{eqnarray}\label{eq:vev}
\Phi_0=
\begin{pmatrix}
0\\ v
\end{pmatrix}\qquad{\rm and}\qquad \Deltab_0=
\begin{pmatrix}
0 & 0 \\
v_\Delta & 0
\end{pmatrix}
\,.
\end{eqnarray}
A strong constraint on the vev of the triplet comes from the contribution to the mass of the $W^{\pm}$ and $Z$ bosons which breaks custodial symmetry and leads to
\begin{align}
\rho=\frac{M_W^2}{M_Z^2\cos^2\theta_W}=\frac{1+2v_\Delta^2/v^2}{1+4v_\Delta^2/v^2}\,,
\end{align}
where $\theta_W$ is the Weinberg angle. The latest constraint on $\rho$ reads \cite{Tanabashi:2018oca}
\begin{align}
\rho=1.00039\pm0.00019\,,
\end{align}
which essentially requires $v_\Delta\ll v$, more precisely $v_\Delta/v<10^{-2}$. For $v_\Delta\ll v$ we can take $v\approx174.08\,{\rm GeV}$ as in the SM. Corrections due to the triplet are tiny. 

Now we turn to the mass of the neutrinos and the seesaw mechanism. While the triplet vev yields a direct Majorana mass for the left-handed neutrinos (seesaw type II), the right-handed neutrinos mix with the left-handed ones through the Higgs vev and contribute a Dirac mass term (seesaw type I). These two contributions can be summarised in a mass matrix for neutrinos given by \cite{Wetterich:1981bx}
\begin{align}
\mathbf{m}_\nu=
\begin{pmatrix}
\sqrt{2} Y_\Delta v_\Delta & Y_\Phi v \\
Y_\Phi v & M_\nu
\end{pmatrix}
\,,
\end{align}
where each entry is a $3\times3$ matrix. From now on we use the approximation of a single leading neutrino, where we assume that there is a neutrino with $m_\nu\sim0.08\,{\rm eV}$ and the two others can be regarded almost massless. In similar lines we employ the approximation where only one entry in the Yukawa coupling matrix is non-vanishing and the right-handed neutrinos have equal mass.
After diagonalisation of the neutrino mass matrix, the mass eigenvalues read
\begin{align}
m_{\nu,\rm light/heavy}=\frac{1}{2}\left\{\sqrt{2} Y_{\Delta} v_\Delta+M_\nu\mp\sqrt{4Y_{\Phi}^2v^2+\left(M_\nu-\sqrt{2}Y_{\Delta}v_\Delta\right)^2}\right\}\,.
\end{align}
Using the constraint on the departure of custodial symmetry, i.e. $v_\Delta\ll v$, and that we are dealing with a high scale seesaw, that is $M_\nu\sim M_\Delta\gg v$, we see that: $(i)$ the heavy neutrinos have a mass $m_{\nu,\rm heavy}\approx M_\nu$ and $(ii)$ the light neutrinos develop a non-zero mass proportional to the Yukawa couplings and suppressed by the mass of the heavy fields, explicitly
\begin{align}
m_{\nu,\rm light}\approx \left|\sqrt{2}Y_{\Delta}v_\Delta-\frac{Y_{\Phi}^2v^2}{M_\nu}\right|\,.
\end{align}
Furthermore, in the approximation where $M_\Delta\gg v$ (with $\lambda_4$ and $\lambda_5$ not excessively large) there is a simple relation between the vev of the doublet and the triplet and the parameters in the potential \eqref{eq:VD}. The joint potential \eqref{eq:VSM} and \eqref{eq:VD} after the SSB reads
\begin{align}\label{eq:potentialafterSSB2}
V=V_{SM}(\Phi)+V(\Phi,\Delta)&=-m^2 v^2+M_\Delta^2 v_\Delta^2 + \frac{\lambda_3}{2}v^4 + \frac{\lambda_1}{2}v_\Delta^4+ 
 v_\Delta v^2 \left( v_\Delta (\lambda_4-\lambda_5)- 2\gamma\right)\,.
 \end{align}
The approximation $M^2_\Delta\gg v^2(\lambda_4-\lambda_5)$ yields the minimum at
\begin{align}\label{eq:vevs}
\frac{v_\Delta}{v}\approx\frac{\gamma}{M_\Delta}\frac{v}{M_\Delta}\quad {\rm and}\quad v\approx\frac{M_H}{\sqrt{2\lambda}}\,.
\end{align}
Here we have defined the effective Higgs quartic coupling below the intermediate scale when the triplet decouples,
\begin{align}\label{eq:lambdaSML6}
\lambda\equiv\lambda_3-2\frac{\gamma^2}{M_\Delta^2}\,.
\end{align}
Indeed, we may evaluate the potential at the relative minimum in the triplet direction, i.e. $\partial V/\partial v_\Delta=0$,
\begin{align}\label{eq:potentialafterSSB}
V\approx -m^2 v^2+\frac{1}{2}v^4\left(\lambda_3-2\frac{\gamma^2}{M_\Delta^2}\right)=-m^2 v^2+\frac{\lambda}{2}v^4\,.
 \end{align}
The value of $\lambda$ from \eqref{eq:lambdaSML6} is the one that should be used when running the SM couplings from $M_\Delta$ down to the Fermi scale. In this respect, we readily have an upper bound on $\gamma^2$ by requiring vacuum stability below the intermediate scale. Imposing that $\lambda(M_\Delta)>0$ in Eq.~\ref{eq:lambdaSML6} and assuming that there is vacuum stability above $M_\Delta$, i.e. $\lambda_3(M_\Delta)>0$ (we explore all the conditions in Sec.~\ref{subsec:vacuumstability}) we find
\begin{align}\label{eq:maximumgamma}
\left|{\gamma}\right|<\gamma_{\rm max}\equiv{M_\Delta}\sqrt{\frac{\lambda_3}{2}}\,.
\end{align}

Before proceeding to the embedding of this model into asymptotic safety, let us have a rough idea of the quantitative values needed for light neutrino masses. Neglecting possible cancellations between the type I and type II seesaw contributions we have two different limiting cases: $(i)$ the triplet dominates the seesaw or $(ii)$ the heavy neutrinos dominate the seesaw. On one hand, using that $m_{\nu,\rm light}\approx 6\times10^{-2}\,{\rm eV}$, we see that $(i)$ leads us to
\begin{align}\label{eq:mnutriplet}
\frac{\gamma}{M_\Delta}\approx 1.4\times 10^{-3} \,Y_\Delta^{-1} \left(\frac{M_\Delta}{10^{12}\,\rm GeV}\right)\,,
\end{align}
which implies that for $Y_\Delta\sim O(1)$ we need $\gamma/M_\Delta\sim 10^{-3}$ for $M_\Delta\sim10^{12}\rm GeV$, and we saturate $\gamma/M_\Delta\sim 1$ for $M_\Delta\sim 5\times 10^{14}\rm GeV$. We can also turn Eq.~\eqref{eq:mnutriplet} into a constraint on the value of $M_\Delta$. For instance, using the constraints on $Y_\Delta$ and $\gamma$ from vacuum stability to be derived in Sec.~\ref{sec:vacuumstability}, roughly given by $Y_\Delta<3$ and $|\gamma|<0.2$, we conclude that $M_\Delta<3\times 10^{15}\,{\rm GeV}$. On the other hand, we find that $(ii)$ requires
\begin{align}\label{eq:mnuneutrino}
Y_\Phi\approx 4\times 10^{-2}\left(\frac{M_\nu}{10^{12}\,\rm GeV}\right)^{1/2}\,.
\end{align}
In this case, we need $Y_\Phi\sim O(1)$ for $M_\nu\sim 10^{16}\,{\rm GeV}$. These order of magnitude estimates will prove useful at the end of the Sec.~\ref{sec:predictions} when we study the predictions from asymptotic safety.

\begin{figure}
\begin{center}
\includegraphics[width=0.5\columnwidth]{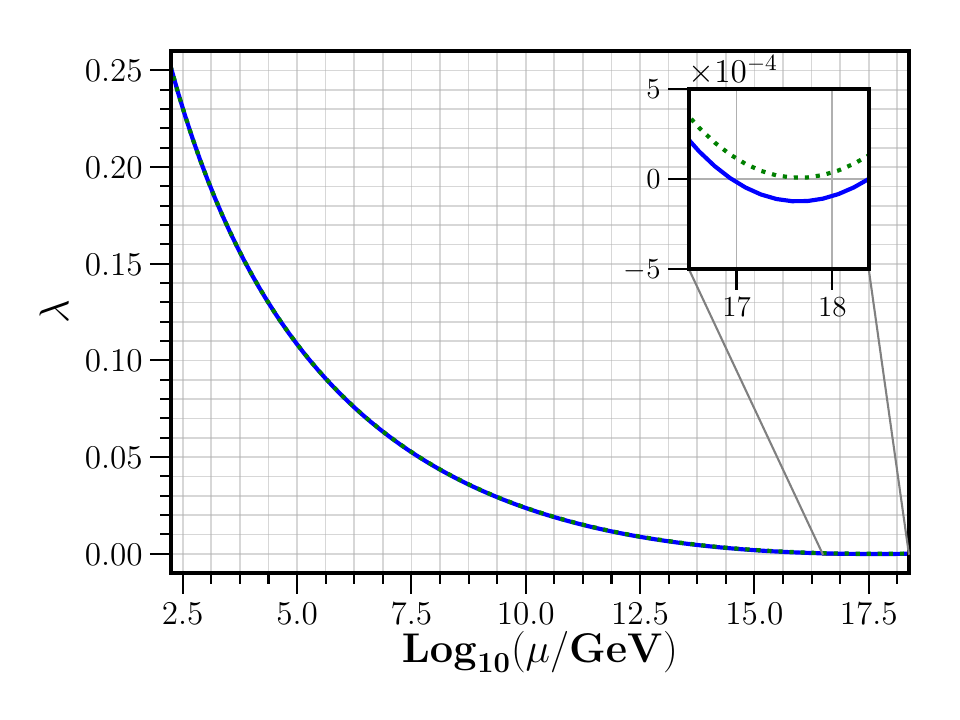}
\caption{Running of the Higgs quartic coupling with the logarithm of the renormalisation scale $\mu$ in the SM using the 2-loop RGEs. We used the values given in Eq.~\eqref{eq:valuesSM} for $M_t=170.97\,{\rm GeV}$ (solid blue line) and $M_t=170.96 \,{\rm GeV}$ (dotted green line). The difference between requiring $\lambda(M_{\rm pl})=0$ or $\lambda\geq0$, respectively the solid blue and dotted green lines, only yields a tiny correction to $\lambda$ or $M_t$.\label{fig:lambdaSM}}
\end{center}
\end{figure}

\section{Vacuum stability \label{sec:vacuumstability}}

Vacuum stability for the Standard Model means that the expectation value of the Higgs doublet given by the Fermi scale corresponds to the absolute minimum of the quantum effective potential (for metastability it is a local minimum). We will discuss this here in a rather simple quartic polynomial approximation to the effective potential with running quartic couplings. In this approximation particular combinations of quartic couplings have to remain positive, such that there is no direction in the field space for which the potential can reach negative large values for large values of the fields. Concerning only the Higgs doublet, the quartic coupling has to be positive. In the spirit of a Coleman-Weinberg potential \cite{Coleman:1973jx} we require this to hold for every value of the renormalisation scale in the range of interest.

Negative directions in the quartic approximation are those for which the potential becomes negative for large fields. In general, a negative direction at scales in the vicinity of the Planck mass does not imply that vacuum stability is lost. For scales close to the Planck scale the effective potential is no longer well approximated by a polynomial \cite{Wetterich:2019rsn}--it typically approaches a constant for large field values. Only for scales sufficiently below the Planck scale, where the graviton fluctuations have already decoupled, a polynomial form becomes a reasonable approximation. Keeping this limitation in mind, we will nevertheless concentrate here on the quartic approximation and investigate the stability issue in terms of the flowing quartic couplings.

\subsection{Flow of couplings \label{sec:flow}}
In this section, we study the 2-loop RGEs with the boundary condition that all quartic couplings vanish at the Planck scale, inspired by asymptotic safety. We derived the 2-loop RGEs for the type I+II seesaw with \texttt{PyR@TE\ 3} \footnote{\url{https://github.com/LSartore/pyrate}} \cite{Lyonnet:2016xiz,Lyonnet:2013dna,Sartore:2020gou} and cross-checked our results using \texttt{SARAH}\footnote{\url{https://sarah.hepforge.org}} \cite{Staub:2008uz,Staub:2010jh,Staub:2012pb,Staub:2013tta,Goodsell:2018tti}. Our results agree with the 1-loop calculations of Ref.~\cite{Schmidt:2007nq} except for an additional term to $\lambda_4$ and $\lambda_5$ proportional to $\tr(Y_e^{\dag}Y_e Y_\Delta^{\dag} Y_\Delta)$ arising from a loop with a right-handed charged lepton and a difference in sign in front of $\lambda_4$ and $\lambda_5$ in the running of the cubic coupling $\gamma$. The 2-loop RGEs are a new result of this work and are presented in detail in App.~\ref{app:RGE2loop}. We also use the 2-loop RGEs for the standard model derived with \texttt{PyR@TE} which coincide with the results in the literature, e.g. in Ref.~\cite{Espinosa:2007qp}. One may use the 3-loop RGEs, e.g. from Ref.~\cite{Buttazzo:2013uya}, but that would only provide a small improvement \cite{Zoller:2014cka} which will not change significantly our results. We stress again that for simplicity we only consider in the main text the gauge bosons, the top quark, the Higgs doublet, the $SU(2)$ triplet and three singlet neutrinos with equal mass and one dominant neutrino yukawa coupling to the doublet and triplet.

As a starting point for the flow from the Planck to the electroweak scale we need the values of gauge and Yukawa couplings at the Planck scale. For this purpose we first run the couplings from $M_t$ up to the triplet scale $M_\Delta$, which we take as a free parameter with values up to the (reduced) Planck scale $M_{\rm pl}=1/\sqrt{8\pi G}\approx 2.4\times 10^{18}{\rm GeV}$. To do so, we use the 2-loop RGEs of the SM with boundary conditions at $M_t$ extracted from Ref.~\cite{Buttazzo:2013uya}, explicitly
\begin{align}\label{eq:valuesSM}
g_1=0.4626\quad,\quad g_2=0.6478 \quad,\quad g_3=1.1666\quad,\quad y_t=0.9369\quad{\rm and}\quad\lambda=0.2521\,,
\end{align}
which correspond to
\begin{align}\label{eq:valuesSM2}
M_t=170.97\,{\rm GeV}\quad,\quad M_H=125.15\,{\rm GeV}\quad,\quad \alpha_3(M_Z)=0.1184\quad {\rm and}\quad M_W=80.384\,{\rm GeV}\,.
\end{align}
In the above equations $y_t$ is the top Yukawa coupling with the Higgs and the QCD-fine structure constant $\alpha_3(M_Z)=g_3^2/(4\pi)$ is evaluated at the scale corresponding to the pole mass of the $Z$ boson. Small changes in $M_H$, $\alpha_3$ and $M_W$ will not change our main conclusion, which will be a prediction for the ratio $M_t/M_H$. In practice, we compute the relative change of the ratio $M_t/M_H$ due to an intermediate scale $M_\Delta<M_{\rm pl}$. This relative change can be read from the figures by comparing with the limit $M_\Delta=M_{\rm pl}$. The relative change is independent of many details of the precise computation of $M_t/M_H$. It is dominated by the one loop contribution. For a given set of particles and couplings at the Planck scale the relative change is a rather robust result. Let us emphasise that the choice of $M_t\approx171\,{\rm GeV}$ is the one that yields a $\lambda(M_{\rm pl})\approx0$ at 2-loop. If we used the 3-loop results we would have found $M_t\approx171.1\,{\rm GeV}$ \cite{Buttazzo:2013uya}. It should be noted that requiring $\lambda(M_{\rm pl})=0$ does not completely solve the issue of vacuum stability, see Fig.~\ref{fig:lambdaSM}. However, any tiny shift of the boundary conditions at the Planck scale, e.g. due to not being exactly at the fixed point or imposing the conditions not exactly at the reduced Planck scale, easily stabilises the vacuum.

Second, we run the couplings from a given $M_\Delta$ to $M_{\rm pl}$ with all new parameters equal to zero, except for the right-handed neutrino Yukawa coupling, as boundary conditions at $M_\Delta$. The reason for this choice is that we are only interested in extrapolating the values of SM-couplings up to the Planck scale. Taking into account the extra triplet scalar only modifies the beta functions of the gauge couplings at 1-loop to
\begin{align}\label{eq:gaugecouplings}
\beta_{g_1}=\frac{47 g_1^3}{10}
\quad,\quad
\beta_{g_2}=-\frac{5 g_2^3}{2}
\quad{\rm and}\quad
\beta_{g_3}=-7 g_3^3\,,
\end{align}
where the running of a parameter, say $X$, is defined as
\begin{align}
\frac{dX}{d\ln\mu}=\frac{1}{16\pi^2}\beta_X\,.
\end{align}
The right handed neutrinos are singlets with respect to the SM-gauge group and do not modify the flow of the gauge couplings.
The beta function for the top Yukawa coupling is left almost unchanged with respect to the SM and it is given by
\begin{align}
\beta_{y_t}=y_t \left(-\frac{17 g_1^2}{20}-\frac{9 g_2^2}{4}-8
   g_3^2+Y_\Phi^2+\frac{9 }{2}y_t^2\right)\,,
\end{align}
where the only difference is an additional interaction with right-handed neutrinos through the Higgs. For investigations of the pure seesaw II mechanism we set $Y_\Phi=0$. This procedure yields the initial values for $g_1$, $g_2$, $g_3$ and $y_t$ at $\mu=M_{\rm pl}$. The neutrino Yukawa couplings have a beta function given by
\begin{align}\label{eq:neutrinoYukawa}
\beta_{Y_\Delta}=Y_\Delta \left(-\frac{9}{10} g_1^2-\frac{9}{2}g_2^2+Y_\Phi^2+8 Y_\Delta^2\right)
\quad{\rm and}\quad
\beta_{Y_\Phi}=Y_\Phi \left(-\frac{9}{20}g_1^2-\frac{9}{4}g_2^2+3Y_\Delta^2+3 y_t^2+\frac{5 }{2}Y_\Phi^2\right)\,.
\end{align}
When $Y_\Phi\gtrsim0.1$ we perform a few iterations to find the values of $Y_\Phi(M_{\Delta})$ and the corresponding values of $Y_\Phi(M_{\rm pl})$ that maintain the value of the top Yukawa coupling $y_t(M_\Delta)$. This procedure will ensure that the asymptotic safety predictions agree with the well measured values of the gauge couplings and the top Yukawa coupling at the EW scale.

\begin{figure}
\begin{center}
\includegraphics[width=0.5\columnwidth]{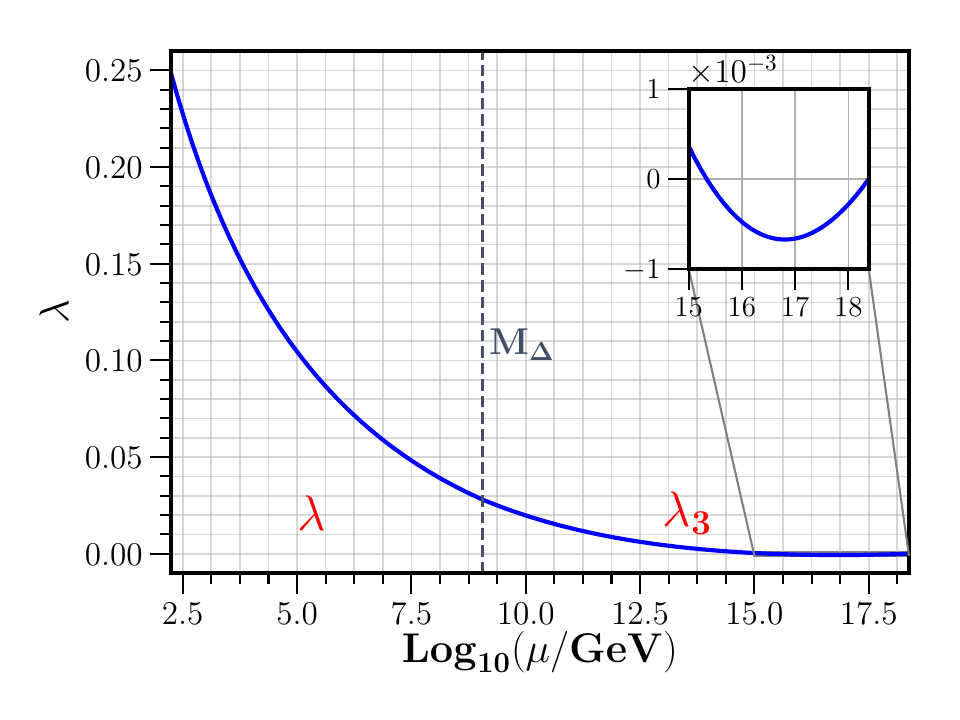}
\caption{Running of the Higgs quartic coupling with the logarithm of the scale in the type II seesaw using the 2-loop RGEs. We used the values given in Eq.~\eqref{eq:valuesSM} for $M_t=170.97\,{\rm GeV}$ as well as the extrapolation to $M_{\rm pl}$ described in the beginning of Sec.~\ref{sec:vacuumstability}. We required that $\lambda_{i}(M_{\rm pl})=\gamma(M_{\rm pl})=0$ with $i={1,2,3,4,5}$ and used $M_{\Delta}=10^9\,{\rm GeV}$ and $Y_\Delta(M_{\rm pl})=1$, which yields a stable triplet potential. The negative value of $\lambda$ between $\mu\sim 10^{15}-10^{18}{\rm GeV}$ makes the stability issue a bit worse than in the SM. Right-handed neutrinos help stabilise the vacuum.\label{fig:lambdaDonly}}
\end{center}
\end{figure}

Now, let us take a closer look at the beta functions for the remaining couplings. We show here only the 1-loop results but we use the 2-loop RGEs for the numerical studies. The quartic couplings run according to 
\begin{align}
\beta_{\lambda_3}&=\frac{27}{100}g_1^4+\frac{9 }{10}g_1^2 g_2^2+\frac{9}{4}g_2^4+\lambda_3  \left(-\frac{9}{5}g_1^2-9 g_2^2+12y_t^2+4 Y_\Phi^2+12 \lambda_3\right)+6 \lambda_4^2+4 \lambda_5^2-12 y_t^4-4Y_\Phi^4\,,\label{eq:lambda}
\\
\beta_{\lambda_1}&=\frac{108}{25}g_1^4+\frac{72}{5}g_1^2g_2^2+18 g_2^4+\lambda_1 \left(-\frac{36}{5}g_1^2-24
   g_2^2+4 \lambda_2+14 \lambda_1+8 Y_\Delta^2\right)+2 \lambda_2^2+4
   \lambda_4^2+4 \lambda_5^2-16 Y_\Delta^4\,,\label{eq:lambda1}
\\
\beta_{\lambda_2}&=12 g_2^4-\frac{144}{5}g_1^2 g_2^2+\lambda_2 \left(-\frac{36 g_1^2}{5}-24 g_2^2+12 \lambda_1+8
   Y_\Delta^2\right)+3
   \lambda_2^2-8 \lambda_5^2+16 Y_\Delta^4\,,\label{eq:lambda2}
\\
\beta_{\lambda_4}&=\frac{27 }{25}g_1^4+6 g_2^4+\lambda_4 \left(-\frac{9}{2}g_1^2-\frac{33}{2}g_2^2+6 \lambda_3 +8 \lambda_1+2 \lambda_2+4 \lambda_4+4 Y_\Delta^2+6
   y_t^2+2 Y_\Phi^2\right)+8 \lambda_5^2-8
   Y_\Delta^2 Y_\Phi^2\,,\label{eq:lambda4}
\\
\beta_{\lambda_5}&=-\frac{18}{5} g_1^2 g_2^2+\lambda_5 \left(-\frac{9}{2}g_1^2-\frac{33}{2}g_2^2+2 \lambda_3 +2
   \lambda_1-2 \lambda_2+8 \lambda_4+4 Y_\Delta^2+6 y_t^2+2
   Y_\Phi^2\right)+8 Y_\Delta^2 Y_\Phi^2\,.\label{eq:lambda5}
\end{align}
Note that the running of the Higgs quartic coupling Eq.~\eqref{eq:lambda} is only modified with respect to the SM by the couplings $Y_\Phi$, $\lambda_4$ and $\lambda_5$. However, we will find that under the asymptotic safety condition, $\lambda_4$ and $\lambda_5$ always remain smaller than or of the order $O(0.1)$. Thus, the correction to the running of $\lambda$ with respect to the SM will mainly be due to the change in running of the gauge couplings and the neutrino Yukawa coupling. Actually, see in Fig.~\ref{fig:lambdaDonly} how the change in the gauge couplings, only considering the effects of the triplet scalar and no right-handed neutrinos, makes the vacuum stability problem slightly worse. This is one of the reasons why we include right-handed neutrinos.

Lastly, the cubic coupling evolves according to
\begin{align}\label{eq:lambda6}
\beta_{\gamma}=\gamma \left(-\frac{27}{10}g_1^2-\frac{21}{2}g_2^2+2 \lambda_3 +4\lambda_4-8 \lambda_5+2Y_\Delta^2+6 y_t^2+2Y_\Phi^2\right)-4\sqrt{2} M_\nu Y_\Delta Y_\Phi^2\,,
\end{align}
together with the running of the right-handed neutrino majorana mass
\begin{align}
\beta_{M_\nu}=2M_\nu Y_\Phi^2\,.
\end{align}
Although $\gamma$ does not appear at 1-loop in any other beta function \eqref{eq:gaugecouplings}-\eqref{eq:neutrinoYukawa}, it plays a very important role. This is due to the shift between $\lambda_3$ and the Higgs quartic coupling $\lambda$ below the triplet scale, which is directly proportional to $\gamma^2/M_\Delta^2$, as given by Eq.~\eqref{eq:lambdaSML6}. Furthermore, since it is a cubic coupling with dimensions of mass, we do not have a direct reason to impose the vanishing of $\gamma$ at $M_{\rm pl}$. Its proper treatment will require functional renormalisation which we leave for future work. For a pure type II seesaw, $\gamma/M_\Delta$ should ultimately be determined from the masses of the light neutrinos.

Before we proceed, it should be noted that perturbative radiative corrections to the Higgs boson mass are in general proportional to the triplet mass and the couplings $\lambda_4$ and $\gamma$ with coefficients depending on the renormalisation procedure \cite{Schmidt:2007nq}. A necessary counterterm of the order of the triplet mass seems to be in conflict with the perturbative notion of ``naturalness'' \cite{tHooft:1979rat,Xing:2009in,Haba:2016zbu,Dev:2017ouk}. In this paper, we will not address naturalness, as we do not attempt to explain the gauge hierarchy. We refer the reader to Ref.~\cite{Wetterich:2019qzx} for a discussion in the context of quantum gravity.

\begin{figure}
\begin{center}
\includegraphics[width=0.49\columnwidth]{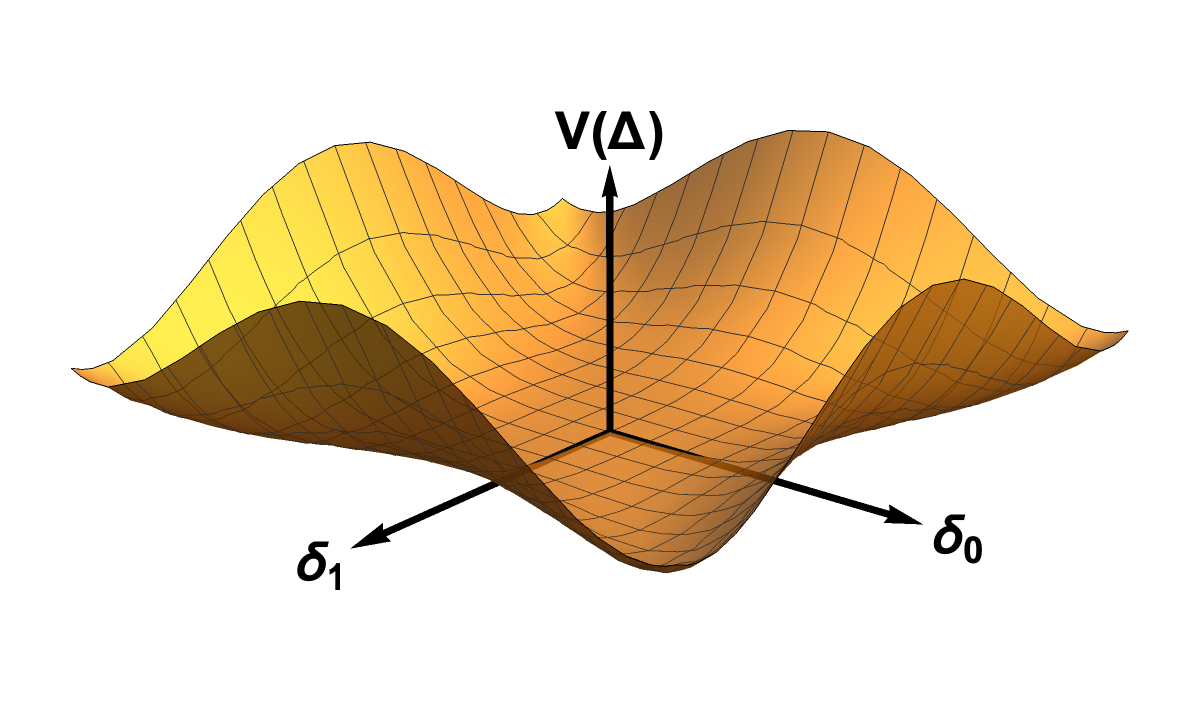}
\caption{Triplet quartic potential with a classically stable vacuum in arbitrary units. We used the values for $\lambda_1$ and $\lambda_2$ at $M_\Delta=10^9\,{\rm GeV}$ in the case of $Y_\Delta=1$. Close to the minimum the mass term dominates and a non-zero expectation value occurs when the interaction with the doublet is taken into account. \label{fig:potential}}
\end{center}
\end{figure}

\subsection{Predictions from asymptotic safety}

With values of the gauge and top Yukawa couplings at the Planck scale that recover the SM values at the EW scale, we are ready to investigate the predictions from asymptotic safe gravity. We use the extrapolations of the gauge and top Yukawa coupling, which depend on the mass of the triplet, as boundary conditions at $M_{\rm pl}$. We impose as well that all quartic couplings vanish, i.e.
\begin{align}\label{eq:boundaryconditions}
\lambda_1(M_{\rm pl})=\lambda_2(M_{\rm pl})=\lambda_3(M_{\rm pl})=\lambda_4(M_{\rm pl})=\lambda_5(M_{\rm pl})=0\,.
\end{align}
At this point, we do not have any condition on the neutrino Yukawa couplings $Y_\Delta$ and $Y_\Phi$ nor on the cubic coupling $\gamma$. Regarding the cubic coupling, it does not significantly affect the running of the other couplings and it only modifies $\lambda$ through Eq.~\eqref{eq:lambdaSML6}. Thus, we shall treat $\gamma$ as a free parameter with a maximum value $\gamma_{\rm max}$ \eqref{eq:maximumgamma} given by the requirement of vacuum stability below the triplet scale and a minimum value of $\gamma(M_\Delta)=0$. Regarding the neutrino Yukawa couplings, we note that even though the quartic couplings vanish at the Planck scale, it does not imply that the doublet-triplet potential will be stable below $M_{\rm pl}$. Some of the couplings may run to a range for which a quartic approximation to the potential develops a negative direction. If this happens at some scale $\bar\mu$, one may suspect spontaneous breaking of $SU(2)$-symmetry by a triplet expectation value $\Delta\approx\bar\mu$. This is incompatible with observation. The neutrino Yukawa couplings will play a crucial role in stabilising the potential. Therefore, as a further condition we will require the stability of the joint doublet-triplet quartic potential. This in turn will place bounds on the values of $Y_\Delta$ and $Y_\Phi$. The study of the full quantum effective potential is left for future work.

After imposing the boundary conditions at the Planck scale, as explained above, we run the couplings down to the the triplet mass scale $M_\Delta$ while checking the vacuum stability. At $M_\Delta$ we integrate out the triplet and run the SM couplings down to the EW scale. Below we present the conditions for the vacuum stability within asymptotically safe gravity and SM plus type I+II seesaw. To understand the different contributions of the triplet and right-handed neutrinos, we will first study the type II seesaw and later turn to type I+II seesaw.

\begin{figure}
\begin{center}
\includegraphics[width=0.49\columnwidth]{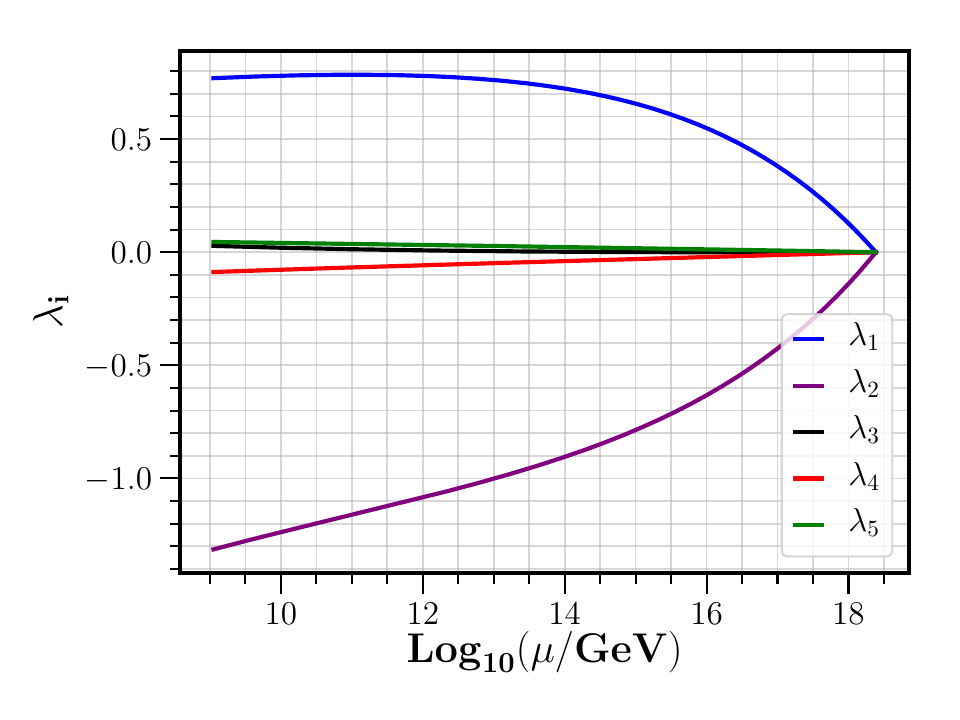}
\includegraphics[width=0.49\columnwidth]{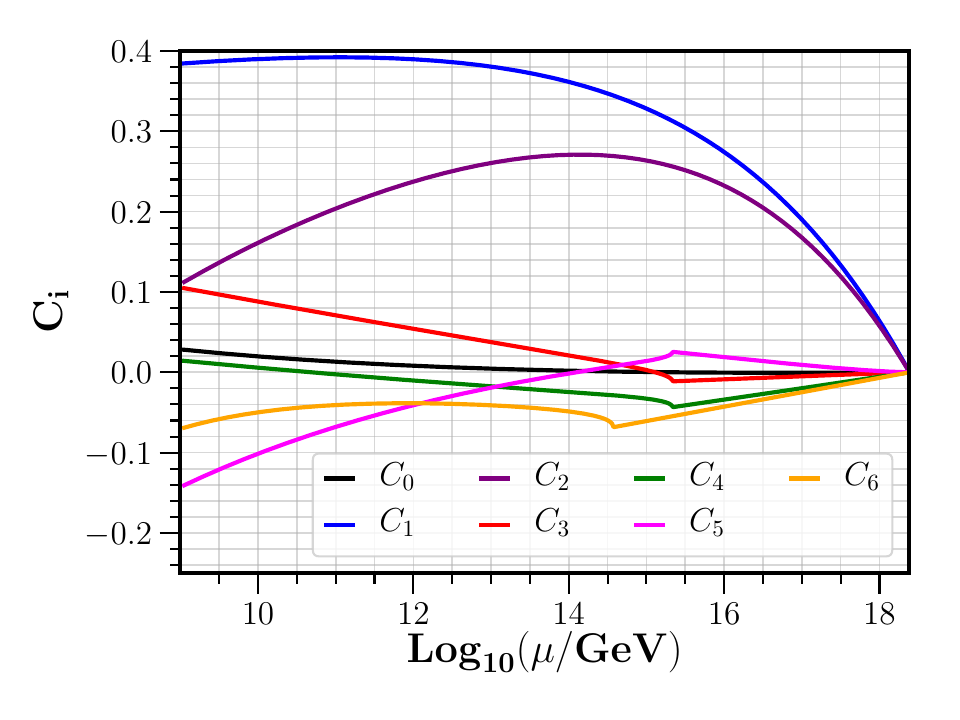}
\caption{Running of the quartic couplings (left figure) and the conditions for the classical stability of the potential (right figure) for $M_\Delta=10^9{\rm GeV}$ and $Y_\Delta(M_{\rm pl})=1$, without right-handed neutrinos. Left: All quartic couplings vanish at the Planck scale. The running of $\lambda_1$ and $\lambda_2$ has opposite trends, and similar for $\lambda_4$ and $\lambda_5$. This is due to the different sign of the contribution of the gauge and neutrino Yukawa couplings in Eq.~\eqref{eq:lambda1}-\eqref{eq:lambda5}. Right: We display the values of the constraints $C_i$ discussed in Sec.~\ref{subsec:vacuumstability}, with a stable potential for positive $C_i$. The interaction of the triplet with left-handed leptons, $Y_\Delta$, stabilises the potential of the triplet alone, that is $C_1,C_2>0$, even though $\lambda_2<0$. The other conditions are in general not met. In particular, we have that $C_5,C_6<0$ evaluated at $M_\Delta$. Also we find that $\lambda_3<0$ for $\mu\gtrsim 10^{15}\,{\rm GeV}$ (see Fig.~\ref{fig:lambdaDonly}). This is the reason why the lines corresponding to $C_{i}$ with $i=3,4,5,6$ change at $\mu\gtrsim 10^{15}\,{\rm GeV}$. For these contributions we employ the real part of the square root of $\lambda_3$ in Eqs.~\eqref{eq:c3c4} and \eqref{eq:c5c6}. \label{fig:cond1}}
\end{center}
\end{figure}

\subsection{Vacuum stability\label{subsec:vacuumstability}} 

The conditions for the (classical) vacuum stability were first studied in Ref.~\cite{Arhrib:2011uy} and then further improved in Ref.~\cite{Bonilla:2015eha}.  Condition number zero, already present in the SM, comes from the quartic interactions of the doublet only and it reads
\begin{align}
C_0\equiv\lambda_3>0\,.
\end{align}
The next two conditions are clearer if we consider for the moment the triplet potential neglecting the doublet. Then we shall use the $SU(2)\times U(1)$ gauge freedom to fix the form of the triplet to
\begin{eqnarray}
{\Deltab}=
\begin{pmatrix}
0& \delta_{1} \\
\delta_{0} & 0
\end{pmatrix}
\,,
\end{eqnarray}
where $\delta_0,\,\delta_{1}$ are real. In this case we find that the quartic terms in the potential are given by
\begin{align}
V(\Deltab)&=\frac{\lambda_1}{2}\left(\delta_0^2-\delta_{1}^2\right)^2+\left(2\lambda_1+\lambda_2\right)\delta_0^2\delta_{1}^2\,.
\end{align}
The stability of the triplet potential requires $\lambda_1>0$ and $2\lambda_1+\lambda_2>0$. For the joint doublet-triplet potential one infers two further conditions \cite{Arhrib:2011uy,Bonilla:2015eha}
\begin{align}\label{eq:c1c2}
 C_1\equiv\lambda_1>0\quad\&\quad C_2\equiv2\lambda_1+\lambda_2>0\,.
\end{align}
A typical triplet potential $V(\delta_0,\delta_1)$ is shown in Fig.~\ref{fig:potential}.

Second, if we take into account the interaction between the doublet and triplet there appear three more conditions. On one hand, we have that\footnote{To translate the results of \cite{Bonilla:2015eha} to our notation we note that $\lambda_H=\lambda_3$, $\lambda_{H\Delta}=\lambda_4+\lambda_5$, $\lambda_{H\Delta}'=-2\lambda_5$, $\lambda_\Delta=\lambda_1+\lambda_2$ and $\lambda_\Delta'=-\lambda_2$.} \cite{Arhrib:2011uy,Bonilla:2015eha} 
\begin{align}\label{eq:c3c4}
C_{3,4}\equiv\lambda_4\pm\lambda_5+\sqrt{\lambda_3\lambda_1}>0\,,
\end{align}
which stops making sense if some of the previous conditions is violated, i.e. $\lambda_3<0$ or $\lambda_1<0$. On the other hand, a detailed analysis of the potential yields that the last condition is given by \cite{Bonilla:2015eha}
\begin{align}\label{eq:c5c6}
C_5\equiv\sqrt{{\lambda_3}}\lambda_2+2\sqrt{\lambda_1}|\lambda_5|>0 \quad{\rm or}\quad
C_6\equiv\lambda_4+\sqrt{\frac{2\lambda_1+\lambda_2}{\lambda_2}\left(\frac{1}{2}\lambda_3\lambda_2+\lambda^2_5\right)}>0\,,
\end{align} 
where either positive $C_5$ or $C_6$ has to be satisfied. We next proceed to study if the predictions from asymptotic safety are compatible with vacuum stability. The issue of this investigation may be phrased differently as the question if asymptotic safety for gravity with a light triplet mass is compatible with $SU(2)\times U(1)$-SSB at the Fermi scale. 

In Fig.~\ref{fig:cond1} we display the flow of the various quartic couplings, together with the corresponding constraints $C_i$. This plot does not include effects from the fluctuations of right handed neutrinos, i.e. we take here $M_\nu=M_{\rm pl}$. The conditions for vacuum stability of the doublet-triplet potential are violated since $C_3$ and $C_4$ get negative. Since both $|C_3|$ and $|C_4|$ are rather small, they are sensitive to small changes around the Planck scale. Nevertheless, small values of $M_\Delta/M_{\rm pl}$ are unlikely to be consistent with vacuum stability if the fluctuation effects of the right handed neutrinos are omitted.

\begin{figure}
\begin{center}
\includegraphics[width=0.49\columnwidth]{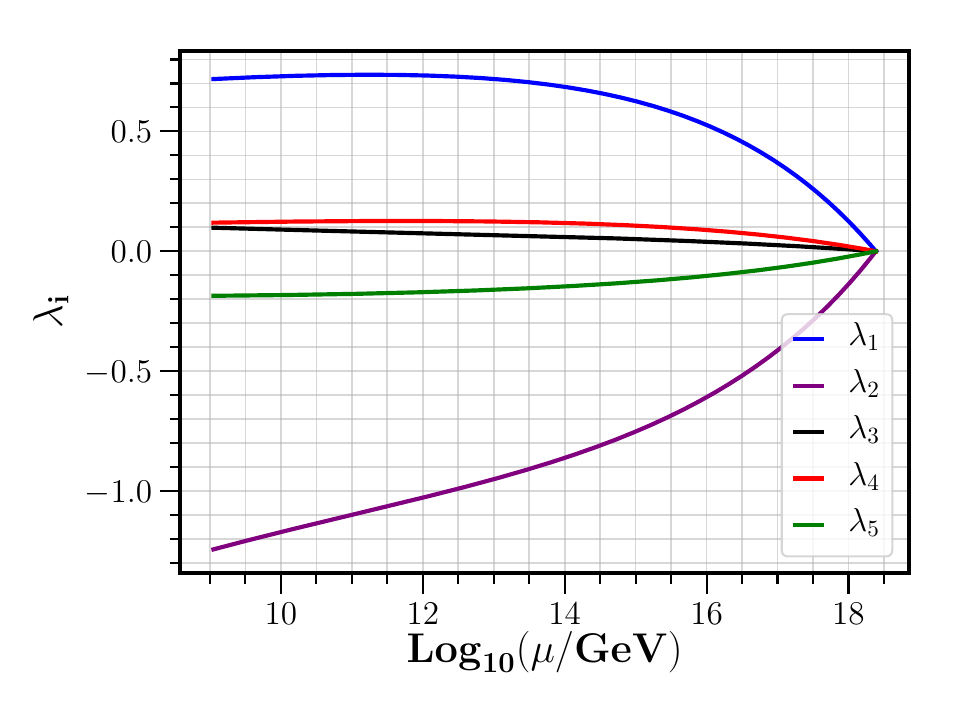}
\includegraphics[width=0.49\columnwidth]{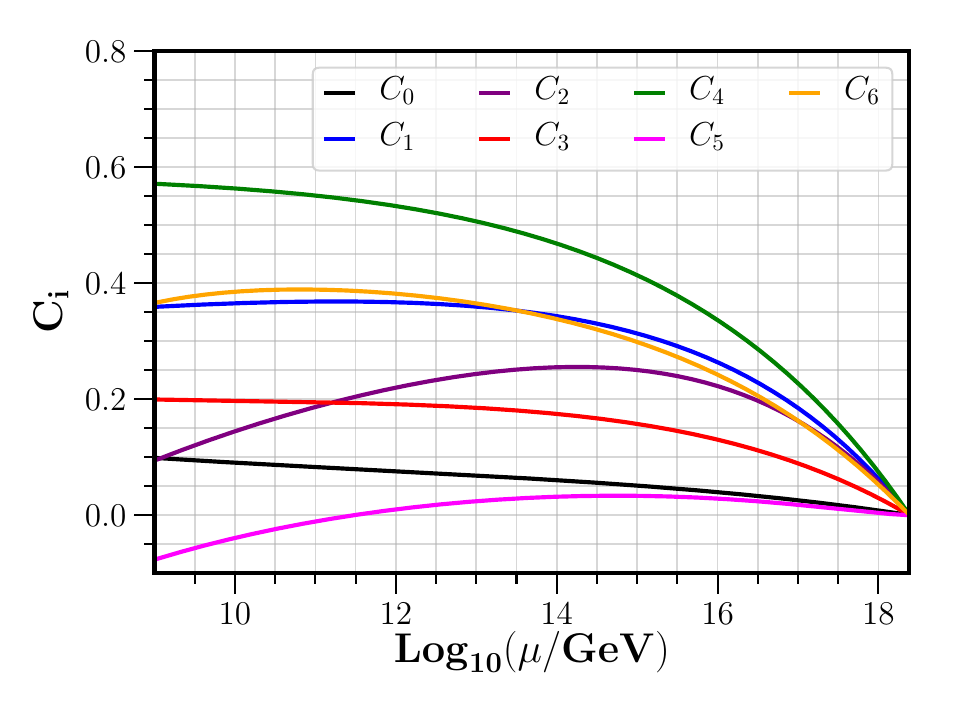}
\caption{Running of the triplet quartic couplings (left figure) and the conditions for the classical stability of the potential (right figure) for $M_\Delta=10^9{\rm GeV}$ and $Y_\Delta=1.3$ considering right-handed neutrinos with $Y_\Phi=0.37$. Left: All quartic couplings vanish at the Planck scale. Also, note how $\lambda_1$ and $\lambda_2$ have opposite trends. The same occurs for $\lambda_4$ and $\lambda_5$. This is due to the difference the sign contribution of the gauge and neutrino Yukawa couplings in Eq.~\eqref{eq:lambda1}-\eqref{eq:lambda5}. Right: See how the Yukawa interaction of the neutrinos with the Higgs, $Y_\Phi$, stabilises the joint doublet-triplet potential with $C_1,C_2,C_3,C_4,C_6>0$ in all the range of scales. Note that $C_5$ is not positive throughout all the scales. \label{fig:cond2}}
\end{center}
\end{figure}

In Fig.~\ref{fig:cond2} we include the effects of the fluctuations of the right handed neutrinos, taking $M_\nu=M_\Delta$, and assuming a substantial Yukawa coupling $Y_\Phi$ between the Higgs doublet and the right- and left-handed neutrinos. In this case vacuum stability of the triplet-doublet potential is realised for the whole range of scales shown. Comparison with Fig.~\ref{fig:cond1} demonstrates that the doublet-neutrino Yukawa coupling can play an important role for stabilising the scalar potential. Asymptotic safety for gravity does not favor a pure seesaw II scenario. Even though the triplet may give the dominant contribution to the light neutrino masses, fluctuation effects of the right handed neutrinos below the Planck scale may be needed to guarantee vacuum stability for $M_\Delta$ sufficiently below $M_{\rm pl}$.

\subsection{Bounds on the triplet-neutrino Yukawa coupling\label{subsec:boundsYukawa}} 

Let us for the moment focus on the vacuum stability of the triplet potential alone, i.e. conditions $C_1$ and $C_2$ \eqref{eq:c1c2} which involve $\lambda_1$ and $\lambda_2$ only. This implies that we can safely neglect the effects of right-handed neutrinos since they do not enter in the 1-loop RGEs of $\lambda_1$ and $\lambda_2$, that is Eqs.\eqref{eq:lambda1} and \eqref{eq:lambda2}. Numerically the critical value $Y_\Delta(M_{\rm pl})=0$ leads in general to a violation of the condition $C_2$. Vacuum stability imposes therefore a lower bound on $Y_\Delta$. We also numerically find that if $Y_\Delta$ is large enough then again $C_2(M_\Delta)<0$. 
Thus, there is also an upper bound on $Y_\Delta$.

\begin{figure}
\begin{center}
\includegraphics[width=0.49\columnwidth]{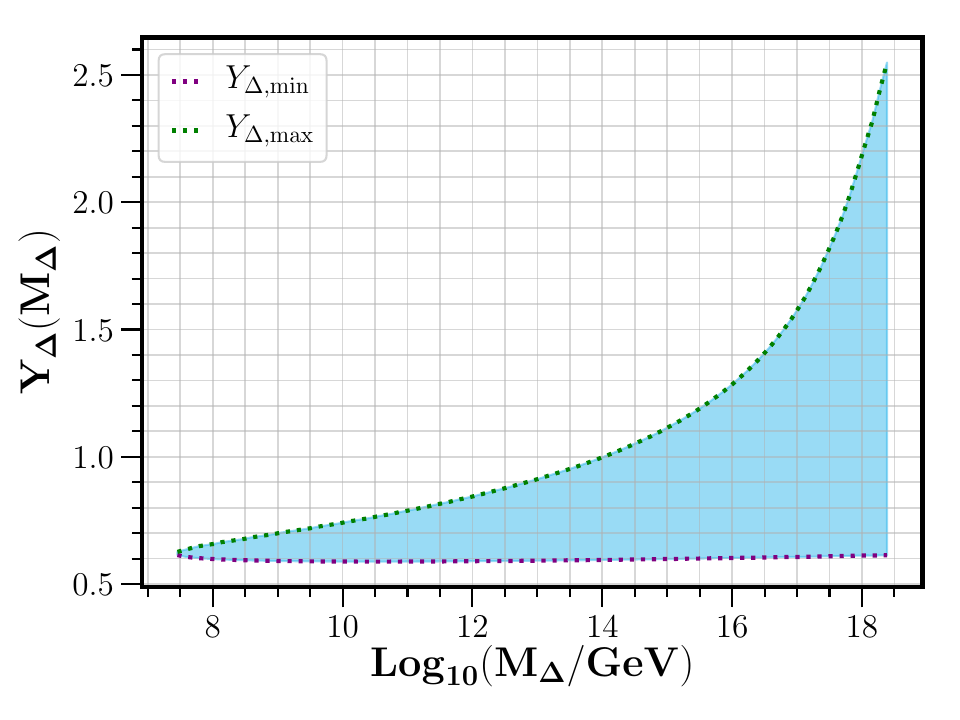}
\includegraphics[width=0.49\columnwidth]{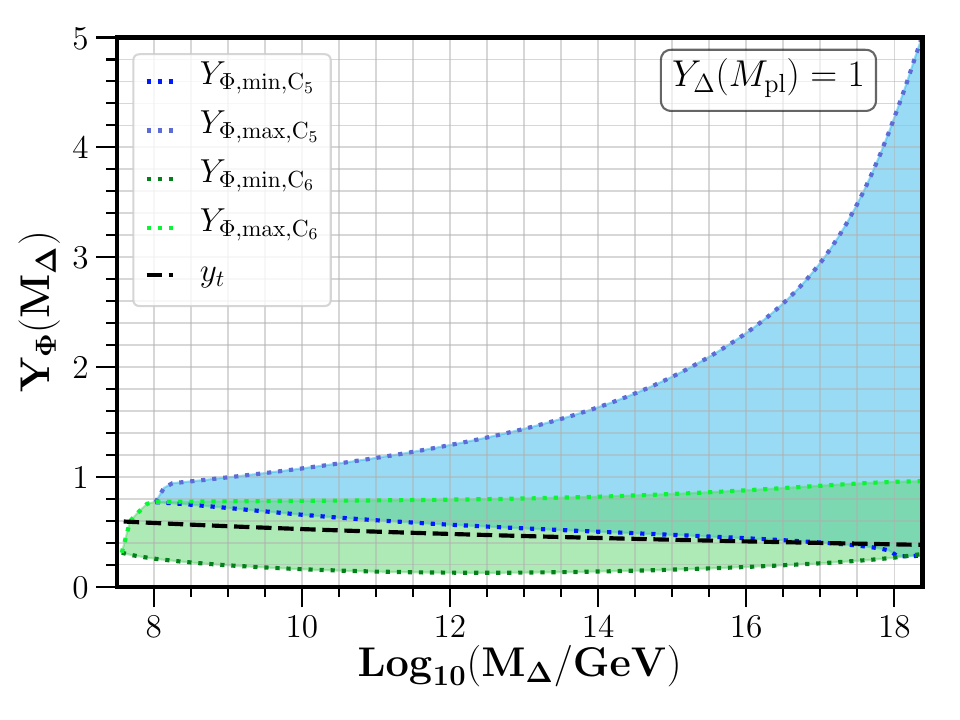}
\caption{Constraints for Yukawa couplings of neutrinos. Left: Allowed parameter space for the triplet Yukawa coupling $Y_\Delta$ by vacuum stability of the triplet potential. The lower bound is roughly at $Y_{\Delta,\rm min}\sim 0.6$. The effect of right-handed neutrinos has been neglected. Right: Allowed parameter region for the doublet Yukawa coupling $Y_\Phi$ by requiring vacuum stability of the doublet-triplet potential, with fixed $Y_\Delta(M_{\rm pl})=1$. The blue and green shaded regions respectively correspond to the case where $C_5$ and $C_6$ are satisfied. The two regions overlap for large enough $M_\Delta$. In black we show the value of the top yukawa coupling.\label{fig:YD}}
\end{center}
\end{figure}

We find the bounds on $Y_\Delta$ with a numerical search by imposing that $C_1,C_2\geq0$ evaluated at $M_\Delta$. Due to the running of $C_1$ and $C_2$ which first increase and then decrease, see Fig.~\ref{fig:cond1}, the condition that $C_1,C_2\geq0$ at $M_\Delta$ is a necessary and sufficient condition for vacuum stability above the triplet scale. We also see that the lower and upper bounds, respectively referred to as $Y_{\Delta,\rm min}$ and $Y_{\Delta,\rm max}$, both come from requiring $C_2(M_\Delta)=0$. We show the results in the left plot in Fig.~\ref{fig:YD}. Interestingly, we find that in this context the neutrino Yukawa coupling should be roughly greater or similar than $0.6$, i.e. $Y_\Delta\gtrsim 0.6$. Furthermore, we find that the lower and upper bounds meet roughly at $M_\Delta\sim 5\times 10^7\,{\rm GeV}$. This implies that there is no vacuum stability for $M_\Delta<5\times 10^7\,{\rm GeV}$ within asymptotic safe gravity in the type II seesaw. We also note that the value of $Y_\Delta(M_{\rm pl})=1$ satisfies $C_1,C_2>0$ for all values of triplet masses $M_\Delta>5\times 10^7\,{\rm GeV}$. For this reason, we use $Y_\Delta(M_{\rm pl})=1$ in our example shown in Fig.~\ref{fig:lambdaDonly} and when considering right-handed neutrinos. We numerically find that the conditions $C_i>0$ with $i=\{0,3,4,5,6\}$ are not satisfied in general and, therefore, we consider next the effects of right-handed neutrinos.

\subsection{Bounds on the doublet-neutrino Yukawa coupling \label{subsec:boundsYukawaHiggs}}

Now that we found non-trivial bounds on the triplet neutrino Yukawa coupling $Y_\Delta$ by requiring stability of the triplet potential, let us explore whether a non-zero value of the doublet-neutrino Yukawa coupling $Y_\Phi$ stabilises the full doublet-triplet potential. It is plausible that a non-zero $Y_\Phi$ renders $C_i>0$ for $i=\{0,3,4,5,6\}$ since it enters directly to the RGEs of $\lambda_3$, $\lambda_4$, $\lambda_5$ (see Eqs.~\eqref{eq:lambda}--\eqref{eq:lambda5}). This is because right-handed neutrinos couple directly to the Higgs field through the Yukawa coupling $Y_\Phi$. One hand hand, looking at the beta functions of $\lambda_4$ \eqref{eq:lambda4} and $\lambda_5$ \eqref{eq:lambda5}, relevant for the conditions $C_3$ and $C_4$ \eqref{eq:c3c4}, we see that a non-zero value of $Y_\Phi$ might lead to $C_3,C_4>0$, in analogy to $Y_\Delta\neq0$ yielding $C_1,C_2>0$. On the other hand, whether a value of $Y_\Phi$ satisfies either the condition $C_5$ or $C_6$ is not directly apparent from the beta functions and, therefore, we have to check it numerically. 

As a proof of principle and to simplify the analysis we choose $Y_\Delta(M_{\rm pl})=1$ in what follows, since such a value has $C_1,C_2>0$ for all triplet scales of interest. Thus, we study the bounds on $Y_\Phi$ by requiring vacuum stability given a fixed value of $Y_\Delta$. We numerically find the bounds on $Y_\Phi$ by requiring that $C_i(M_\Delta)\geq0$ with $i=[0-6]$. Contrary to the analysis of $Y_\Delta$ in Sec.~\ref{subsec:boundsYukawa}, we find that the condition $C_i(M_\Delta)\geq0$ is a necessary condition but not sufficient in general. By construction, one has $C_i(M_{\rm pl})=0$ for all $i$. Even, if we find $Y_\Delta$ and $Y_\Phi$ such that $C_i(M_{\Delta})\geq0$, it is possible that one condition, say $C_j$, decreases enough with increasing scale $\mu$ and becomes negative $C_j(\mu)<0$ at some point. However, as it must reach $C_j(M_{\rm pl})=0$ the maximum negative value for $C_j$ cannot be too large. In fact, one can easily find a small shift in $Y_\Phi$ that has $C_j>0$ for all scales. This means that the lower and upper bounds derived by requiring stability at all scales would be slightly stricter. Nevertheless, the derived values give a good order of magnitude estimate.

We find that the requirement that $C_5(M_\Delta)\geq0$ or $C_6(M_\Delta)\geq0$ yields two different parameter regions with a smooth overlap, see right Fig.~\ref{fig:YD}. The lower bounds, called $Y_{\Phi,\rm min}$, have either $C_5(M_\Delta)=0$ or $C_6(M_\Delta)=0$. The upper bounds, called $Y_{\Phi,\rm max}$, are mainly due to $C_5(M_\Delta)=0$, $C_6(M_\Delta)=0$ or $C_3(M_\Delta)=0$ for $Y_{\Phi}>Y_{\Phi,\rm min}$. Interestingly, we find that for $Y_\Delta(M_\Delta)=1$, the Higgs neutrino yukawa coupling $Y_\Phi$ is in general $Y_\Phi\gtrsim 0.2$. We also note that the value of $Y_\Phi(M_\Delta)=y_t(M_\Delta)$ is within the allowed parameter range for $Y_\Phi$.

\subsection{Bounds on the cubic coupling\label{subsec:boundscubic}} 

Before ending this section, let us explore possible bounds on the cubic coupling $\gamma$. We recall that $\gamma$ only plays an important role in the shift of the effective Higgs quartic coupling Eq.~\ref{eq:LSM}. In other words, it does not modify any of the beta functions for the quartic couplings Eqs.~\eqref{eq:lambda}--\eqref{eq:lambda5}. Therefore, $\gamma$ does not change the constraints from vacuum stability above the triplet scale in our approximation of the effective potential. However, it may change the vacuum stability below $M_\Delta$, as discussed near Eq.~\eqref{eq:maximumgamma}. It can also modify predictions for the top mass substantially by lowering the value of $\lambda/y_t^2$ at the Fermi scale.

\begin{figure}
\begin{center}
\includegraphics[width=0.5\columnwidth]{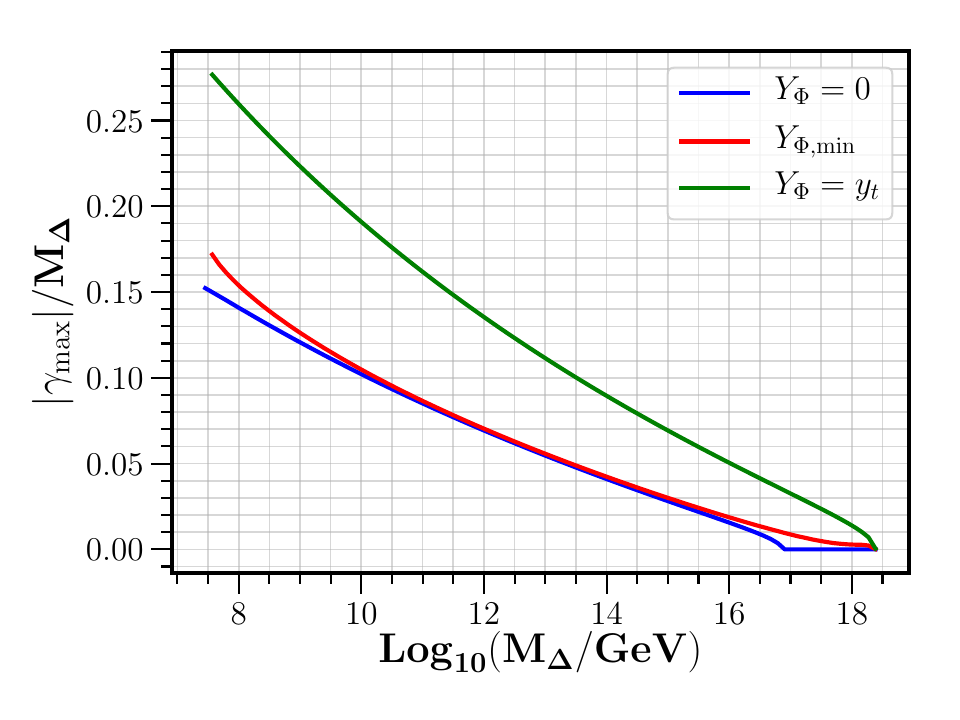}
\caption{Maximum bound on the cubic coupling $|\gamma|$ by requiring vacuum stability below $M_\Delta$, i.e. $\lambda(M_\Delta)>0$, as a function of the triplet mass. In blue, red and green we respectively show the cases of $Y_\Phi(M_\Delta)=\{0,Y_{\Phi,\rm min},y_t\}$. The larger the Yukawa coupling $Y_\Phi$, the larger the maximum bound of $|\gamma|$. This is due to the fact that $\lambda_3$ grows faster with decreasing scale for larger $Y_\Phi$, see Eq.~\eqref{eq:lambda}. \label{fig:gamma}}
\end{center}
\end{figure}

From the effective Higgs quartic coupling below the triplet scale in Eq.~\ref{eq:LSM}, and the requirement of vacuum stability above and below $M_\Delta$, we derived an upper bound $\gamma_{\rm max}$ in Eq.~\eqref{eq:maximumgamma}, given by ${\gamma_{\rm max}}(M_\Delta)={M_\Delta}\sqrt{{\lambda_3(M_\Delta)}/{2}}\,$. Since we use the boundary condition $\lambda_3(M_{\rm pl})=0$, the value $\lambda_3(M_\Delta)$ is not large unless $Y_\Phi$ is large. In this way, the bound on $\gamma$ depends on the value of $Y_\Phi$. For example, for $Y_\Delta(M_{\rm pl})=1$ and $Y_\Phi(M_\Delta)=y_t(M_\Delta)\sim 0.55 $ at $M_\Delta\sim 10^9\,{\rm GeV}$ we have that $\left|{\gamma}/{M_\Delta}\right|<0.22$. In Fig.~\ref{fig:gamma} we show the dependence of $\gamma_{\rm max}$ on $M_\Delta$ for different values of $Y_\Phi$. The value $Y_{\Phi,\rm min}$ corresponds to the constraint $C_6>0$. We see that the lower the $M_\Delta$ the higher $\gamma_{\rm max}$. Also, the larger $Y_\Phi$, the larger the $\gamma_{\rm max}$, as expected. We use such upper bounds on $\gamma$ to get an impression for uncertainties in the predictions of the top mass. In App.~\ref{app:unitarity}, we study the bounds on $\gamma$ due to unitarity and obtain that in general $|\gamma|<1.6 M_\Delta$. The bound from unitarity is compatible with the cases studied in this paper.

\section{Predictions for the top quark mass\label{sec:predictions}}

Let us now turn to the predictions for the top mass from asymptotic safety. As explained in Sec.~\ref{sec:vacuumstability}, we impose boundary conditions at the Planck scale inspired by asymptotic safety, i.e. $\lambda_i(M_{\rm pl})=0$ for all $i$. We employ initial values of the gauge and top yukawa couplings at $M_{\rm pl}$ that recover the SM predictions at the EW scale. Then we run the couplings from $M_{\rm pl}$ down to $M_\Delta$, integrate out the triplet and flow to the Fermi scale. By construction, this procedure gives very similar values of the gauge and top Yukawa couplings to those of the SM. However, it changes the prediction from asymptotic safety for the Higgs quartic coupling compared to the case of the SM alone. It should be noted that this prediction is sensitive to small shifts in the boundary conditions set by a full study in functional renormalisation. Thus, in order to simplify the analysis, we can give a fair estimate on the predictions for the mass ratio $M_t/M_H$ by studying the ratio between the top Yukawa and the Higgs quartic coupling and compare it with the predictions of asymptotic safe gravity and SM. In other words, we expect that the relative change in the ratio of couplings given by
\begin{align}\label{eq:ratio}
\frac{(y_t\big/\sqrt{2\lambda})_{\rm SM\,\&\,type\,I+II}}{({y_t}\big/{\sqrt{2\lambda}})_{\rm SM}}\bigg|_{M_t}\approx\frac{\left(M_t/M_H\right)_{\rm SM\,\&\,type\,I+II}}{\left(M_t/M_H\right)_{\rm SM}}\bigg|_{M_t}\,,
\end{align}
gives a good estimate in the relative change of the prediction for the mass ratio.

\begin{figure}
\begin{center}
\includegraphics[width=0.49\columnwidth]{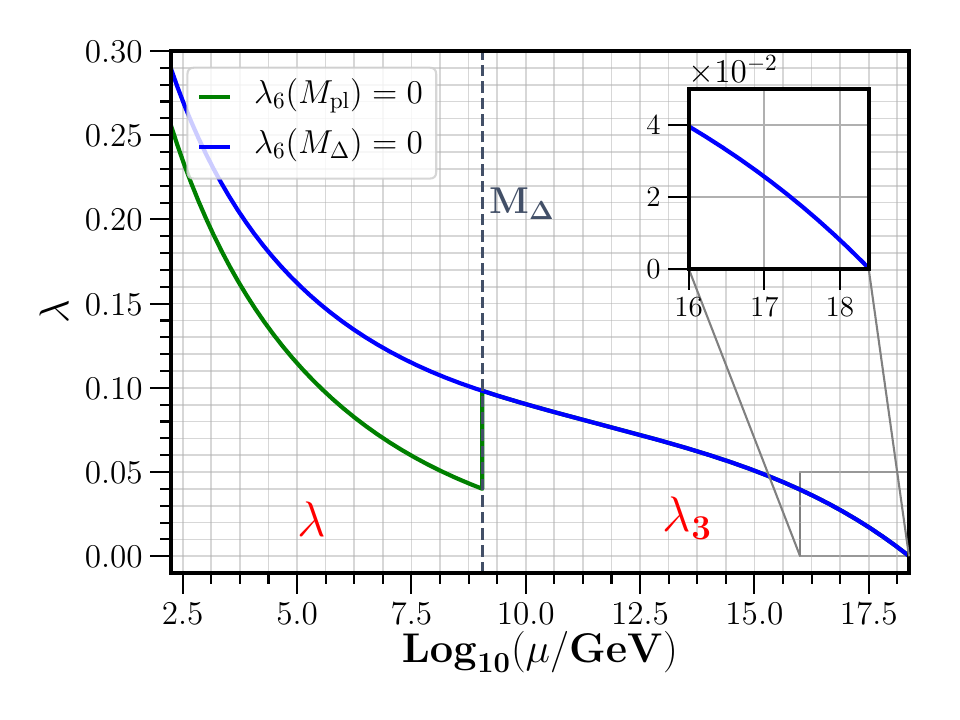}
\includegraphics[width=0.49\columnwidth]{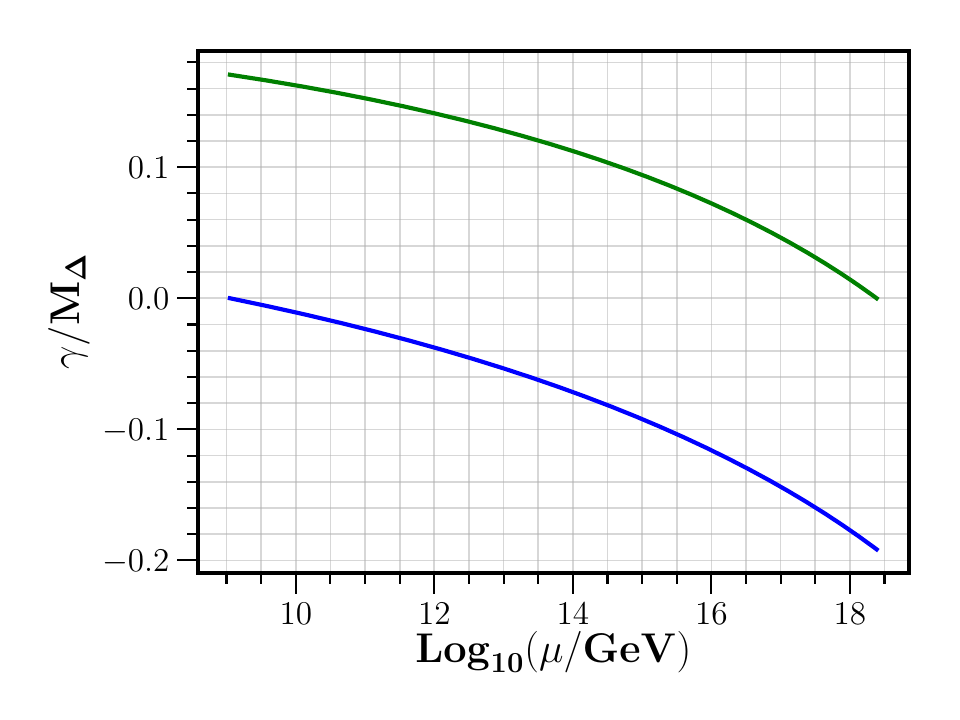}
\caption{Left: Running of the Higgs quartic coupling with the logarithm of the renormalisation scale in the type I+II seesaw using the 2-loop RGEs. We used the values given in Eq.~\eqref{eq:valuesSM} for $M_t=170.97\,{\rm GeV}$ as well as the extrapolation to $M_{\rm pl}$ described in the beginning of Sec.~\ref{sec:vacuumstability}. We required that $\lambda_{i}(M_{\rm pl})=0$ with $i={1,2,3,4,5}$ and used $M_{\Delta}=10^9\,{\rm GeV}$, $Y_\Delta(M_{\rm pl})=1$ and $Y_\Phi(M_\Delta)=y_t(M_\Delta)$ which yields a stable doublet-triplet potential. The solid blue line corresponds to $\gamma(M_{\Delta})=0$ and the solid green line to $\gamma(M_{\rm pl})=0$. One finds $\lambda,\lambda_3>0$ for all $\mu$, leading to a stable vacuum. For $\gamma(M_{\rm pl})=0$ (solid blue) there is a step between $\lambda_3$ and the effective Higgs quartic coupling $\lambda$ at $M_\Delta$ due to Eq.~\eqref{eq:lambdaSML6}. Right: Running of the cubic coupling $\gamma$ for the same example as in the left figure. The green and blue lines respectively corresponds to $\gamma(M_{\rm pl})=0$ and $\gamma(M_{\Delta})=0$. \label{fig:L6}}
\end{center}
\end{figure}

\subsection{Influence of the cubic scalar coupling}

Before going to the predictions for the top mass, let us discuss the relevant values of the cubic coupling. We note from the RGE of $\gamma$ Eq.~\eqref{eq:lambda6} that $\gamma$ only runs if both $Y_\Delta$ and $Y_\Phi$ are non-zero. When we study the effects of the triplet alone and we set $Y_\Phi=0$, $\gamma$ is constant. Contrariwise, when we include the right-handed neutrinos with non-vanishing $Y_\Phi$, $\gamma$ runs with the scale. The influence of $\gamma$ on the prediction of the top quark mass arises from the shift \eqref{eq:lambdaSML6} between $\lambda$ and $\lambda_3$. This effect can be substantial, and the lack of knowledge for $\gamma$ constitutes one of the main uncertainties for the prediction of $M_t$. Treating $\gamma$ as a free parameter, we will study three cases of interest: $(i)$ when there is no effect of $\gamma$ for the effective theory below the triplet scale, i.e. $\gamma(M_\Delta)=0$ or $\lambda(M_\Delta)=\lambda_3(M_\Delta)$, $(ii)$ when $\gamma$ attains the maximum value allowed by vacuum stability below the triplet scale $\gamma_{\rm max}$ from Eq.~\eqref{eq:maximumgamma} which corresponds to $\lambda(M_\Delta)=0$ and $(iii)$ when $\gamma(M_{\rm pl})=0$. Case $(iii)$ is inspired by a left-right symmetric generalisation, to be discussed later in Sec.~\ref{sec:conclusions}, where $\gamma$ arises from a quartic coupling. Then, treated as a quartic coupling we may impose $\gamma(M_{\rm pl})=0$. Note that cases $(i)$ and $(ii)$ comprise the possible uncertainties in the prediction of the top mass from Eq.~\eqref{eq:ratio} due to the cubic coupling.  More concretely, since $\lambda$ is in the denominator in Eq.~\eqref{eq:ratio}, $\gamma(M_\Delta)=0$ corresponds to the lower bound on the ration Eq.~\eqref{eq:ratio} and  $\gamma_{\rm max}$ to the upper bound. It should be noted that while $\lambda(M_\Delta)=\lambda_3(M_\Delta)$ depends on the value of $Y_\Phi$ through the running of $\lambda_3$, the upper bound on the ratio given by $\lambda(M_\Delta)=0$ is independent of $Y_\Delta$ and $Y_\Phi$. Thus, the upper bound on the top quark mass shown in Fig.~\ref{fig:mt} is robust. We display in Fig.~\ref{fig:L6} the running of $\lambda$ and $\lambda_3$ (left part), as well as the running of $\gamma$ (right part). The two curves correspond to the maximal and minimal values of $|\gamma|$. They are shown for given values of $Y_\Delta$ and $Y_\Phi$ that are consistent with vacuum stability. 

\subsection{Quantitative predictions for the top quark mass}

\begin{figure}
\begin{center}
\includegraphics[width=0.49\columnwidth]{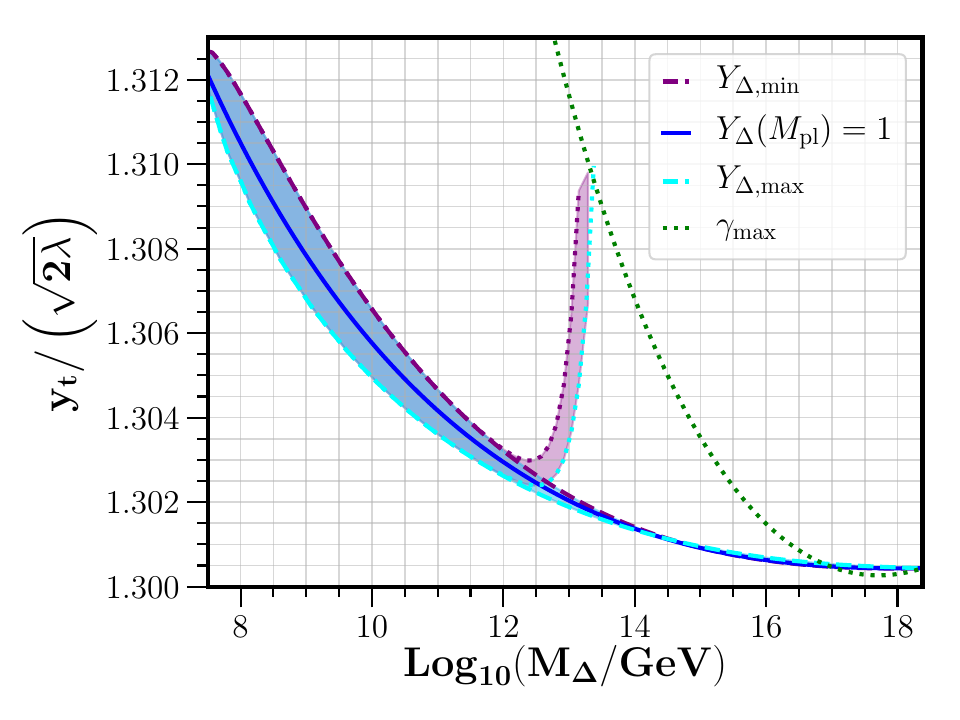}
\includegraphics[width=0.49\columnwidth]{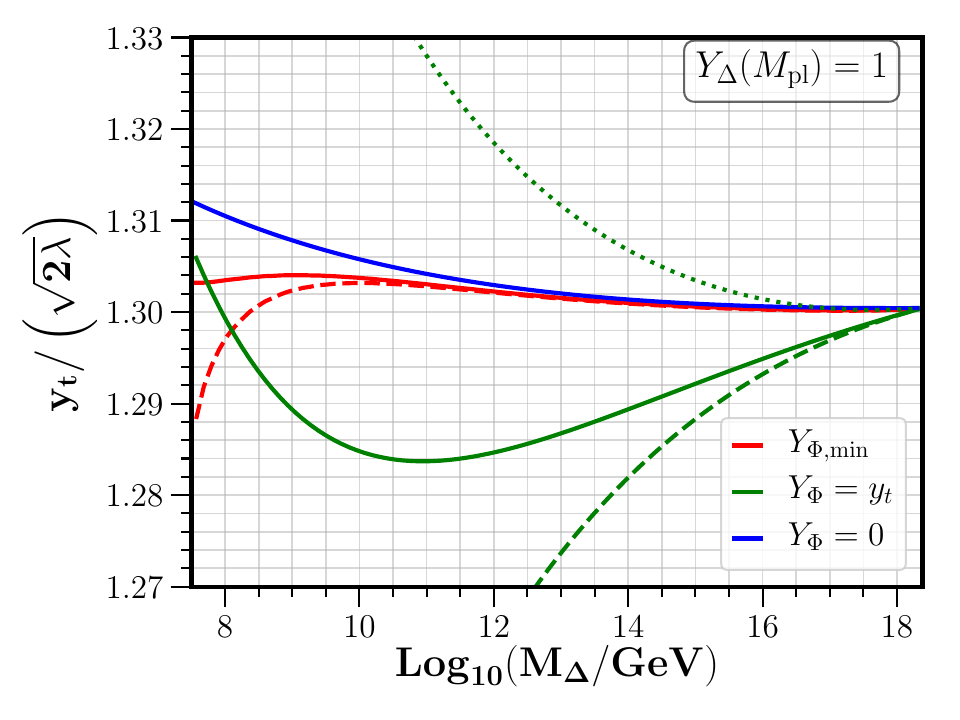}
\caption{ Predictions on the top Yukawa-Higgs quartic coupling ratio as a function of the intermediate scale $M_\Delta$. The relative change with $M_\Delta$ can be directly translated to a prediction for the top-Higgs mass ratio if compared with the ratio for $M_\Delta=M_{\rm pl}$, which is given by $y_t\big/\sqrt{2\lambda}\sim 1.3004$. Left: Change in the predictions due to the triplet only. The solid blue line refers to the value $Y_\Delta(M_{\rm pl})=1$ and the dotted lines are the predictions corresponding to the lower and upper bounds on $Y_\Delta$ derived from vacuum stability in Sec.~\ref{subsec:boundsYukawa}. The blue shaded region is the range of predictions allowed by vacuum stability with vanishing cubic coupling $\gamma$. The purple shadded region shows the prediction when $\gamma$ is given by Eq.~\eqref{eq:mnutriplet} and the masses of the light neutrinos are due to a seesaw type II. The dotted green line correspond to $|\gamma|=\gamma_{\rm max}$. Right: Change in the predictions when right-handed neutrinos are taken into account. We fixed $Y_\Delta(M_{\rm pl})=1$. The red lines are the predictions corresponding to the lowest bound on $Y_{\Phi,{\rm min}}$ derived from vacuum stability respectively in Sec.~\ref{subsec:boundsYukawaHiggs}. For the solid red line we use $\gamma(M_{\rm pl})=0$ and for the dashed red line we set $\gamma(M_\Delta)=0$. The green lines are the predictions corresponding to $Y_\Phi(M_\Delta)=y_t(M_\Delta)$. The solid green line obtains for $\gamma(M_{\rm pl})=0$, the dashed blue line uses $\gamma(M_\Delta)=0$ and  the dotted green line corresponds to the maximum bound $\gamma_{\rm max}$ in Eq.~\eqref{eq:maximumgamma}. The latter would be exactly the same for the lowest bound $Y_{\Phi,{\rm min}}$ since it corresponds to the choice $\lambda(M_\Delta)=0$ independent of $Y_\Phi$. The blue solid line corresponds to the case without right-handed neutrinos, as in the left figure. \label{fig:ytlYv}}
\end{center}
\end{figure}

The predictions for the ratio of top Yukawa and Higgs quartic coupling evaluated at the EW scale in terms of the triplet mass are shown in Fig.~\ref{fig:ytlYv}. On the left of Fig.~\ref{fig:ytlYv}, we show the changes in the prediction of the top yukawa and Higgs quartic coupling due to the triplet alone. We find that the ratio increases with respect to the limiting value $M_\Delta=M_{\rm pl}$, which corresponds to asymptotic safe gravity and SM, up to roughly $1\%$ at $M_\Delta=5\times 10^7\,{\rm GeV}$. We also note that the lower and upper bounds of the neutrino Yukawa coupling $Y_\Delta$ have a mild impact to the coupling ratio, as $Y_\Delta$ does not affect $\lambda_3$ directly, see Eq.~\eqref{eq:lambda}. We conclude that including a $SU(2)$ triplet scalar may increase the prediction of asymptotic safe gravity on the ratio up to $1\%$. For instance, this means that given a Higgs mass of $M_H\sim125\,{\rm GeV}$ the top quark mass is between $M_t\sim171\,{\rm GeV}$ and $M_t\sim172.6\,{\rm GeV}$, respectively for no triplet and for a triplet with $M_\Delta=5\times10^{7}\,{\rm GeV}$. At this stage, we may say that better measurements of the top quark mass will determine the mass scale of the triplet scalar, assuming quantum gravity is asymptotically safe. The current constraint given by Refs.~\cite{Sirunyan:2019zvx,Aad:2019mkw} tells us that the triplet cannot be much smaller than $M_\Delta\gtrsim 3\times 10^{11}\,{\rm GeV}$, if we use that the upper $1\sigma$ bound of Ref.~\cite{Sirunyan:2019zvx} is around $M_t\sim 171.3\,{\rm GeV}$.

In the presence of right-handed neutrinos at the intermediate scale the prediction from asymptotic safety derived may vary substantially depending on the magnitude of $Y_\Phi$ and the value of $\gamma(M_\Delta)$. Since $Y_\Phi$ directly enters the beta function of $\lambda_3$ \eqref{eq:lambda}, a larger $Y_\Phi$ implies a larger $\lambda$ at the EW scale, assuming $\gamma(M_\Delta)=0$, and the prediction on the ratio \eqref{eq:ratio} is lowered. Second, a larger value of $\gamma$ leads to a smaller value of $\lambda$ at the EW scale. In that case, the ratio \eqref{eq:ratio} increases. Thus, we have two competing effects from $Y_\Phi$ and $\gamma$. The numerical results for the ratio of the top Yukawa and the Higgs quartic coupling including right-handed neutrinos are shown on the right of Fig.~\ref{fig:ytlYv}. We plot the results for the lower bound of $Y_{\Phi,{\rm min}}$ according to the constraint $C_6$ and the value of $Y_\Phi(M_\Delta)=y_t(M_\Delta)$, also consistent with vacuum stability for the values of $M_\Delta$ of interest, e.g. see Fig.~\ref{fig:YD}. For $Y_\Phi=y_t$ both $\gamma(M_\Delta)=0$ or $\gamma(M_{\rm pl})=0$ substantially lower the prediction as compared to asymptotic safety with the triplet only. This effect is much smaller for $Y_\Phi=Y_{\Phi,{\rm min}}$. This case may lead to a small increase with respect to the prediction of the SM. In particular, for the case where $\gamma(M_{\rm pl})=0$ and $Y_\Phi=Y_{\Phi,{\rm min}}$ the predicted ratio increases up to $0.3\%$ at $M_\Delta\sim 10^9{\rm GeV}$. The upper bound on the ratio \eqref{eq:ratio} corresponds to the maximum value of $\gamma$ \eqref{eq:maximumgamma}. This effect is able to compensate the effects of $Y_\Phi$.

The upper and lower bounds for $y_t/\sqrt{2\lambda}$ in Fig.~\ref{fig:ytlYv} are rather conservative. They assume $M_\nu=M_\Delta$, while the allowed interval shrinks for $M_\nu/M_\Delta$ increasing. For the boundary $M_\nu=M_{\rm pl}$ one recovers the result for the pure seesaw II scenario shown in the left panel of Fig.~\ref{fig:ytlYv}. A rather likely boundary value for the cubic coupling is $\gamma(M_{\rm pl})=0$. The corresponding allowed interval for $y_t/\sqrt{2\lambda}$ lies in between the red and green solid lines in Fig.~\ref{fig:ytlYv}. This interval also shrinks if $M_\nu/M_\Delta$ is larger than one.

At the present stage the experimental bounds on $M_t$ rule out a certain range in the space of neutrino Yukawa couplings and cubic scalar coupling. With further information on the values of the couplings at the Planck scale the uncertainty of the prediction for $M_t$ will get smaller. Certain scenarios can then be falsified by future precision measurements of the top quark mass.

\subsection{Implications from neutrino masses}

After studying the general predictions for the top-Higgs mass ratio in the type I+II seesaw within asymptotic safe gravity, let us look more specifically to the implications from the value of light neutrino masses. First, it is interesting to focus on the case of the $SU(2)$ triplet with no right-handed neutrinos. In this case, $\gamma$ is fixed by Eq.~\eqref{eq:mnutriplet} to explain the light neutrino masses. The results are shown on the left of Fig.~\ref{fig:ytlYv} with a purple shaded region. We see that as $M_\Delta$ grows so does $\gamma$ and, therefore, the prediction for the ratio Eq.~\eqref{eq:ratio} increases. We also see that this time, the prediction is bounded from below leading to a prediction for the top mass $171.25\,{\rm GeV}\lesssim M_t\lesssim 172.6\,{\rm GeV}$ which can be falsified by experiment. The mass of the triplet is also bounded by $5\times 10^{7}\,{\rm GeV}\lesssim M_\Delta\lesssim 5\times 10^{13}\,{\rm GeV}$. This shows how precise measurements of the mass of the neutrinos and the top mass might tell us about the intermediate scale, assuming that the vacuum can be stabilised by other means without the need for right-handed neutrinos or taking right-handed neutrinos to be heavier than the triplet.

Let us now consider that the vacuum is stabilised by the Yukawa coupling of right-handed neutrinos. We found in Sec.~\ref{subsec:boundsYukawaHiggs} a lower bound on the neutrino-Higgs Yukawa coupling given by $Y_\Phi>0.2$. This implies that if the masses of the light neutrinos come from a type I seesaw, the Majorana mass of the right-handed neutrinos should be $M_\nu\gtrsim 10^{12}\, {\rm GeV}$ from Eq.~\eqref{eq:mnuneutrino}. In our simplistic set up, it implies that $M_\Delta\gtrsim10^{12}\,{\rm GeV}$ and also narrows the predictions for the top mass. It also tells us that, in the present case, it is unlikely to find a case where the type II seesaw completely dominates since it requires $M_\Delta<5\times 10^{13}\,{\rm GeV}$. A detailed numerical inspection shows that one example could be $Y_\Phi(M_{\Delta})=0.13$, $Y_\Delta(M_\Delta)=0.71$, $\gamma(M_{\rm pl})=0$ and $M_\Delta=6\times 10^{12}\,{\rm GeV}$. In this particular example, the prediction for the top quark mass would be around $171.1\,{\rm GeV}$. Another example would be $Y_\Phi(M_{\Delta})=y_t(M_\Delta)=0.45$, $Y_\Delta(M_\Delta)=0.74$, $\gamma(M_\Delta)=\gamma_{\rm max}$ and $M_\Delta=7\times 10^{13}\,{\rm GeV}$. In this case, the prediction for the top quark mass would be around $171.9\,{\rm GeV}$. In both cases the heaviest of the light neutrinos has a mass $m_\nu\sim 0.08\,{\rm eV}$. However, in the first case the mass of the light neutrinos is mainly coming from the type I seesaw contribution while in the second example it is a combination of both type I+II seesaw.

\section{Discussion and conclusions \label{sec:conclusions}}

The proposal of asymptotic safety for quantum gravity is rather predictive. This concerns, in particular, the shape of the effective potential for scalar fields. Due to an anomalous dimension induced by fluctuations of the metric, all quartic couplings are predicted to be (almost) zero for momentum scales beyond the Planck mass $M_{\rm pl}$. They start flowing away from these fixed point values only once the momentum drops sufficiently below $M_{\rm pl}$ such that the fluctuations of the metric decouple effectively. Below $M_{\rm pl}$ the running of the couplings is induced by the fluctuations of fields for the particles in the effective theory below $M_{\rm pl}$.

With given initial conditions near $M_{\rm pl}$, the flow towards the Fermi scale yields a prediction for the ratio between the quartic coupling of the Higgs doublet and the squared Yukawa coupling of the top quark. This turns to a prediction for the ratio between the Higgs boson mass and the top quark mass \cite{Shaposhnikov:2009pv}. If the effective theory below $M_{\rm pl}$ is the Standard Model, with no further particles, the predicted value of $M_H$ was around $126\,{\rm GeV}$ with a few GeV uncertainty, as indeed found later at the LHC. With the present rather precise measurement of the mass of the Higgs boson, the prediction of the mass ratio turns now to a prediction of the top quark mass, which can possibly be falsified by experiment. In view of the large efforts to measure the pole mass of the top quark precisely it becomes important to investigate the precision of the asymptotic safety prediction for the top quark mass.

There are theoretical uncertainties from three sources. The first concerns the physics at and beyond the Planck scale. This concerns the precise value of the fixed point and the threshold value for the decoupling of gravity. Initial investigations \cite{Wetterich:2019qzx,Wetterich:2019rsn} suggest that the resulting uncertainty is rather small. The second concerns the particle content of the effective theory below $M_{\rm pl}$. Additional particles will modify the flow between $M_{\rm pl}$ and the Fermi scale, modifying the asymptotic safety prediction for $M_t$. This has been the topic of the present paper. Finally, a third uncertainty arises from the translation between the values of the running couplings at the Fermi scale and the measurable quantities. This concerns physics at the Fermi scale and a precise understanding of experimental setups.

In the present paper we have investigated the impact of an intermediate scale which is relevant for the understanding of the masses of neutrinos. The new particles beyond the standard model are a complex triplet of scalar fields, as well as three generations of right-handed (singlet) neutrinos. They play a role for the flow of couplings between $M_{\rm pl}$ and the intermediate scale, which we identify here with the mass of the scalar triplet. In turn, this modifies the prediction of $M_t/M_H$.

The new particles come along with new couplings, as quartic couplings for the triplet or a cubic coupling $\gamma$ between the triplet and the Higgs doublet, as well as a Yukawa coupling $Y_\Delta$ between the triplet and the leptons, or a Yukawa coupling $Y_\Phi$ involving the Higgs doublet and the right-and left-handed neutrinos. While the initial values of the quartic couplings near the Planck scale are fixed to be almost zero by asymptotic safety, the unknown values of $Y_\Delta$, $Y_\Phi$ and $\gamma$ induce new sources of uncertainties. 

We find that the possible values of the new couplings are restricted by vacuum stability for the combined triplet-doublet potential. Not only should the absolute minimum of the combined effective potential occur for a doublet expectation value at the Fermi scale, but also for a much smaller expectation value of the triplet. For a large mass of the triplet this is guaranteed by the seesaw II mechanism, provided that suitable combinations of quartic couplings involving the triplet remain positive. In turn, this restricts the allowed range of values for the new Yukawa couplings. For the allowed range the problem of vacuum stability is not only solved for the Higgs doublet, but also for the full doublet-triplet system with an intermediate scale. 

Our main findings for the prediction of the mass of the top quark are the following: if the right-handed neutrinos are heavy and the cubic coupling $\gamma$ is small, the scalar triplet has only a mild influence, increasing the predicted value of the top quark from $171\,{\rm GeV}$ for $M_\Delta=M_{\rm pl}$ to $171.3\, {\rm GeV}$ for $M_\Delta\sim10^{12}\,{\rm GeV}$ and $172.6\, {\rm GeV}$ for $M_\Delta\sim 5\times10^{7}\,{\rm GeV}$. For right-handed neutrino mass around $M_\Delta$ and sizable Yukawa couplings the prediction for $M_t$ is lowered. It could be as low as $154\,{\rm GeV}$, which is excluded by observations. On the other hand, the presence of a cubic coupling $\gamma$ enhances the predicted value of $M_t$ and can compensate for the decrease due to the Yukawa couplings of right-handed neutrinos. As long as all these couplings are taken as free parameters the resulting uncertainty exceeds the presently allowed experimental range. In this case experiment can be used to restrict the neutrino sector at the intermediate scale.

This situation may change if we embed the triplet into an enlarged gauge symmetry, such as the left-right symmetric subgroups $SU(2)_L\times SU(2)_R\times U(1)\times SU(3)_C$ or $SU(2)_L\times SU(2)_R\times SU(4)_C$ of a $SO(10)$ grand unification. In this context the cubic coupling $\gamma$ arises from a quartic coupling multiplied with the expectation value breaking left-right symmetry. This quartic coupling is again predicted to vanish at $M_{\rm pl}$ by asymptotic safety, such that $\gamma$ is no longer a free parameter. Furthermore, for $SU(4)_C$ the coupling $Y_\Phi$ equals the Yukawa coupling of the top quark. It is a very interesting question as to whether an intermediate scale is compatible with vacuum stability for such theories, and what would be the precise prediction for $M_t$.

Throughout we have used the two-loop RGEs to derive precise predictions for the top-quark mass. At the electroweak scale, we took the values for the SM-parameters extracted at two loops, but at our intermediate scale we have been limited to leading-order matching, when formally we should include one-loop corrections for consistency. Since the quartic couplings at the matching scale are small, we expect the loop corrections to also be very modest, unless the parameter $\gamma$ or the neutrino Yukawa couplings are large. It would be interesting to explore the impact of improved precision at the matching scale, and to thereby obtain accurate estimates of the uncertainties, in future work.

\acknowledgments

We would like to thank J.~Rubio, L.~Sartore and M.~A.~Schmidt for helpful correspondence. C.W. and G.D. were was partially supported by DFG Collaborative Research center SFB 1225 (ISOQUANT). This work used the \texttt{Python} package \texttt{PyR@TE\ 3} and the \texttt{Mathematica} package \texttt{SARAH} to compute the RGEs. MDG acknowledges the support of the Agence Nationale de Recherche grant ANR-15-CE31-0002 ``HiggsAutomator,''  and the Labex ``Institut Lagrange de Paris'' (ANR-11-IDEX-0004-02,  ANR-10-LABX-63). 

\appendix

\section{Bounds on the cubic coupling from unitarity\label{app:unitarity}} 

In this appendix, we explore possible bounds on the cubic coupling $\gamma$, in addition to the constraint in the theory below $M_\Delta$ coming from the  redefinition of the effective Higgs quartic coupling Eq.~\ref{eq:maximumgamma}. Above  $M_\Delta$, $\gamma$  does not modify any of the beta functions for the quartic couplings Eqs.~\eqref{eq:lambda}--\eqref{eq:lambda5}, and hence does not change the constraints from vacuum stability (at least in our approximation) through running. However, it can affect the stability conditions regarding the presence of additional charge-breaking minima; and a large $\gamma$ can also violate the constraints from perturbative unitarity. While we leave an investigation of the former to future work, here we shall investigate the latter. 



\begin{figure}
\begin{center}
\includegraphics[width=0.49\columnwidth]{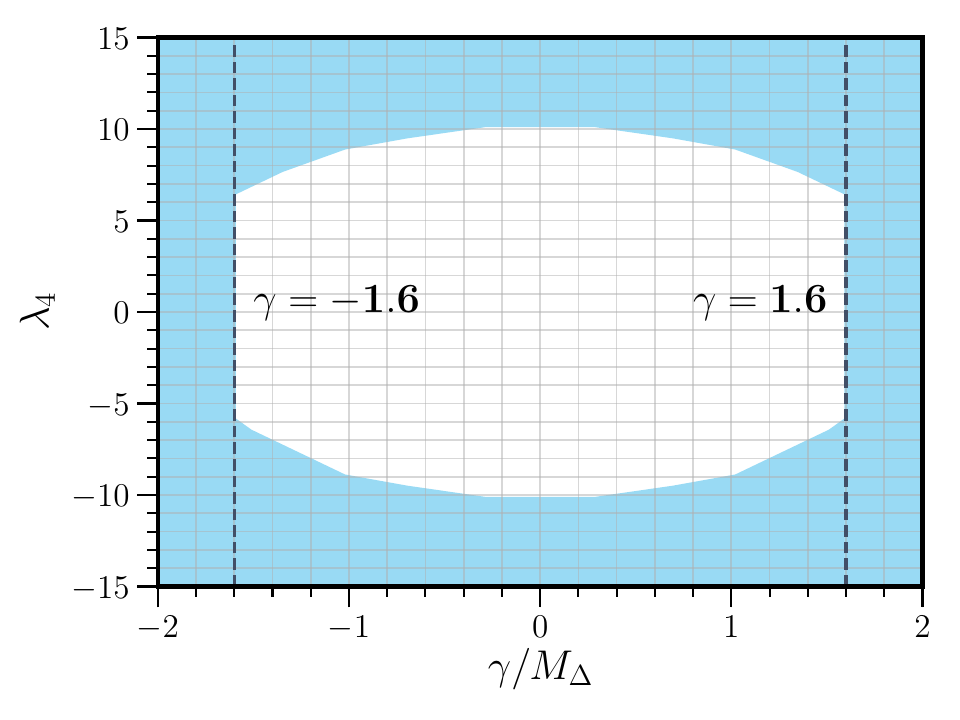}
\includegraphics[width=0.49\columnwidth]{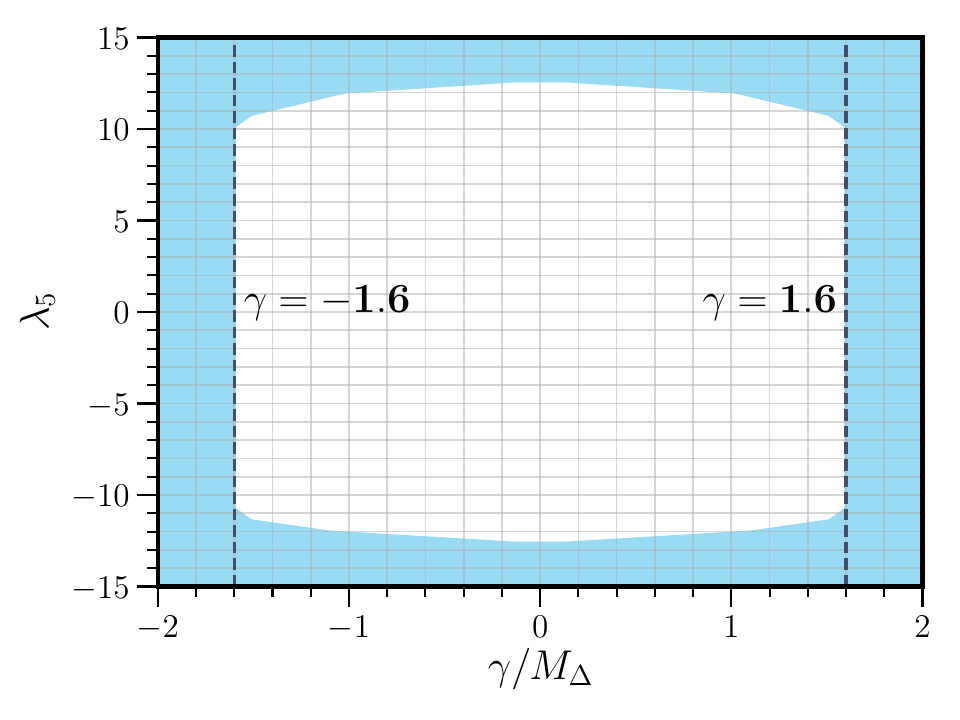}
\caption{ Unitarity bounds on the ratio $\gamma/M_\Delta$ against $\lambda_4$ and $\lambda_5$, including finite scattering energy. \label{fig:unitarity}}
\end{center}
\end{figure}

In the regime $M_\Delta \gg v$ we can only constrain the ratio $\gamma/M_\Delta$ rather than the absolute value due to the absence of other physical scales. We therefore investigated the upper bound on $\gamma/M_\Delta$ from unitarity using the inbuilt routines in \texttt{SARAH} described in \cite{Goodsell:2018tti}. These routines scan over the centre of mass energy in $2\rightarrow 2$ scalar scattering processes, and find the largest value of the largest eigenvalue of the scattering matrix. They are currently restricted to states with trivial $SU(2)$ gauge structure, and so the calculation is performed in the \emph{broken} electroweak phase: this means that we define
\begin{align}
\gamma \equiv& y M_\Delta, \qquad v_\Delta \equiv \frac{v^2 y}{ M_\Delta}, \qquad \lambda_3 \equiv \lambda + y^2,
\end{align}
and scan over scattering energies from well below to well above $M_\Delta$. We verified that the constraints are independent to a high degree of the value of $\lambda \in [0, m_h^2/2v^2] $ and independent of the actual value of $M_\Delta \gg v$ chosen, and find $|y| \lesssim 1.6$ i.e. 
\begin{align}
|\gamma| \lesssim 1.6 M_\Delta .
\end{align}
Note that there are two contributing factors here: the ``conventional'' scattering involving only the quartic couplings (since we must include a large $\lambda_3$ when $y$ is large), and also the scattering involving $s$--, $t$-- and $u$--channel processes and the genuine cubic couplings. However, if we consider only the conventional constraints, the bound becomes much weaker; we have $\lambda_3 \lesssim 8.4$, showing that we have a genuine constraint on the cubic coupling. While the constraint on $y$ is almost independent of $\lambda_{1}$ and $\lambda_2$, there is a non-trivial dependence on $\lambda_4$ and $\lambda_5$, which we show in Fig.~\ref{fig:unitarity}.

\section{2-loop Renormalisation group equations of Y=1 triplet and right handed neutrinos.\label{app:RGE2loop}}

In this appendix we present the RGEs for the type I+II seesaw up to 2-loop extending the results of Ref.~\cite{Schmidt:2007nq}. Our definitions of the beta functions are given by
\begin{align}
\beta\left(X\right) \equiv \mu \frac{d X}{d \mu}\equiv\frac{1}{\left(4 \pi\right)^{2}}\beta^{(1)}(X)+\frac{1}{\left(4 \pi\right)^{4}}\beta^{(2)}(X)\,,
\end{align}
where $X$ would be either a gauge, Yukawa, cubic, quartic or mass coupling.

\subsection{Gauge couplings}
{\allowdisplaybreaks

\begin{align*}
\begin{autobreak}
\beta^{(1)}(g_1) =\frac{47}{10} g_1^{3}
\end{autobreak}
\end{align*}
\begin{align*}
\begin{autobreak}
\beta^{(2)}(g_1) =

 \frac{83}{10} g_1^{5}

+ \frac{171}{10} g_1^{3} g_2^{2}

+ \frac{44}{5} g_1^{3} g_3^{2}

-  \frac{17}{10} g_1^{3} \tr\left(Y_u Y_u^{\dagger} \right)

-  \frac{1}{2} g_1^{3} \tr\left(Y_d Y_d^{\dagger} \right)

-  \frac{3}{2} g_1^{3} \tr\left(Y_e Y_e^{\dagger} \right)

-  \frac{3}{10} g_1^{3} \tr\left(Y_\Phi Y_\Phi^{\dagger} \right)

-  \frac{9}{5} g_1^{3} \tr\left(Y_\Delta Y_\Delta^{*} \right)
\end{autobreak}
\end{align*}
\begin{align*}
\begin{autobreak}
\beta^{(1)}(g_2) =- \frac{5}{2} g_2^{3}
\end{autobreak}
\end{align*}
\begin{align*}
\begin{autobreak}
\beta^{(2)}(g_2) =

 \frac{57}{10} g_1^{2} g_2^{3}

+ \frac{49}{2} g_2^{5}

+ 12 g_2^{3} g_3^{2}

-  \frac{3}{2} g_2^{3} \tr\left(Y_u Y_u^{\dagger} \right)

-  \frac{3}{2} g_2^{3} \tr\left(Y_d Y_d^{\dagger} \right)

-  \frac{1}{2} g_2^{3} \tr\left(Y_e Y_e^{\dagger} \right)

-  \frac{1}{2} g_2^{3} \tr\left(Y_\Phi Y_\Phi^{\dagger} \right)

- 3 g_2^{3} \tr\left(Y_\Delta Y_\Delta^{*} \right)
\end{autobreak}
\end{align*}
\begin{align*}
\begin{autobreak}
\beta^{(1)}(g_3) =-7 g_3^{3}
\end{autobreak}
\end{align*}
\begin{align*}
\begin{autobreak}
\beta^{(2)}(g_3) =

 \frac{11}{10} g_1^{2} g_3^{3}

+ \frac{9}{2} g_2^{2} g_3^{3}

- 26 g_3^{5}

- 2 g_3^{3} \tr\left(Y_u Y_u^{\dagger} \right)

- 2 g_3^{3} \tr\left(Y_d Y_d^{\dagger} \right)
\end{autobreak}
\end{align*}
}

\subsection{Yukawa couplings}
{\allowdisplaybreaks

\begin{align*}
\begin{autobreak}
\beta^{(1)}(Y_u) =

 \frac{3}{2} Y_u Y_u^{\dagger} Y_u

-  \frac{3}{2} Y_d Y_d^{\dagger} Y_u

+ 3 \tr\left(Y_u Y_u^{\dagger} \right) Y_u

+ 3 \tr\left(Y_d Y_d^{\dagger} \right) Y_u

+ \tr\left(Y_e Y_e^{\dagger} \right) Y_u

+ \tr\left(Y_\Phi Y_\Phi^{\dagger} \right) Y_u

-  \frac{17}{20} g_1^{2} Y_u

-  \frac{9}{4} g_2^{2} Y_u

- 8 g_3^{2} Y_u
\end{autobreak}
\end{align*}
\begin{align*}
\begin{autobreak}
\beta^{(2)}(Y_u) =

\frac{3}{2} Y_u Y_u^{\dagger} Y_u Y_u^{\dagger} Y_u

-  \frac{1}{4} Y_u Y_u^{\dagger} Y_d Y_d^{\dagger} Y_u

-  Y_d Y_d^{\dagger} Y_u Y_u^{\dagger} Y_u

+ \frac{11}{4} Y_d Y_d^{\dagger} Y_d Y_d^{\dagger} Y_u

-  \frac{27}{4} \tr\left(Y_u Y_u^{\dagger} Y_u Y_u^{\dagger} \right) Y_u

-  \frac{27}{4} \tr\left(Y_u Y_u^{\dagger} \right) Y_u Y_u^{\dagger} Y_u

+ \frac{3}{2} \tr\left(Y_u Y_u^{\dagger} Y_d Y_d^{\dagger} \right) Y_u

+ \frac{15}{4} \tr\left(Y_u Y_u^{\dagger} \right) Y_d Y_d^{\dagger} Y_u

-  \frac{27}{4} \tr\left(Y_d Y_d^{\dagger} \right) Y_u Y_u^{\dagger} Y_u

-  \frac{27}{4} \tr\left(Y_d Y_d^{\dagger} Y_d Y_d^{\dagger} \right) Y_u

+ \frac{15}{4} \tr\left(Y_d Y_d^{\dagger} \right) Y_d Y_d^{\dagger} Y_u

-  \frac{9}{4} \tr\left(Y_e Y_e^{\dagger} \right) Y_u Y_u^{\dagger} Y_u

+ \frac{5}{4} \tr\left(Y_e Y_e^{\dagger} \right) Y_d Y_d^{\dagger} Y_u

-  \frac{9}{4} \tr\left(Y_e Y_e^{\dagger} Y_e Y_e^{\dagger} \right) Y_u

+ \frac{1}{2} \tr\left(Y_e Y_e^{\dagger} Y_\Phi^{*} Y_\Phi^{\trans} \right) Y_u

-  \frac{9}{2} \tr\left(Y_e Y_e^{\dagger} Y_\Delta^{*} Y_\Delta \right) Y_u

-  \frac{9}{4} \tr\left(Y_\Phi Y_\Phi^{\dagger} \right) Y_u Y_u^{\dagger} Y_u

+ \frac{5}{4} \tr\left(Y_\Phi Y_\Phi^{\dagger} \right) Y_d Y_d^{\dagger} Y_u

-  \frac{9}{4} \tr\left(Y_\Phi Y_\Phi^{\dagger} Y_\Phi Y_\Phi^{\dagger} \right) Y_u

-  \frac{9}{2} \tr\left(Y_\Phi Y_\Phi^{\dagger} Y_\Delta Y_\Delta^{*} \right) Y_u

- 6 \lambda_3 Y_u Y_u^{\dagger} Y_u

+ \frac{3}{2} \lambda_3^{2} Y_u

+ \frac{3}{2} \lambda_4^{2} Y_u

+ 3 \lambda_5^{2} Y_u

+ \frac{223}{80} g_1^{2} Y_u Y_u^{\dagger} Y_u

+ \frac{135}{16} g_2^{2} Y_u Y_u^{\dagger} Y_u

+ 16 g_3^{2} Y_u Y_u^{\dagger} Y_u

-  \frac{43}{80} g_1^{2} Y_d Y_d^{\dagger} Y_u

+ \frac{9}{16} g_2^{2} Y_d Y_d^{\dagger} Y_u

- 16 g_3^{2} Y_d Y_d^{\dagger} Y_u

+ \frac{17}{8} g_1^{2} \tr\left(Y_u Y_u^{\dagger} \right) Y_u

+ \frac{45}{8} g_2^{2} \tr\left(Y_u Y_u^{\dagger} \right) Y_u

+ 20 g_3^{2} \tr\left(Y_u Y_u^{\dagger} \right) Y_u

+ \frac{5}{8} g_1^{2} \tr\left(Y_d Y_d^{\dagger} \right) Y_u

+ \frac{45}{8} g_2^{2} \tr\left(Y_d Y_d^{\dagger} \right) Y_u

+ 20 g_3^{2} \tr\left(Y_d Y_d^{\dagger} \right) Y_u

+ \frac{15}{8} g_1^{2} \tr\left(Y_e Y_e^{\dagger} \right) Y_u

+ \frac{15}{8} g_2^{2} \tr\left(Y_e Y_e^{\dagger} \right) Y_u

+ \frac{3}{8} g_1^{2} \tr\left(Y_\Phi Y_\Phi^{\dagger} \right) Y_u

+ \frac{15}{8} g_2^{2} \tr\left(Y_\Phi Y_\Phi^{\dagger} \right) Y_u

+ \frac{1667}{600} g_1^{4} Y_u

-  \frac{9}{20} g_1^{2} g_2^{2} Y_u

+ \frac{19}{15} g_1^{2} g_3^{2} Y_u

-  \frac{15}{4} g_2^{4} Y_u

+ 9 g_2^{2} g_3^{2} Y_u

- 108 g_3^{4} Y_u
\end{autobreak}
\end{align*}
\begin{align*}
\begin{autobreak}
\beta^{(1)}(Y_d) =

-  \frac{3}{2} Y_u Y_u^{\dagger} Y_d

+ \frac{3}{2} Y_d Y_d^{\dagger} Y_d

+ 3 \tr\left(Y_u Y_u^{\dagger} \right) Y_d

+ 3 \tr\left(Y_d Y_d^{\dagger} \right) Y_d

+ \tr\left(Y_e Y_e^{\dagger} \right) Y_d

+ \tr\left(Y_\Phi Y_\Phi^{\dagger} \right) Y_d

-  \frac{1}{4} g_1^{2} Y_d

-  \frac{9}{4} g_2^{2} Y_d

- 8 g_3^{2} Y_d
\end{autobreak}
\end{align*}
\begin{align*}
\begin{autobreak}
\beta^{(2)}(Y_d) =

\frac{11}{4} Y_u Y_u^{\dagger} Y_u Y_u^{\dagger} Y_d

-  Y_u Y_u^{\dagger} Y_d Y_d^{\dagger} Y_d

-  \frac{1}{4} Y_d Y_d^{\dagger} Y_u Y_u^{\dagger} Y_d

+ \frac{3}{2} Y_d Y_d^{\dagger} Y_d Y_d^{\dagger} Y_d

-  \frac{27}{4} \tr\left(Y_u Y_u^{\dagger} Y_u Y_u^{\dagger} \right) Y_d

+ \frac{15}{4} \tr\left(Y_u Y_u^{\dagger} \right) Y_u Y_u^{\dagger} Y_d

+ \frac{3}{2} \tr\left(Y_u Y_u^{\dagger} Y_d Y_d^{\dagger} \right) Y_d

-  \frac{27}{4} \tr\left(Y_u Y_u^{\dagger} \right) Y_d Y_d^{\dagger} Y_d

+ \frac{15}{4} \tr\left(Y_d Y_d^{\dagger} \right) Y_u Y_u^{\dagger} Y_d

-  \frac{27}{4} \tr\left(Y_d Y_d^{\dagger} Y_d Y_d^{\dagger} \right) Y_d

-  \frac{27}{4} \tr\left(Y_d Y_d^{\dagger} \right) Y_d Y_d^{\dagger} Y_d

+ \frac{5}{4} \tr\left(Y_e Y_e^{\dagger} \right) Y_u Y_u^{\dagger} Y_d

-  \frac{9}{4} \tr\left(Y_e Y_e^{\dagger} \right) Y_d Y_d^{\dagger} Y_d

-  \frac{9}{4} \tr\left(Y_e Y_e^{\dagger} Y_e Y_e^{\dagger} \right) Y_d

+ \frac{1}{2} \tr\left(Y_e Y_e^{\dagger} Y_\Phi^{*} Y_\Phi^{\trans} \right) Y_d

-  \frac{9}{2} \tr\left(Y_e Y_e^{\dagger} Y_\Delta^{*} Y_\Delta \right) Y_d

+ \frac{5}{4} \tr\left(Y_\Phi Y_\Phi^{\dagger} \right) Y_u Y_u^{\dagger} Y_d

-  \frac{9}{4} \tr\left(Y_\Phi Y_\Phi^{\dagger} \right) Y_d Y_d^{\dagger} Y_d

-  \frac{9}{4} \tr\left(Y_\Phi Y_\Phi^{\dagger} Y_\Phi Y_\Phi^{\dagger} \right) Y_d

-  \frac{9}{2} \tr\left(Y_\Phi Y_\Phi^{\dagger} Y_\Delta Y_\Delta^{*} \right) Y_d

- 6 \lambda_3 Y_d Y_d^{\dagger} Y_d

+ \frac{3}{2} \lambda_3^{2} Y_d

+ \frac{3}{2} \lambda_4^{2} Y_d

+ 3 \lambda_5^{2} Y_d

-  \frac{79}{80} g_1^{2} Y_u Y_u^{\dagger} Y_d

+ \frac{9}{16} g_2^{2} Y_u Y_u^{\dagger} Y_d

- 16 g_3^{2} Y_u Y_u^{\dagger} Y_d

+ \frac{187}{80} g_1^{2} Y_d Y_d^{\dagger} Y_d

+ \frac{135}{16} g_2^{2} Y_d Y_d^{\dagger} Y_d

+ 16 g_3^{2} Y_d Y_d^{\dagger} Y_d

+ \frac{17}{8} g_1^{2} \tr\left(Y_u Y_u^{\dagger} \right) Y_d

+ \frac{45}{8} g_2^{2} \tr\left(Y_u Y_u^{\dagger} \right) Y_d

+ 20 g_3^{2} \tr\left(Y_u Y_u^{\dagger} \right) Y_d

+ \frac{5}{8} g_1^{2} \tr\left(Y_d Y_d^{\dagger} \right) Y_d

+ \frac{45}{8} g_2^{2} \tr\left(Y_d Y_d^{\dagger} \right) Y_d

+ 20 g_3^{2} \tr\left(Y_d Y_d^{\dagger} \right) Y_d

+ \frac{15}{8} g_1^{2} \tr\left(Y_e Y_e^{\dagger} \right) Y_d

+ \frac{15}{8} g_2^{2} \tr\left(Y_e Y_e^{\dagger} \right) Y_d

+ \frac{3}{8} g_1^{2} \tr\left(Y_\Phi Y_\Phi^{\dagger} \right) Y_d

+ \frac{15}{8} g_2^{2} \tr\left(Y_\Phi Y_\Phi^{\dagger} \right) Y_d

-  \frac{43}{600} g_1^{4} Y_d

-  \frac{27}{20} g_1^{2} g_2^{2} Y_d

+ \frac{31}{15} g_1^{2} g_3^{2} Y_d

-  \frac{15}{4} g_2^{4} Y_d

+ 9 g_2^{2} g_3^{2} Y_d

- 108 g_3^{4} Y_d
\end{autobreak}
\end{align*}
\begin{align*}
\begin{autobreak}
\beta^{(1)}(Y_e) =

\frac{3}{2} Y_e Y_e^{\dagger} Y_e

-  \frac{3}{2} Y_\Phi^{*} Y_\Phi^{\trans} Y_e

+ 3 Y_\Delta^{*} Y_\Delta Y_e

+ 3 \tr\left(Y_u Y_u^{\dagger} \right) Y_e

+ 3 \tr\left(Y_d Y_d^{\dagger} \right) Y_e

+ \tr\left(Y_e Y_e^{\dagger} \right) Y_e

+ \tr\left(Y_\Phi Y_\Phi^{\dagger} \right) Y_e

-  \frac{9}{4} g_1^{2} Y_e

-  \frac{9}{4} g_2^{2} Y_e
\end{autobreak}
\end{align*}
\begin{align*}
\begin{autobreak}
\beta^{(2)}(Y_e) =

+ \frac{3}{2} Y_e Y_e^{\dagger} Y_e Y_e^{\dagger} Y_e

-  \frac{1}{4} Y_e Y_e^{\dagger} Y_\Phi^{*} Y_\Phi^{\trans} Y_e

-  \frac{3}{2} Y_e Y_e^{\dagger} Y_\Delta^{*} Y_\Delta Y_e

-  Y_\Phi^{*} Y_\Phi^{\trans} Y_e Y_e^{\dagger} Y_e

+ \frac{11}{4} Y_\Phi^{*} Y_\Phi^{\trans} Y_\Phi^{*} Y_\Phi^{\trans} Y_e

+ 6 Y_\Phi^{*} Y_\Phi^{\trans} Y_\Delta^{*} Y_\Delta Y_e

+ \frac{45}{4} Y_\Delta^{*} Y_e^{*} Y_e^{\trans} Y_\Delta Y_e

-  \frac{3}{4} Y_\Delta^{*} Y_\Phi Y_\Phi^{\dagger} Y_\Delta Y_e

-  \frac{9}{2} Y_\Delta^{*} Y_\Delta Y_\Delta^{*} Y_\Delta Y_e

-  \frac{27}{4} \tr\left(Y_u Y_u^{\dagger} Y_u Y_u^{\dagger} \right) Y_e

+ \frac{3}{2} \tr\left(Y_u Y_u^{\dagger} Y_d Y_d^{\dagger} \right) Y_e

-  \frac{27}{4} \tr\left(Y_u Y_u^{\dagger} \right) Y_e Y_e^{\dagger} Y_e

+ \frac{15}{4} \tr\left(Y_u Y_u^{\dagger} \right) Y_\Phi^{*} Y_\Phi^{\trans} Y_e

-  \frac{27}{4} \tr\left(Y_d Y_d^{\dagger} Y_d Y_d^{\dagger} \right) Y_e

-  \frac{27}{4} \tr\left(Y_d Y_d^{\dagger} \right) Y_e Y_e^{\dagger} Y_e

+ \frac{15}{4} \tr\left(Y_d Y_d^{\dagger} \right) Y_\Phi^{*} Y_\Phi^{\trans} Y_e

-  \frac{9}{4} \tr\left(Y_e Y_e^{\dagger} Y_e Y_e^{\dagger} \right) Y_e

-  \frac{9}{4} \tr\left(Y_e Y_e^{\dagger} \right) Y_e Y_e^{\dagger} Y_e

+ \frac{1}{2} \tr\left(Y_e Y_e^{\dagger} Y_\Phi^{*} Y_\Phi^{\trans} \right) Y_e

+ \frac{5}{4} \tr\left(Y_e Y_e^{\dagger} \right) Y_\Phi^{*} Y_\Phi^{\trans} Y_e

-  \frac{9}{2} \tr\left(Y_e Y_e^{\dagger} Y_\Delta^{*} Y_\Delta \right) Y_e

-  \frac{9}{4} \tr\left(Y_\Phi Y_\Phi^{\dagger} \right) Y_e Y_e^{\dagger} Y_e

-  \frac{9}{4} \tr\left(Y_\Phi Y_\Phi^{\dagger} Y_\Phi Y_\Phi^{\dagger} \right) Y_e

+ \frac{5}{4} \tr\left(Y_\Phi Y_\Phi^{\dagger} \right) Y_\Phi^{*} Y_\Phi^{\trans} Y_e

-  \frac{9}{2} \tr\left(Y_\Phi Y_\Phi^{\dagger} Y_\Delta Y_\Delta^{*} \right) Y_e

- 9 \tr\left(Y_\Delta Y_\Delta^{*} \right) Y_\Delta^{*} Y_\Delta Y_e

- 6 \lambda_3 Y_e Y_e^{\dagger} Y_e

- 12 \lambda_4 Y_\Delta^{*} Y_\Delta Y_e

- 24 \lambda_5 Y_\Delta^{*} Y_\Delta Y_e

+ \frac{3}{2} \lambda_3^{2} Y_e

+ \frac{3}{2} \lambda_4^{2} Y_e

+ 3 \lambda_5^{2} Y_e

+ \frac{387}{80} g_1^{2} Y_e Y_e^{\dagger} Y_e

+ \frac{135}{16} g_2^{2} Y_e Y_e^{\dagger} Y_e

-  \frac{27}{16} g_1^{2} Y_\Phi^{*} Y_\Phi^{\trans} Y_e

+ \frac{9}{16} g_2^{2} Y_\Phi^{*} Y_\Phi^{\trans} Y_e

+ \frac{18}{5} g_1^{2} Y_\Delta^{*} Y_\Delta Y_e

+ 45 g_2^{2} Y_\Delta^{*} Y_\Delta Y_e

+ \frac{17}{8} g_1^{2} \tr\left(Y_u Y_u^{\dagger} \right) Y_e

+ \frac{45}{8} g_2^{2} \tr\left(Y_u Y_u^{\dagger} \right) Y_e

+ 20 g_3^{2} \tr\left(Y_u Y_u^{\dagger} \right) Y_e

+ \frac{5}{8} g_1^{2} \tr\left(Y_d Y_d^{\dagger} \right) Y_e

+ \frac{45}{8} g_2^{2} \tr\left(Y_d Y_d^{\dagger} \right) Y_e

+ 20 g_3^{2} \tr\left(Y_d Y_d^{\dagger} \right) Y_e

+ \frac{15}{8} g_1^{2} \tr\left(Y_e Y_e^{\dagger} \right) Y_e

+ \frac{15}{8} g_2^{2} \tr\left(Y_e Y_e^{\dagger} \right) Y_e

+ \frac{3}{8} g_1^{2} \tr\left(Y_\Phi Y_\Phi^{\dagger} \right) Y_e

+ \frac{15}{8} g_2^{2} \tr\left(Y_\Phi Y_\Phi^{\dagger} \right) Y_e

+ \frac{1839}{200} g_1^{4} Y_e

+ \frac{27}{20} g_1^{2} g_2^{2} Y_e

-  \frac{15}{4} g_2^{4} Y_e
\end{autobreak}
\end{align*}
\begin{align*}
\begin{autobreak}
\beta^{(1)}(Y_\Phi) =

-  \frac{3}{2} Y_e^{*} Y_e^{\trans} Y_\Phi

+ \frac{3}{2} Y_\Phi Y_\Phi^{\dagger} Y_\Phi

+ 3 Y_\Delta Y_\Delta^{*} Y_\Phi

+ 3 \tr\left(Y_u Y_u^{\dagger} \right) Y_\Phi

+ 3 \tr\left(Y_d Y_d^{\dagger} \right) Y_\Phi

+ \tr\left(Y_e Y_e^{\dagger} \right) Y_\Phi

+ \tr\left(Y_\Phi Y_\Phi^{\dagger} \right) Y_\Phi

-  \frac{9}{20} g_1^{2} Y_\Phi

-  \frac{9}{4} g_2^{2} Y_\Phi
\end{autobreak}
\end{align*}
\begin{align*}
\begin{autobreak}
\beta^{(2)}(Y_\Phi) =

\frac{11}{4} Y_e^{*} Y_e^{\trans} Y_e^{*} Y_e^{\trans} Y_\Phi

-  Y_e^{*} Y_e^{\trans} Y_\Phi Y_\Phi^{\dagger} Y_\Phi

+ 6 Y_e^{*} Y_e^{\trans} Y_\Delta Y_\Delta^{*} Y_\Phi

-  \frac{1}{4} Y_\Phi Y_\Phi^{\dagger} Y_e^{*} Y_e^{\trans} Y_\Phi

+ \frac{3}{2} Y_\Phi Y_\Phi^{\dagger} Y_\Phi Y_\Phi^{\dagger} Y_\Phi

-  \frac{3}{2} Y_\Phi Y_\Phi^{\dagger} Y_\Delta Y_\Delta^{*} Y_\Phi

-  \frac{3}{4} Y_\Delta Y_e Y_e^{\dagger} Y_\Delta^{*} Y_\Phi

+ \frac{45}{4} Y_\Delta Y_\Phi^{*} Y_\Phi^{\trans} Y_\Delta^{*} Y_\Phi

-  \frac{9}{2} Y_\Delta Y_\Delta^{*} Y_\Delta Y_\Delta^{*} Y_\Phi

-  \frac{27}{4} \tr\left(Y_u Y_u^{\dagger} Y_u Y_u^{\dagger} \right) Y_\Phi

+ \frac{3}{2} \tr\left(Y_u Y_u^{\dagger} Y_d Y_d^{\dagger} \right) Y_\Phi

+ \frac{15}{4} \tr\left(Y_u Y_u^{\dagger} \right) Y_e^{*} Y_e^{\trans} Y_\Phi

-  \frac{27}{4} \tr\left(Y_u Y_u^{\dagger} \right) Y_\Phi Y_\Phi^{\dagger} Y_\Phi

-  \frac{27}{4} \tr\left(Y_d Y_d^{\dagger} Y_d Y_d^{\dagger} \right) Y_\Phi

+ \frac{15}{4} \tr\left(Y_d Y_d^{\dagger} \right) Y_e^{*} Y_e^{\trans} Y_\Phi

-  \frac{27}{4} \tr\left(Y_d Y_d^{\dagger} \right) Y_\Phi Y_\Phi^{\dagger} Y_\Phi

-  \frac{9}{4} \tr\left(Y_e Y_e^{\dagger} Y_e Y_e^{\dagger} \right) Y_\Phi

+ \frac{5}{4} \tr\left(Y_e Y_e^{\dagger} \right) Y_e^{*} Y_e^{\trans} Y_\Phi

+ \frac{1}{2} \tr\left(Y_e Y_e^{\dagger} Y_\Phi^{*} Y_\Phi^{\trans} \right) Y_\Phi

-  \frac{9}{4} \tr\left(Y_e Y_e^{\dagger} \right) Y_\Phi Y_\Phi^{\dagger} Y_\Phi

-  \frac{9}{2} \tr\left(Y_e Y_e^{\dagger} Y_\Delta^{*} Y_\Delta \right) Y_\Phi

+ \frac{5}{4} \tr\left(Y_\Phi Y_\Phi^{\dagger} \right) Y_e^{*} Y_e^{\trans} Y_\Phi

-  \frac{9}{4} \tr\left(Y_\Phi Y_\Phi^{\dagger} Y_\Phi Y_\Phi^{\dagger} \right) Y_\Phi

-  \frac{9}{4} \tr\left(Y_\Phi Y_\Phi^{\dagger} \right) Y_\Phi Y_\Phi^{\dagger} Y_\Phi

-  \frac{9}{2} \tr\left(Y_\Phi Y_\Phi^{\dagger} Y_\Delta Y_\Delta^{*} \right) Y_\Phi

- 9 \tr\left(Y_\Delta Y_\Delta^{*} \right) Y_\Delta Y_\Delta^{*} Y_\Phi

- 6 \lambda_3 Y_\Phi Y_\Phi^{\dagger} Y_\Phi

- 12 \lambda_4 Y_\Delta Y_\Delta^{*} Y_\Phi

+ 24 \lambda_5 Y_\Delta Y_\Delta^{*} Y_\Phi

+ \frac{3}{2} \lambda_3^{2} Y_\Phi

+ \frac{3}{2} \lambda_4^{2} Y_\Phi

+ 3 \lambda_5^{2} Y_\Phi

-  \frac{243}{80} g_1^{2} Y_e^{*} Y_e^{\trans} Y_\Phi

+ \frac{9}{16} g_2^{2} Y_e^{*} Y_e^{\trans} Y_\Phi

+ \frac{279}{80} g_1^{2} Y_\Phi Y_\Phi^{\dagger} Y_\Phi

+ \frac{135}{16} g_2^{2} Y_\Phi Y_\Phi^{\dagger} Y_\Phi

+ \frac{72}{5} g_1^{2} Y_\Delta Y_\Delta^{*} Y_\Phi

+ 45 g_2^{2} Y_\Delta Y_\Delta^{*} Y_\Phi

+ \frac{17}{8} g_1^{2} \tr\left(Y_u Y_u^{\dagger} \right) Y_\Phi

+ \frac{45}{8} g_2^{2} \tr\left(Y_u Y_u^{\dagger} \right) Y_\Phi

+ 20 g_3^{2} \tr\left(Y_u Y_u^{\dagger} \right) Y_\Phi

+ \frac{5}{8} g_1^{2} \tr\left(Y_d Y_d^{\dagger} \right) Y_\Phi

+ \frac{45}{8} g_2^{2} \tr\left(Y_d Y_d^{\dagger} \right) Y_\Phi

+ 20 g_3^{2} \tr\left(Y_d Y_d^{\dagger} \right) Y_\Phi

+ \frac{15}{8} g_1^{2} \tr\left(Y_e Y_e^{\dagger} \right) Y_\Phi

+ \frac{15}{8} g_2^{2} \tr\left(Y_e Y_e^{\dagger} \right) Y_\Phi

+ \frac{3}{8} g_1^{2} \tr\left(Y_\Phi Y_\Phi^{\dagger} \right) Y_\Phi

+ \frac{15}{8} g_2^{2} \tr\left(Y_\Phi Y_\Phi^{\dagger} \right) Y_\Phi

+ \frac{177}{200} g_1^{4} Y_\Phi

-  \frac{27}{20} g_1^{2} g_2^{2} Y_\Phi

-  \frac{15}{4} g_2^{4} Y_\Phi
\end{autobreak}
\end{align*}
\begin{align*}
\begin{autobreak}
\beta^{(1)}(Y_\Delta) =

 \frac{1}{2} Y_e^{*} Y_e^{\trans} Y_\Delta

+ \frac{1}{2} Y_\Phi Y_\Phi^{\dagger} Y_\Delta

+ \frac{1}{2} Y_\Delta Y_e Y_e^{\dagger}

+ \frac{1}{2} Y_\Delta Y_\Phi^{*} Y_\Phi^{\trans}

+ 6 Y_\Delta Y_\Delta^{*} Y_\Delta

+ 2 \tr\left(Y_\Delta Y_\Delta^{*} \right) Y_\Delta

-  \frac{9}{10} g_1^{2} Y_\Delta

-  \frac{9}{2} g_2^{2} Y_\Delta
\end{autobreak}
\end{align*}
\begin{align*}
\begin{autobreak}
\beta^{(2)}(Y_\Delta) =

-  \frac{1}{4} Y_e^{*} Y_e^{\trans} Y_e^{*} Y_e^{\trans} Y_\Delta

+ 2 Y_e^{*} Y_e^{\trans} Y_\Delta Y_e Y_e^{\dagger}

-  \frac{1}{4} Y_\Phi Y_\Phi^{\dagger} Y_\Phi Y_\Phi^{\dagger} Y_\Delta

+ 2 Y_\Phi Y_\Phi^{\dagger} Y_\Delta Y_\Phi^{*} Y_\Phi^{\trans}

-  \frac{1}{4} Y_\Delta Y_e Y_e^{\dagger} Y_e Y_e^{\dagger}

-  \frac{3}{4} Y_\Delta Y_e Y_e^{\dagger} Y_\Delta^{*} Y_\Delta

-  \frac{1}{4} Y_\Delta Y_\Phi^{*} Y_\Phi^{\trans} Y_\Phi^{*} Y_\Phi^{\trans}

-  \frac{3}{4} Y_\Delta Y_\Phi^{*} Y_\Phi^{\trans} Y_\Delta^{*} Y_\Delta

-  \frac{3}{4} Y_\Delta Y_\Delta^{*} Y_e^{*} Y_e^{\trans} Y_\Delta

-  \frac{3}{4} Y_\Delta Y_\Delta^{*} Y_\Phi Y_\Phi^{\dagger} Y_\Delta

+ 31 Y_\Delta Y_\Delta^{*} Y_\Delta Y_\Delta^{*} Y_\Delta

-  \frac{9}{4} \tr\left(Y_u Y_u^{\dagger} \right) Y_e^{*} Y_e^{\trans} Y_\Delta

-  \frac{9}{4} \tr\left(Y_u Y_u^{\dagger} \right) Y_\Phi Y_\Phi^{\dagger} Y_\Delta

-  \frac{9}{4} \tr\left(Y_u Y_u^{\dagger} \right) Y_\Delta Y_e Y_e^{\dagger}

-  \frac{9}{4} \tr\left(Y_u Y_u^{\dagger} \right) Y_\Delta Y_\Phi^{*} Y_\Phi^{\trans}

-  \frac{9}{4} \tr\left(Y_d Y_d^{\dagger} \right) Y_e^{*} Y_e^{\trans} Y_\Delta

-  \frac{9}{4} \tr\left(Y_d Y_d^{\dagger} \right) Y_\Phi Y_\Phi^{\dagger} Y_\Delta

-  \frac{9}{4} \tr\left(Y_d Y_d^{\dagger} \right) Y_\Delta Y_e Y_e^{\dagger}

-  \frac{9}{4} \tr\left(Y_d Y_d^{\dagger} \right) Y_\Delta Y_\Phi^{*} Y_\Phi^{\trans}

-  \frac{3}{4} \tr\left(Y_e Y_e^{\dagger} \right) Y_e^{*} Y_e^{\trans} Y_\Delta

-  \frac{3}{4} \tr\left(Y_e Y_e^{\dagger} \right) Y_\Phi Y_\Phi^{\dagger} Y_\Delta

-  \frac{3}{4} \tr\left(Y_e Y_e^{\dagger} \right) Y_\Delta Y_e Y_e^{\dagger}

-  \frac{3}{4} \tr\left(Y_e Y_e^{\dagger} \right) Y_\Delta Y_\Phi^{*} Y_\Phi^{\trans}

- 3 \tr\left(Y_e Y_e^{\dagger} Y_\Delta^{*} Y_\Delta \right) Y_\Delta

-  \frac{3}{4} \tr\left(Y_\Phi Y_\Phi^{\dagger} \right) Y_e^{*} Y_e^{\trans} Y_\Delta

-  \frac{3}{4} \tr\left(Y_\Phi Y_\Phi^{\dagger} \right) Y_\Phi Y_\Phi^{\dagger} Y_\Delta

-  \frac{3}{4} \tr\left(Y_\Phi Y_\Phi^{\dagger} \right) Y_\Delta Y_e Y_e^{\dagger}

-  \frac{3}{4} \tr\left(Y_\Phi Y_\Phi^{\dagger} \right) Y_\Delta Y_\Phi^{*} Y_\Phi^{\trans}

- 3 \tr\left(Y_\Phi Y_\Phi^{\dagger} Y_\Delta Y_\Delta^{*} \right) Y_\Delta

- 18 \tr\left(Y_\Delta Y_\Delta^{*} Y_\Delta Y_\Delta^{*} \right) Y_\Delta

- 18 \tr\left(Y_\Delta Y_\Delta^{*} \right) Y_\Delta Y_\Delta^{*} Y_\Delta

- 2 \lambda_4 Y_e^{*} Y_e^{\trans} Y_\Delta

- 4 \lambda_5 Y_e^{*} Y_e^{\trans} Y_\Delta

- 2 \lambda_4 Y_\Phi Y_\Phi^{\dagger} Y_\Delta

+ 4 \lambda_5 Y_\Phi Y_\Phi^{\dagger} Y_\Delta

- 2 \lambda_4 Y_\Delta Y_e Y_e^{\dagger}

- 4 \lambda_5 Y_\Delta Y_e Y_e^{\dagger}

- 2 \lambda_4 Y_\Delta Y_\Phi^{*} Y_\Phi^{\trans}

+ 4 \lambda_5 Y_\Delta Y_\Phi^{*} Y_\Phi^{\trans}

- 24 \lambda_1 Y_\Delta Y_\Delta^{*} Y_\Delta

+ 4 \lambda_2 Y_\Delta Y_\Delta^{*} Y_\Delta

+ 2 \lambda_1^{2} Y_\Delta

+ \lambda_1 \lambda_2 Y_\Delta

+ \frac{3}{4} \lambda_2^{2} Y_\Delta

+ \lambda_4^{2} Y_\Delta

+ 2 \lambda_5^{2} Y_\Delta

-  \frac{123}{80} g_1^{2} Y_e^{*} Y_e^{\trans} Y_\Delta

+ \frac{33}{16} g_2^{2} Y_e^{*} Y_e^{\trans} Y_\Delta

+ \frac{33}{80} g_1^{2} Y_\Phi Y_\Phi^{\dagger} Y_\Delta

+ \frac{33}{16} g_2^{2} Y_\Phi Y_\Phi^{\dagger} Y_\Delta

-  \frac{123}{80} g_1^{2} Y_\Delta Y_e Y_e^{\dagger}

+ \frac{33}{16} g_2^{2} Y_\Delta Y_e Y_e^{\dagger}

+ \frac{33}{80} g_1^{2} Y_\Delta Y_\Phi^{*} Y_\Phi^{\trans}

+ \frac{33}{16} g_2^{2} Y_\Delta Y_\Phi^{*} Y_\Phi^{\trans}

+ \frac{198}{5} g_1^{2} Y_\Delta Y_\Delta^{*} Y_\Delta

+ 96 g_2^{2} Y_\Delta Y_\Delta^{*} Y_\Delta

+ \frac{3}{2} g_1^{2} \tr\left(Y_\Delta Y_\Delta^{*} \right) Y_\Delta

+ \frac{15}{2} g_2^{2} \tr\left(Y_\Delta Y_\Delta^{*} \right) Y_\Delta

-  \frac{1137}{400} g_1^{4} Y_\Delta

-  \frac{135}{8} g_1^{2} g_2^{2} Y_\Delta

-  \frac{285}{16} g_2^{4} Y_\Delta
\end{autobreak}
\end{align*}
}

\subsection{cubic couplings}
{\allowdisplaybreaks

\begin{align*}
\begin{autobreak}
\beta^{(1)}(\gamma) =

-  \frac{27}{10} g_1^{2} \gamma

-  \frac{21}{2} g_2^{2} \gamma

+ 2 \lambda_3 \gamma

+ 4 \lambda_4 \gamma

- 8 \lambda_5 \gamma

+ 6 \gamma \tr\left(Y_u Y_u^{\dagger} \right)

+ 6 \gamma \tr\left(Y_d Y_d^{\dagger} \right)

+ 2 \gamma \tr\left(Y_e Y_e^{\dagger} \right)

+ 2 \gamma \tr\left(Y_\Phi Y_\Phi^{\dagger} \right)

+ 2 \gamma \tr\left(Y_\Delta Y_\Delta^{*} \right)

-  2 \sqrt{2} \tr\left(Y_\Phi M_\nu^{\dagger} Y_\Phi^{\trans} Y_\Delta^{*} \right)

-  2 \sqrt{2} \tr\left(Y_\Delta Y_\Phi^{*} M_\nu Y_\Phi^{\dagger} \right)
\end{autobreak}
\end{align*}
\begin{align*}
\begin{autobreak}
\beta^{(2)}(\gamma) =

\frac{5391}{400} g_1^{4} \gamma

+ \frac{273}{40} g_1^{2} g_2^{2} \gamma

-  \frac{329}{16} g_2^{4} \gamma

-  \frac{6}{5} \lambda_3 g_1^{2} \gamma

+ \frac{51}{5} \lambda_4 g_1^{2} \gamma

-  \frac{102}{5} \lambda_5 g_1^{2} \gamma

+ 35 \lambda_4 g_2^{2} \gamma

- 70 \lambda_5 g_2^{2} \gamma

- 7 \lambda_3^{2} \gamma

- 20 \lambda_3 \lambda_4 \gamma

+ 24 \lambda_3 \lambda_5 \gamma

+ 2 \lambda_1^{2} \gamma

+ \lambda_1 \lambda_2 \gamma

- 16 \lambda_1 \lambda_4 \gamma

+ 8 \lambda_1 \lambda_5 \gamma

+ \frac{3}{4} \lambda_2^{2} \gamma

- 4 \lambda_2 \lambda_4 \gamma

- 8 \lambda_2 \lambda_5 \gamma

- 10 \lambda_4^{2} \gamma

+ 32 \lambda_4 \lambda_5 \gamma

- 12 \lambda_5^{2} \gamma

+ \frac{17}{4} g_1^{2} \gamma \tr\left(Y_u Y_u^{\dagger} \right)

+ \frac{5}{4} g_1^{2} \gamma \tr\left(Y_d Y_d^{\dagger} \right)

+ \frac{15}{4} g_1^{2} \gamma \tr\left(Y_e Y_e^{\dagger} \right)

+ \frac{3}{4} g_1^{2} \gamma \tr\left(Y_\Phi Y_\Phi^{\dagger} \right)

+ \frac{3}{2} g_1^{2} \gamma \tr\left(Y_\Delta Y_\Delta^{*} \right)

+ \frac{45}{4} g_2^{2} \gamma \tr\left(Y_u Y_u^{\dagger} \right)

+ \frac{45}{4} g_2^{2} \gamma \tr\left(Y_d Y_d^{\dagger} \right)

+ \frac{15}{4} g_2^{2} \gamma \tr\left(Y_e Y_e^{\dagger} \right)

+ \frac{15}{4} g_2^{2} \gamma \tr\left(Y_\Phi Y_\Phi^{\dagger} \right)

+ \frac{15}{2} g_2^{2} \gamma \tr\left(Y_\Delta Y_\Delta^{*} \right)

+ 40 g_3^{2} \gamma \tr\left(Y_u Y_u^{\dagger} \right)

+ 40 g_3^{2} \gamma \tr\left(Y_d Y_d^{\dagger} \right)

- 12 \lambda_3 \gamma \tr\left(Y_u Y_u^{\dagger} \right)

- 12 \lambda_3 \gamma \tr\left(Y_d Y_d^{\dagger} \right)

- 4 \lambda_3 \gamma \tr\left(Y_e Y_e^{\dagger} \right)

- 4 \lambda_3 \gamma \tr\left(Y_\Phi Y_\Phi^{\dagger} \right)

- 12 \lambda_4 \gamma \tr\left(Y_u Y_u^{\dagger} \right)

- 12 \lambda_4 \gamma \tr\left(Y_d Y_d^{\dagger} \right)

- 4 \lambda_4 \gamma \tr\left(Y_e Y_e^{\dagger} \right)

- 4 \lambda_4 \gamma \tr\left(Y_\Phi Y_\Phi^{\dagger} \right)

- 8 \lambda_4 \gamma \tr\left(Y_\Delta Y_\Delta^{*} \right)

+ 24 \lambda_5 \gamma \tr\left(Y_u Y_u^{\dagger} \right)

+ 24 \lambda_5 \gamma \tr\left(Y_d Y_d^{\dagger} \right)

+ 8 \lambda_5 \gamma \tr\left(Y_e Y_e^{\dagger} \right)

+ 8 \lambda_5 \gamma \tr\left(Y_\Phi Y_\Phi^{\dagger} \right)

+ 16 \lambda_5 \gamma \tr\left(Y_\Delta Y_\Delta^{*} \right)

-  \frac{3}{2} \gamma \tr\left(Y_u Y_u^{\dagger} Y_u Y_u^{\dagger} \right)

+ 3 \gamma \tr\left(Y_u Y_u^{\dagger} Y_d Y_d^{\dagger} \right)

-  \frac{3}{2} \gamma \tr\left(Y_d Y_d^{\dagger} Y_d Y_d^{\dagger} \right)

-  \frac{1}{2} \gamma \tr\left(Y_e Y_e^{\dagger} Y_e Y_e^{\dagger} \right)

+ \gamma \tr\left(Y_e Y_e^{\dagger} Y_\Phi^{*} Y_\Phi^{\trans} \right)

- 28 \gamma \tr\left(Y_e Y_e^{\dagger} Y_\Delta^{*} Y_\Delta \right)

-  \frac{1}{2} \gamma \tr\left(Y_\Phi Y_\Phi^{\dagger} Y_\Phi Y_\Phi^{\dagger} \right)

- 12 \gamma \tr\left(Y_\Phi Y_\Phi^{\dagger} Y_\Delta Y_\Delta^{*} \right)

- 18 \gamma \tr\left(Y_\Delta Y_\Delta^{*} Y_\Delta Y_\Delta^{*} \right)

+ \frac{3 \sqrt{2}}{10} g_1^{2} \tr\left(Y_\Phi M_\nu^{\dagger} Y_\Phi^{\trans} Y_\Delta^{*} \right)

+ \frac{3 \sqrt{2}}{10} g_1^{2} \tr\left(Y_\Delta Y_\Phi^{*} M_\nu Y_\Phi^{\dagger} \right)

+ \frac{\sqrt{2}}{2} g_2^{2} \tr\left(Y_\Phi M_\nu^{\dagger} Y_\Phi^{\trans} Y_\Delta^{*} \right)

+ \frac{\sqrt{2}}{2} g_2^{2} \tr\left(Y_\Delta Y_\Phi^{*} M_\nu Y_\Phi^{\dagger} \right)

+ 2 \sqrt{2} \lambda_3 \tr\left(Y_\Phi M_\nu^{\dagger} Y_\Phi^{\trans} Y_\Delta^{*} \right)

+ 2 \sqrt{2} \lambda_3 \tr\left(Y_\Delta Y_\Phi^{*} M_\nu Y_\Phi^{\dagger} \right)

-  2 \sqrt{2} \tr\left(Y_e Y_e^{\dagger} Y_\Phi^{*} M_\nu^{\trans} Y_\Phi^{\dagger} Y_\Delta \right)

-  2 \sqrt{2} \tr\left(Y_e Y_e^{\dagger} Y_\Delta^{*} Y_\Phi M_\nu^{\dagger} Y_\Phi^{\trans} \right)

+ 4 \sqrt{2} \tr\left(Y_\Phi Y_\Phi^{\dagger} Y_\Phi M_\nu^{\dagger} Y_\Phi^{\trans} Y_\Delta^{*} \right)

+ 6 \sqrt{2} \tr\left(Y_\Phi Y_\Phi^{\dagger} Y_\Phi M_\nu^{*} Y_\Phi^{\trans} Y_\Delta^{*} \right)

+ 4 \sqrt{2} \tr\left(Y_\Phi Y_\Phi^{\dagger} Y_\Delta Y_\Phi^{*} M_\nu^{\trans} Y_\Phi^{\dagger} \right)

+ 6 \sqrt{2} \tr\left(Y_\Phi Y_\Phi^{\dagger} Y_\Delta Y_\Phi^{*} M_\nu Y_\Phi^{\dagger} \right)

+ 12 \sqrt{2} \tr\left(Y_\Phi M_\nu^{\dagger} Y_\Phi^{\trans} Y_\Delta^{*} Y_\Delta Y_\Delta^{*} \right)

+ 12 \sqrt{2} \tr\left(Y_\Delta Y_\Phi^{*} M_\nu Y_\Phi^{\dagger} Y_\Delta Y_\Delta^{*} \right)
\end{autobreak}
\end{align*}
}

\subsection{Quartic couplings}
{\allowdisplaybreaks

\begin{align*}
\begin{autobreak}
\beta^{(1)}(\lambda_3) =

 12 \lambda_3^{2}

+ 6 \lambda_4^{2}

+ 4 \lambda_5^{2}

-  \frac{9}{5} \lambda_3 g_1^{2}

- 9 \lambda_3 g_2^{2}

+ \frac{27}{100} g_1^{4}

+ \frac{9}{10} g_1^{2} g_2^{2}

+ \frac{9}{4} g_2^{4}

+ 12 \lambda_3 \tr\left(Y_u Y_u^{\dagger} \right)

+ 12 \lambda_3 \tr\left(Y_d Y_d^{\dagger} \right)

+ 4 \lambda_3 \tr\left(Y_e Y_e^{\dagger} \right)

+ 4 \lambda_3 \tr\left(Y_\Phi Y_\Phi^{\dagger} \right)

- 12 \tr\left(Y_u Y_u^{\dagger} Y_u Y_u^{\dagger} \right)

- 12 \tr\left(Y_d Y_d^{\dagger} Y_d Y_d^{\dagger} \right)

- 4 \tr\left(Y_e Y_e^{\dagger} Y_e Y_e^{\dagger} \right)

- 4 \tr\left(Y_\Phi Y_\Phi^{\dagger} Y_\Phi Y_\Phi^{\dagger} \right)
\end{autobreak}
\end{align*}
\begin{align*}
\begin{autobreak}
\beta^{(2)}(\lambda_3) =

- 78 \lambda_3^{3}

- 30 \lambda_3 \lambda_4^{2}

- 28 \lambda_3 \lambda_5^{2}

- 24 \lambda_4^{3}

- 80 \lambda_4 \lambda_5^{2}

+ \frac{54}{5} \lambda_3^{2} g_1^{2}

+ 54 \lambda_3^{2} g_2^{2}

+ \frac{144}{5} \lambda_4^{2} g_1^{2}

+ 96 \lambda_4^{2} g_2^{2}

+ \frac{96}{5} \lambda_5^{2} g_1^{2}

+ 40 \lambda_5^{2} g_2^{2}

+ \frac{2283}{200} \lambda_3 g_1^{4}

+ \frac{117}{20} \lambda_3 g_1^{2} g_2^{2}

+ \frac{15}{8} \lambda_3 g_2^{4}

+ \frac{54}{5} \lambda_4 g_1^{4}

+ 60 \lambda_4 g_2^{4}

- 24 \lambda_5 g_1^{2} g_2^{2}

-  \frac{4167}{1000} g_1^{6}

-  \frac{1929}{200} g_1^{4} g_2^{2}

-  \frac{69}{8} g_1^{2} g_2^{4}

+ \frac{249}{8} g_2^{6}

- 72 \lambda_3^{2} \tr\left(Y_u Y_u^{\dagger} \right)

- 72 \lambda_3^{2} \tr\left(Y_d Y_d^{\dagger} \right)

- 24 \lambda_3^{2} \tr\left(Y_e Y_e^{\dagger} \right)

- 24 \lambda_3^{2} \tr\left(Y_\Phi Y_\Phi^{\dagger} \right)

- 24 \lambda_4^{2} \tr\left(Y_\Delta Y_\Delta^{*} \right)

- 16 \lambda_5^{2} \tr\left(Y_\Delta Y_\Delta^{*} \right)

+ \frac{17}{2} \lambda_3 g_1^{2} \tr\left(Y_u Y_u^{\dagger} \right)

+ \frac{5}{2} \lambda_3 g_1^{2} \tr\left(Y_d Y_d^{\dagger} \right)

+ \frac{15}{2} \lambda_3 g_1^{2} \tr\left(Y_e Y_e^{\dagger} \right)

+ \frac{3}{2} \lambda_3 g_1^{2} \tr\left(Y_\Phi Y_\Phi^{\dagger} \right)

+ \frac{45}{2} \lambda_3 g_2^{2} \tr\left(Y_u Y_u^{\dagger} \right)

+ \frac{45}{2} \lambda_3 g_2^{2} \tr\left(Y_d Y_d^{\dagger} \right)

+ \frac{15}{2} \lambda_3 g_2^{2} \tr\left(Y_e Y_e^{\dagger} \right)

+ \frac{15}{2} \lambda_3 g_2^{2} \tr\left(Y_\Phi Y_\Phi^{\dagger} \right)

+ 80 \lambda_3 g_3^{2} \tr\left(Y_u Y_u^{\dagger} \right)

+ 80 \lambda_3 g_3^{2} \tr\left(Y_d Y_d^{\dagger} \right)

-  \frac{171}{50} g_1^{4} \tr\left(Y_u Y_u^{\dagger} \right)

+ \frac{9}{10} g_1^{4} \tr\left(Y_d Y_d^{\dagger} \right)

-  \frac{9}{2} g_1^{4} \tr\left(Y_e Y_e^{\dagger} \right)

-  \frac{9}{50} g_1^{4} \tr\left(Y_\Phi Y_\Phi^{\dagger} \right)

+ \frac{63}{5} g_1^{2} g_2^{2} \tr\left(Y_u Y_u^{\dagger} \right)

+ \frac{27}{5} g_1^{2} g_2^{2} \tr\left(Y_d Y_d^{\dagger} \right)

+ \frac{33}{5} g_1^{2} g_2^{2} \tr\left(Y_e Y_e^{\dagger} \right)

-  \frac{3}{5} g_1^{2} g_2^{2} \tr\left(Y_\Phi Y_\Phi^{\dagger} \right)

-  \frac{9}{2} g_2^{4} \tr\left(Y_u Y_u^{\dagger} \right)

-  \frac{9}{2} g_2^{4} \tr\left(Y_d Y_d^{\dagger} \right)

-  \frac{3}{2} g_2^{4} \tr\left(Y_e Y_e^{\dagger} \right)

-  \frac{3}{2} g_2^{4} \tr\left(Y_\Phi Y_\Phi^{\dagger} \right)

- 3 \lambda_3 \tr\left(Y_u Y_u^{\dagger} Y_u Y_u^{\dagger} \right)

- 42 \lambda_3 \tr\left(Y_u Y_u^{\dagger} Y_d Y_d^{\dagger} \right)

- 3 \lambda_3 \tr\left(Y_d Y_d^{\dagger} Y_d Y_d^{\dagger} \right)

-  \lambda_3 \tr\left(Y_e Y_e^{\dagger} Y_e Y_e^{\dagger} \right)

- 14 \lambda_3 \tr\left(Y_e Y_e^{\dagger} Y_\Phi^{*} Y_\Phi^{\trans} \right)

- 18 \lambda_3 \tr\left(Y_e Y_e^{\dagger} Y_\Delta^{*} Y_\Delta \right)

-  \lambda_3 \tr\left(Y_\Phi Y_\Phi^{\dagger} Y_\Phi Y_\Phi^{\dagger} \right)

- 18 \lambda_3 \tr\left(Y_\Phi Y_\Phi^{\dagger} Y_\Delta Y_\Delta^{*} \right)

-  \frac{16}{5} g_1^{2} \tr\left(Y_u Y_u^{\dagger} Y_u Y_u^{\dagger} \right)

+ \frac{8}{5} g_1^{2} \tr\left(Y_d Y_d^{\dagger} Y_d Y_d^{\dagger} \right)

-  \frac{24}{5} g_1^{2} \tr\left(Y_e Y_e^{\dagger} Y_e Y_e^{\dagger} \right)

- 64 g_3^{2} \tr\left(Y_u Y_u^{\dagger} Y_u Y_u^{\dagger} \right)

- 64 g_3^{2} \tr\left(Y_d Y_d^{\dagger} Y_d Y_d^{\dagger} \right)

+ 60 \tr\left(Y_u Y_u^{\dagger} Y_u Y_u^{\dagger} Y_u Y_u^{\dagger} \right)

- 12 \tr\left(Y_u Y_u^{\dagger} Y_u Y_u^{\dagger} Y_d Y_d^{\dagger} \right)

- 12 \tr\left(Y_u Y_u^{\dagger} Y_d Y_d^{\dagger} Y_d Y_d^{\dagger} \right)

+ 60 \tr\left(Y_d Y_d^{\dagger} Y_d Y_d^{\dagger} Y_d Y_d^{\dagger} \right)

+ 20 \tr\left(Y_e Y_e^{\dagger} Y_e Y_e^{\dagger} Y_e Y_e^{\dagger} \right)

- 4 \tr\left(Y_e Y_e^{\dagger} Y_e Y_e^{\dagger} Y_\Phi^{*} Y_\Phi^{\trans} \right)

+ 24 \tr\left(Y_e Y_e^{\dagger} Y_e Y_e^{\dagger} Y_\Delta^{*} Y_\Delta \right)

+ 16 \tr\left(Y_e Y_e^{\dagger} Y_\Delta^{*} Y_e^{*} Y_e^{\trans} Y_\Delta \right)

+ 16 \tr\left(Y_e Y_e^{\dagger} Y_\Delta^{*} Y_\Phi Y_\Phi^{\dagger} Y_\Delta \right)

- 4 \tr\left(Y_\Phi Y_\Phi^{\dagger} Y_\Phi Y_\Phi^{\dagger} Y_e^{*} Y_e^{\trans} \right)

+ 20 \tr\left(Y_\Phi Y_\Phi^{\dagger} Y_\Phi Y_\Phi^{\dagger} Y_\Phi Y_\Phi^{\dagger} \right)

+ 24 \tr\left(Y_\Phi Y_\Phi^{\dagger} Y_\Phi Y_\Phi^{\dagger} Y_\Delta Y_\Delta^{*} \right)

+ 16 \tr\left(Y_\Phi Y_\Phi^{\dagger} Y_\Delta Y_\Phi^{*} Y_\Phi^{\trans} Y_\Delta^{*} \right)
\end{autobreak}
\end{align*}
\begin{align*}
\begin{autobreak}
\beta^{(1)}(\lambda_1) =

14 \lambda_1^{2}

+ 4 \lambda_1 \lambda_2

+ 2 \lambda_2^{2}

+ 4 \lambda_4^{2}

+ 4 \lambda_5^{2}

-  \frac{36}{5} \lambda_1 g_1^{2}

- 24 \lambda_1 g_2^{2}

+ \frac{108}{25} g_1^{4}

+ \frac{72}{5} g_1^{2} g_2^{2}

+ 18 g_2^{4}

+ 8 \lambda_1 \tr\left(Y_\Delta Y_\Delta^{*} \right)

- 32 \tr\left(Y_\Delta Y_\Delta^{*} Y_\Delta Y_\Delta^{*} \right)
\end{autobreak}
\end{align*}
\begin{align*}
\begin{autobreak}
\beta^{(2)}(\lambda_1) =

- 96 \lambda_1^{3}

- 39 \lambda_1 \lambda_2^{2}

- 20 \lambda_1 \lambda_4^{2}

- 24 \lambda_1 \lambda_5^{2}

- 44 \lambda_1^{2} \lambda_2

- 10 \lambda_2^{3}

- 16 \lambda_4^{3}

- 64 \lambda_4 \lambda_5^{2}

+ \frac{264}{5} \lambda_1^{2} g_1^{2}

+ 176 \lambda_1^{2} g_2^{2}

+ \frac{96}{5} \lambda_1 \lambda_2 g_1^{2}

+ 64 \lambda_1 \lambda_2 g_2^{2}

+ \frac{48}{5} \lambda_2^{2} g_1^{2}

+ 20 \lambda_2^{2} g_2^{2}

+ \frac{24}{5} \lambda_4^{2} g_1^{2}

+ 24 \lambda_4^{2} g_2^{2}

+ \frac{24}{5} \lambda_5^{2} g_1^{2}

+ \frac{2433}{25} \lambda_1 g_1^{4}

+ \frac{408}{5} \lambda_1 g_1^{2} g_2^{2}

+ 198 \lambda_1 g_2^{4}

+ \frac{72}{5} \lambda_2 g_1^{4}

- 48 \lambda_2 g_1^{2} g_2^{2}

+ 64 \lambda_2 g_2^{4}

+ \frac{36}{5} \lambda_4 g_1^{4}

+ 40 \lambda_4 g_2^{4}

- 24 \lambda_5 g_1^{2} g_2^{2}

-  \frac{10764}{125} g_1^{6}

-  \frac{6828}{25} g_1^{4} g_2^{2}

- 372 g_1^{2} g_2^{4}

- 30 g_2^{6}

- 56 \lambda_1^{2} \tr\left(Y_\Delta Y_\Delta^{*} \right)

- 16 \lambda_1 \lambda_2 \tr\left(Y_\Delta Y_\Delta^{*} \right)

- 8 \lambda_2^{2} \tr\left(Y_\Delta Y_\Delta^{*} \right)

- 24 \lambda_4^{2} \tr\left(Y_u Y_u^{\dagger} \right)

- 24 \lambda_4^{2} \tr\left(Y_d Y_d^{\dagger} \right)

- 8 \lambda_4^{2} \tr\left(Y_e Y_e^{\dagger} \right)

- 8 \lambda_4^{2} \tr\left(Y_\Phi Y_\Phi^{\dagger} \right)

- 24 \lambda_5^{2} \tr\left(Y_u Y_u^{\dagger} \right)

- 24 \lambda_5^{2} \tr\left(Y_d Y_d^{\dagger} \right)

- 8 \lambda_5^{2} \tr\left(Y_e Y_e^{\dagger} \right)

- 8 \lambda_5^{2} \tr\left(Y_\Phi Y_\Phi^{\dagger} \right)

+ 6 \lambda_1 g_1^{2} \tr\left(Y_\Delta Y_\Delta^{*} \right)

+ 30 \lambda_1 g_2^{2} \tr\left(Y_\Delta Y_\Delta^{*} \right)

+ \frac{288}{25} g_1^{4} \tr\left(Y_\Delta Y_\Delta^{*} \right)

+ \frac{192}{5} g_1^{2} g_2^{2} \tr\left(Y_\Delta Y_\Delta^{*} \right)

+ 24 g_2^{4} \tr\left(Y_\Delta Y_\Delta^{*} \right)

- 12 \lambda_1 \tr\left(Y_e Y_e^{\dagger} Y_\Delta^{*} Y_\Delta \right)

- 12 \lambda_1 \tr\left(Y_\Phi Y_\Phi^{\dagger} Y_\Delta Y_\Delta^{*} \right)

- 8 \lambda_1 \tr\left(Y_\Delta Y_\Delta^{*} Y_\Delta Y_\Delta^{*} \right)

+ \frac{96}{5} g_1^{2} \tr\left(Y_\Delta Y_\Delta^{*} Y_\Delta Y_\Delta^{*} \right)

+ 32 g_2^{2} \tr\left(Y_\Delta Y_\Delta^{*} Y_\Delta Y_\Delta^{*} \right)

+ 64 \tr\left(Y_e Y_e^{\dagger} Y_\Delta^{*} Y_\Delta Y_\Delta^{*} Y_\Delta \right)

+ 64 \tr\left(Y_\Phi Y_\Phi^{\dagger} Y_\Delta Y_\Delta^{*} Y_\Delta Y_\Delta^{*} \right)

+ 640 \tr\left(Y_\Delta Y_\Delta^{*} Y_\Delta Y_\Delta^{*} Y_\Delta Y_\Delta^{*} \right)
\end{autobreak}
\end{align*}
\begin{align*}
\begin{autobreak}
\beta^{(1)}(\lambda_2) =

12 \lambda_1 \lambda_2

+ 3 \lambda_2^{2}

- 8 \lambda_5^{2}

-  \frac{36}{5} \lambda_2 g_1^{2}

- 24 \lambda_2 g_2^{2}

-  \frac{144}{5} g_1^{2} g_2^{2}

+ 12 g_2^{4}

+ 8 \lambda_2 \tr\left(Y_\Delta Y_\Delta^{*} \right)

+ 32 \tr\left(Y_\Delta Y_\Delta^{*} Y_\Delta Y_\Delta^{*} \right)
\end{autobreak}
\end{align*}
\begin{align*}
\begin{autobreak}
\beta^{(2)}(\lambda_2) =

- 56 \lambda_1 \lambda_2^{2}

+ 32 \lambda_1 \lambda_5^{2}

- 112 \lambda_1^{2} \lambda_2

-  \lambda_2^{3}

- 20 \lambda_2 \lambda_4^{2}

+ 24 \lambda_2 \lambda_5^{2}

+ 64 \lambda_4 \lambda_5^{2}

+ \frac{144}{5} \lambda_1 \lambda_2 g_1^{2}

+ 96 \lambda_1 \lambda_2 g_2^{2}

-  \frac{36}{5} \lambda_2^{2} g_1^{2}

+ 24 \lambda_2^{2} g_2^{2}

-  \frac{48}{5} \lambda_5^{2} g_1^{2}

-  \frac{576}{5} \lambda_1 g_1^{2} g_2^{2}

+ 48 \lambda_1 g_2^{4}

+ \frac{993}{25} \lambda_2 g_1^{4}

+ \frac{504}{5} \lambda_2 g_1^{2} g_2^{2}

+ 14 \lambda_2 g_2^{4}

+ 48 \lambda_5 g_1^{2} g_2^{2}

+ \frac{9336}{25} g_1^{4} g_2^{2}

+ 456 g_1^{2} g_2^{4}

- 308 g_2^{6}

- 48 \lambda_1 \lambda_2 \tr\left(Y_\Delta Y_\Delta^{*} \right)

- 12 \lambda_2^{2} \tr\left(Y_\Delta Y_\Delta^{*} \right)

+ 48 \lambda_5^{2} \tr\left(Y_u Y_u^{\dagger} \right)

+ 48 \lambda_5^{2} \tr\left(Y_d Y_d^{\dagger} \right)

+ 16 \lambda_5^{2} \tr\left(Y_e Y_e^{\dagger} \right)

+ 16 \lambda_5^{2} \tr\left(Y_\Phi Y_\Phi^{\dagger} \right)

+ 6 \lambda_2 g_1^{2} \tr\left(Y_\Delta Y_\Delta^{*} \right)

+ 30 \lambda_2 g_2^{2} \tr\left(Y_\Delta Y_\Delta^{*} \right)

-  \frac{384}{5} g_1^{2} g_2^{2} \tr\left(Y_\Delta Y_\Delta^{*} \right)

+ 80 g_2^{4} \tr\left(Y_\Delta Y_\Delta^{*} \right)

- 64 \lambda_1 \tr\left(Y_\Delta Y_\Delta^{*} Y_\Delta Y_\Delta^{*} \right)

- 12 \lambda_2 \tr\left(Y_e Y_e^{\dagger} Y_\Delta^{*} Y_\Delta \right)

- 12 \lambda_2 \tr\left(Y_\Phi Y_\Phi^{\dagger} Y_\Delta Y_\Delta^{*} \right)

- 104 \lambda_2 \tr\left(Y_\Delta Y_\Delta^{*} Y_\Delta Y_\Delta^{*} \right)

-  \frac{96}{5} g_1^{2} \tr\left(Y_\Delta Y_\Delta^{*} Y_\Delta Y_\Delta^{*} \right)

- 32 g_2^{2} \tr\left(Y_\Delta Y_\Delta^{*} Y_\Delta Y_\Delta^{*} \right)

- 64 \tr\left(Y_e Y_e^{\dagger} Y_\Delta^{*} Y_\Delta Y_\Delta^{*} Y_\Delta \right)

- 64 \tr\left(Y_\Phi Y_\Phi^{\dagger} Y_\Delta Y_\Delta^{*} Y_\Delta Y_\Delta^{*} \right)

- 512 \tr\left(Y_\Delta Y_\Delta^{*} Y_\Delta Y_\Delta^{*} Y_\Delta Y_\Delta^{*} \right)
\end{autobreak}
\end{align*}
\begin{align*}
\begin{autobreak}
\beta^{(1)}(\lambda_4) =

6 \lambda_3 \lambda_4

+ 8 \lambda_1 \lambda_4

+ 2 \lambda_2 \lambda_4

+ 4 \lambda_4^{2}

+ 8 \lambda_5^{2}

-  \frac{9}{2} \lambda_4 g_1^{2}

-  \frac{33}{2} \lambda_4 g_2^{2}

+ \frac{27}{25} g_1^{4}

+ 6 g_2^{4}

+ 6 \lambda_4 \tr\left(Y_u Y_u^{\dagger} \right)

+ 6 \lambda_4 \tr\left(Y_d Y_d^{\dagger} \right)

+ 2 \lambda_4 \tr\left(Y_e Y_e^{\dagger} \right)

+ 2 \lambda_4 \tr\left(Y_\Phi Y_\Phi^{\dagger} \right)

+ 4 \lambda_4 \tr\left(Y_\Delta Y_\Delta^{*} \right)

- 8 \tr\left(Y_e Y_e^{\dagger} Y_\Delta^{*} Y_\Delta \right)

- 8 \tr\left(Y_\Phi Y_\Phi^{\dagger} Y_\Delta Y_\Delta^{*} \right)
\end{autobreak}
\end{align*}
\begin{align*}
\begin{autobreak}
\beta^{(2)}(\lambda_4) =

- 36 \lambda_3 \lambda_4^{2}

- 40 \lambda_3 \lambda_5^{2}

- 10 \lambda_1 \lambda_2 \lambda_4

- 48 \lambda_1 \lambda_4^{2}

- 48 \lambda_1 \lambda_5^{2}

- 12 \lambda_2 \lambda_4^{2}

+ 8 \lambda_2 \lambda_5^{2}

- 15 \lambda_3^{2} \lambda_4

- 20 \lambda_1^{2} \lambda_4

-  \frac{15}{2} \lambda_2^{2} \lambda_4

- 13 \lambda_4^{3}

- 58 \lambda_4 \lambda_5^{2}

+ \frac{36}{5} \lambda_3 \lambda_4 g_1^{2}

+ 36 \lambda_3 \lambda_4 g_2^{2}

+ \frac{192}{5} \lambda_1 \lambda_4 g_1^{2}

+ 128 \lambda_1 \lambda_4 g_2^{2}

+ \frac{48}{5} \lambda_2 \lambda_4 g_1^{2}

+ 32 \lambda_2 \lambda_4 g_2^{2}

+ 3 \lambda_4^{2} g_1^{2}

+ 11 \lambda_4^{2} g_2^{2}

+ 6 \lambda_5^{2} g_1^{2}

+ 46 \lambda_5^{2} g_2^{2}

+ \frac{27}{5} \lambda_3 g_1^{4}

+ 30 \lambda_3 g_2^{4}

+ \frac{36}{5} \lambda_1 g_1^{4}

+ 40 \lambda_1 g_2^{4}

+ \frac{9}{5} \lambda_2 g_1^{4}

+ 10 \lambda_2 g_2^{4}

+ \frac{18363}{400} \lambda_4 g_1^{4}

+ \frac{105}{8} \lambda_4 g_1^{2} g_2^{2}

+ \frac{1015}{16} \lambda_4 g_2^{4}

-  \frac{24}{5} \lambda_5 g_1^{2} g_2^{2}

-  \frac{9549}{500} g_1^{6}

-  \frac{297}{20} g_1^{4} g_2^{2}

-  \frac{45}{2} g_1^{2} g_2^{4}

+ \frac{91}{2} g_2^{6}

- 36 \lambda_3 \lambda_4 \tr\left(Y_u Y_u^{\dagger} \right)

- 36 \lambda_3 \lambda_4 \tr\left(Y_d Y_d^{\dagger} \right)

- 12 \lambda_3 \lambda_4 \tr\left(Y_e Y_e^{\dagger} \right)

- 12 \lambda_3 \lambda_4 \tr\left(Y_\Phi Y_\Phi^{\dagger} \right)

- 32 \lambda_1 \lambda_4 \tr\left(Y_\Delta Y_\Delta^{*} \right)

- 8 \lambda_2 \lambda_4 \tr\left(Y_\Delta Y_\Delta^{*} \right)

- 12 \lambda_4^{2} \tr\left(Y_u Y_u^{\dagger} \right)

- 12 \lambda_4^{2} \tr\left(Y_d Y_d^{\dagger} \right)

- 4 \lambda_4^{2} \tr\left(Y_e Y_e^{\dagger} \right)

- 4 \lambda_4^{2} \tr\left(Y_\Phi Y_\Phi^{\dagger} \right)

- 8 \lambda_4^{2} \tr\left(Y_\Delta Y_\Delta^{*} \right)

- 24 \lambda_5^{2} \tr\left(Y_u Y_u^{\dagger} \right)

- 24 \lambda_5^{2} \tr\left(Y_d Y_d^{\dagger} \right)

- 8 \lambda_5^{2} \tr\left(Y_e Y_e^{\dagger} \right)

- 8 \lambda_5^{2} \tr\left(Y_\Phi Y_\Phi^{\dagger} \right)

- 16 \lambda_5^{2} \tr\left(Y_\Delta Y_\Delta^{*} \right)

+ \frac{17}{4} \lambda_4 g_1^{2} \tr\left(Y_u Y_u^{\dagger} \right)

+ \frac{5}{4} \lambda_4 g_1^{2} \tr\left(Y_d Y_d^{\dagger} \right)

+ \frac{15}{4} \lambda_4 g_1^{2} \tr\left(Y_e Y_e^{\dagger} \right)

+ \frac{3}{4} \lambda_4 g_1^{2} \tr\left(Y_\Phi Y_\Phi^{\dagger} \right)

+ 3 \lambda_4 g_1^{2} \tr\left(Y_\Delta Y_\Delta^{*} \right)

+ \frac{45}{4} \lambda_4 g_2^{2} \tr\left(Y_u Y_u^{\dagger} \right)

+ \frac{45}{4} \lambda_4 g_2^{2} \tr\left(Y_d Y_d^{\dagger} \right)

+ \frac{15}{4} \lambda_4 g_2^{2} \tr\left(Y_e Y_e^{\dagger} \right)

+ \frac{15}{4} \lambda_4 g_2^{2} \tr\left(Y_\Phi Y_\Phi^{\dagger} \right)

+ 15 \lambda_4 g_2^{2} \tr\left(Y_\Delta Y_\Delta^{*} \right)

+ 40 \lambda_4 g_3^{2} \tr\left(Y_u Y_u^{\dagger} \right)

+ 40 \lambda_4 g_3^{2} \tr\left(Y_d Y_d^{\dagger} \right)

-  \frac{171}{25} g_1^{4} \tr\left(Y_u Y_u^{\dagger} \right)

+ \frac{9}{5} g_1^{4} \tr\left(Y_d Y_d^{\dagger} \right)

- 9 g_1^{4} \tr\left(Y_e Y_e^{\dagger} \right)

-  \frac{9}{25} g_1^{4} \tr\left(Y_\Phi Y_\Phi^{\dagger} \right)

+ \frac{36}{25} g_1^{4} \tr\left(Y_\Delta Y_\Delta^{*} \right)

- 6 g_2^{4} \tr\left(Y_u Y_u^{\dagger} \right)

- 6 g_2^{4} \tr\left(Y_d Y_d^{\dagger} \right)

- 2 g_2^{4} \tr\left(Y_e Y_e^{\dagger} \right)

- 2 g_2^{4} \tr\left(Y_\Phi Y_\Phi^{\dagger} \right)

+ 2 g_2^{4} \tr\left(Y_\Delta Y_\Delta^{*} \right)

-  \frac{27}{2} \lambda_4 \tr\left(Y_u Y_u^{\dagger} Y_u Y_u^{\dagger} \right)

- 21 \lambda_4 \tr\left(Y_u Y_u^{\dagger} Y_d Y_d^{\dagger} \right)

-  \frac{27}{2} \lambda_4 \tr\left(Y_d Y_d^{\dagger} Y_d Y_d^{\dagger} \right)

-  \frac{9}{2} \lambda_4 \tr\left(Y_e Y_e^{\dagger} Y_e Y_e^{\dagger} \right)

- 7 \lambda_4 \tr\left(Y_e Y_e^{\dagger} Y_\Phi^{*} Y_\Phi^{\trans} \right)

+ \lambda_4 \tr\left(Y_e Y_e^{\dagger} Y_\Delta^{*} Y_\Delta \right)

-  \frac{9}{2} \lambda_4 \tr\left(Y_\Phi Y_\Phi^{\dagger} Y_\Phi Y_\Phi^{\dagger} \right)

+ \lambda_4 \tr\left(Y_\Phi Y_\Phi^{\dagger} Y_\Delta Y_\Delta^{*} \right)

- 36 \lambda_4 \tr\left(Y_\Delta Y_\Delta^{*} Y_\Delta Y_\Delta^{*} \right)

+ 32 \lambda_5 \tr\left(Y_e Y_e^{\dagger} Y_\Delta^{*} Y_\Delta \right)

- 32 \lambda_5 \tr\left(Y_\Phi Y_\Phi^{\dagger} Y_\Delta Y_\Delta^{*} \right)

-  \frac{12}{5} g_1^{2} \tr\left(Y_e Y_e^{\dagger} Y_\Delta^{*} Y_\Delta \right)

+ \frac{12}{5} g_1^{2} \tr\left(Y_\Phi Y_\Phi^{\dagger} Y_\Delta Y_\Delta^{*} \right)

+ 4 g_2^{2} \tr\left(Y_e Y_e^{\dagger} Y_\Delta^{*} Y_\Delta \right)

+ 4 g_2^{2} \tr\left(Y_\Phi Y_\Phi^{\dagger} Y_\Delta Y_\Delta^{*} \right)

+ 32 \tr\left(Y_e Y_e^{\dagger} Y_e Y_e^{\dagger} Y_\Delta^{*} Y_\Delta \right)

- 8 \tr\left(Y_e Y_e^{\dagger} Y_\Phi^{*} Y_\Phi^{\trans} Y_\Delta^{*} Y_\Delta \right)

+ 12 \tr\left(Y_e Y_e^{\dagger} Y_\Delta^{*} Y_e^{*} Y_e^{\trans} Y_\Delta \right)

+ 8 \tr\left(Y_e Y_e^{\dagger} Y_\Delta^{*} Y_\Phi Y_\Phi^{\dagger} Y_\Delta \right)

- 8 \tr\left(Y_e Y_e^{\dagger} Y_\Delta^{*} Y_\Delta Y_\Phi^{*} Y_\Phi^{\trans} \right)

+ 120 \tr\left(Y_e Y_e^{\dagger} Y_\Delta^{*} Y_\Delta Y_\Delta^{*} Y_\Delta \right)

+ 32 \tr\left(Y_\Phi Y_\Phi^{\dagger} Y_\Phi Y_\Phi^{\dagger} Y_\Delta Y_\Delta^{*} \right)

+ 12 \tr\left(Y_\Phi Y_\Phi^{\dagger} Y_\Delta Y_\Phi^{*} Y_\Phi^{\trans} Y_\Delta^{*} \right)

+ 120 \tr\left(Y_\Phi Y_\Phi^{\dagger} Y_\Delta Y_\Delta^{*} Y_\Delta Y_\Delta^{*} \right)
\end{autobreak}
\end{align*}
\begin{align*}
\begin{autobreak}
\beta^{(1)}(\lambda_5) =

2 \lambda_3 \lambda_5

+ 2 \lambda_1 \lambda_5

- 2 \lambda_2 \lambda_5

+ 8 \lambda_4 \lambda_5

-  \frac{9}{2} \lambda_5 g_1^{2}

-  \frac{33}{2} \lambda_5 g_2^{2}

-  \frac{18}{5} g_1^{2} g_2^{2}

+ 6 \lambda_5 \tr\left(Y_u Y_u^{\dagger} \right)

+ 6 \lambda_5 \tr\left(Y_d Y_d^{\dagger} \right)

+ 2 \lambda_5 \tr\left(Y_e Y_e^{\dagger} \right)

+ 2 \lambda_5 \tr\left(Y_\Phi Y_\Phi^{\dagger} \right)

+ 4 \lambda_5 \tr\left(Y_\Delta Y_\Delta^{*} \right)

- 8 \tr\left(Y_e Y_e^{\dagger} Y_\Delta^{*} Y_\Delta \right)

+ 8 \tr\left(Y_\Phi Y_\Phi^{\dagger} Y_\Delta Y_\Delta^{*} \right)
\end{autobreak}
\end{align*}
\begin{align*}
\begin{autobreak}
\beta^{(2)}(\lambda_5) =

- 40 \lambda_3 \lambda_4 \lambda_5

+ 6 \lambda_1 \lambda_2 \lambda_5

- 48 \lambda_1 \lambda_4 \lambda_5

+ 8 \lambda_2 \lambda_4 \lambda_5

- 7 \lambda_3^{2} \lambda_5

- 8 \lambda_1^{2} \lambda_5

+ \frac{9}{2} \lambda_2^{2} \lambda_5

- 29 \lambda_4^{2} \lambda_5

+ 2 \lambda_5^{3}

+ \frac{12}{5} \lambda_3 \lambda_5 g_1^{2}

+ \frac{48}{5} \lambda_1 \lambda_5 g_1^{2}

+ 20 \lambda_1 \lambda_5 g_2^{2}

-  \frac{48}{5} \lambda_2 \lambda_5 g_1^{2}

- 20 \lambda_2 \lambda_5 g_2^{2}

+ 6 \lambda_4 \lambda_5 g_1^{2}

+ 46 \lambda_4 \lambda_5 g_2^{2}

-  \frac{36}{5} \lambda_5^{2} g_1^{2}

- 6 \lambda_3 g_1^{2} g_2^{2}

- 6 \lambda_1 g_1^{2} g_2^{2}

+ 6 \lambda_2 g_1^{2} g_2^{2}

-  \frac{12}{5} \lambda_4 g_1^{2} g_2^{2}

+ \frac{9363}{400} \lambda_5 g_1^{4}

+ \frac{489}{8} \lambda_5 g_1^{2} g_2^{2}

-  \frac{465}{16} \lambda_5 g_2^{4}

+ \frac{1929}{50} g_1^{4} g_2^{2}

+ \frac{87}{2} g_1^{2} g_2^{4}

- 12 \lambda_3 \lambda_5 \tr\left(Y_u Y_u^{\dagger} \right)

- 12 \lambda_3 \lambda_5 \tr\left(Y_d Y_d^{\dagger} \right)

- 4 \lambda_3 \lambda_5 \tr\left(Y_e Y_e^{\dagger} \right)

- 4 \lambda_3 \lambda_5 \tr\left(Y_\Phi Y_\Phi^{\dagger} \right)

- 8 \lambda_1 \lambda_5 \tr\left(Y_\Delta Y_\Delta^{*} \right)

+ 8 \lambda_2 \lambda_5 \tr\left(Y_\Delta Y_\Delta^{*} \right)

- 24 \lambda_4 \lambda_5 \tr\left(Y_u Y_u^{\dagger} \right)

- 24 \lambda_4 \lambda_5 \tr\left(Y_d Y_d^{\dagger} \right)

- 8 \lambda_4 \lambda_5 \tr\left(Y_e Y_e^{\dagger} \right)

- 8 \lambda_4 \lambda_5 \tr\left(Y_\Phi Y_\Phi^{\dagger} \right)

- 16 \lambda_4 \lambda_5 \tr\left(Y_\Delta Y_\Delta^{*} \right)

+ \frac{17}{4} \lambda_5 g_1^{2} \tr\left(Y_u Y_u^{\dagger} \right)

+ \frac{5}{4} \lambda_5 g_1^{2} \tr\left(Y_d Y_d^{\dagger} \right)

+ \frac{15}{4} \lambda_5 g_1^{2} \tr\left(Y_e Y_e^{\dagger} \right)

+ \frac{3}{4} \lambda_5 g_1^{2} \tr\left(Y_\Phi Y_\Phi^{\dagger} \right)

+ 3 \lambda_5 g_1^{2} \tr\left(Y_\Delta Y_\Delta^{*} \right)

+ \frac{45}{4} \lambda_5 g_2^{2} \tr\left(Y_u Y_u^{\dagger} \right)

+ \frac{45}{4} \lambda_5 g_2^{2} \tr\left(Y_d Y_d^{\dagger} \right)

+ \frac{15}{4} \lambda_5 g_2^{2} \tr\left(Y_e Y_e^{\dagger} \right)

+ \frac{15}{4} \lambda_5 g_2^{2} \tr\left(Y_\Phi Y_\Phi^{\dagger} \right)

+ 15 \lambda_5 g_2^{2} \tr\left(Y_\Delta Y_\Delta^{*} \right)

+ 40 \lambda_5 g_3^{2} \tr\left(Y_u Y_u^{\dagger} \right)

+ 40 \lambda_5 g_3^{2} \tr\left(Y_d Y_d^{\dagger} \right)

-  \frac{126}{5} g_1^{2} g_2^{2} \tr\left(Y_u Y_u^{\dagger} \right)

-  \frac{54}{5} g_1^{2} g_2^{2} \tr\left(Y_d Y_d^{\dagger} \right)

-  \frac{66}{5} g_1^{2} g_2^{2} \tr\left(Y_e Y_e^{\dagger} \right)

+ \frac{6}{5} g_1^{2} g_2^{2} \tr\left(Y_\Phi Y_\Phi^{\dagger} \right)

-  \frac{24}{5} g_1^{2} g_2^{2} \tr\left(Y_\Delta Y_\Delta^{*} \right)

+ 16 \lambda_4 \tr\left(Y_e Y_e^{\dagger} Y_\Delta^{*} Y_\Delta \right)

- 16 \lambda_4 \tr\left(Y_\Phi Y_\Phi^{\dagger} Y_\Delta Y_\Delta^{*} \right)

-  \frac{27}{2} \lambda_5 \tr\left(Y_u Y_u^{\dagger} Y_u Y_u^{\dagger} \right)

+ 27 \lambda_5 \tr\left(Y_u Y_u^{\dagger} Y_d Y_d^{\dagger} \right)

-  \frac{27}{2} \lambda_5 \tr\left(Y_d Y_d^{\dagger} Y_d Y_d^{\dagger} \right)

-  \frac{9}{2} \lambda_5 \tr\left(Y_e Y_e^{\dagger} Y_e Y_e^{\dagger} \right)

+ 9 \lambda_5 \tr\left(Y_e Y_e^{\dagger} Y_\Phi^{*} Y_\Phi^{\trans} \right)

- 15 \lambda_5 \tr\left(Y_e Y_e^{\dagger} Y_\Delta^{*} Y_\Delta \right)

-  \frac{9}{2} \lambda_5 \tr\left(Y_\Phi Y_\Phi^{\dagger} Y_\Phi Y_\Phi^{\dagger} \right)

- 15 \lambda_5 \tr\left(Y_\Phi Y_\Phi^{\dagger} Y_\Delta Y_\Delta^{*} \right)

- 36 \lambda_5 \tr\left(Y_\Delta Y_\Delta^{*} Y_\Delta Y_\Delta^{*} \right)

-  \frac{12}{5} g_1^{2} \tr\left(Y_e Y_e^{\dagger} Y_\Delta^{*} Y_\Delta \right)

-  \frac{12}{5} g_1^{2} \tr\left(Y_\Phi Y_\Phi^{\dagger} Y_\Delta Y_\Delta^{*} \right)

+ 4 g_2^{2} \tr\left(Y_e Y_e^{\dagger} Y_\Delta^{*} Y_\Delta \right)

- 4 g_2^{2} \tr\left(Y_\Phi Y_\Phi^{\dagger} Y_\Delta Y_\Delta^{*} \right)

+ 16 \tr\left(Y_e Y_e^{\dagger} Y_e Y_e^{\dagger} Y_\Delta^{*} Y_\Delta \right)

+ 12 \tr\left(Y_e Y_e^{\dagger} Y_\Delta^{*} Y_e^{*} Y_e^{\trans} Y_\Delta \right)

+ 88 \tr\left(Y_e Y_e^{\dagger} Y_\Delta^{*} Y_\Delta Y_\Delta^{*} Y_\Delta \right)

- 16 \tr\left(Y_\Phi Y_\Phi^{\dagger} Y_\Phi Y_\Phi^{\dagger} Y_\Delta Y_\Delta^{*} \right)

- 12 \tr\left(Y_\Phi Y_\Phi^{\dagger} Y_\Delta Y_\Phi^{*} Y_\Phi^{\trans} Y_\Delta^{*} \right)

- 88 \tr\left(Y_\Phi Y_\Phi^{\dagger} Y_\Delta Y_\Delta^{*} Y_\Delta Y_\Delta^{*} \right)
\end{autobreak}
\end{align*}
}

\subsection{Scalar mass couplings}
{\allowdisplaybreaks

\begin{align*}
\begin{autobreak}
\beta^{(1)}(m^2) =

-  \frac{9}{10} g_1^{2} m^2

-  \frac{9}{2} g_2^{2} m^2

- 12 \gamma^{2}

+ 6 \lambda_3 m^2

- 12 \lambda_4 M_\Delta^2

+ 6 m^2 \tr\left(Y_u Y_u^{\dagger} \right)

+ 6 m^2 \tr\left(Y_d Y_d^{\dagger} \right)

+ 2 m^2 \tr\left(Y_e Y_e^{\dagger} \right)

+ 2 m^2 \tr\left(Y_\Phi Y_\Phi^{\dagger} \right)

+ 4 \tr\left(M_\nu M_\nu^{\dagger} Y_\Phi^{\trans} Y_\Phi^{*} \right)

+ 4 \tr\left(Y_\Phi M_\nu^{\dagger} M_\nu Y_\Phi^{\dagger} \right)
\end{autobreak}
\end{align*}
\begin{align*}
\begin{autobreak}
\beta^{(2)}(m^2) =

\frac{2067}{400} g_1^{4} m^2

+ \frac{9}{8} g_1^{2} g_2^{2} m^2

-  \frac{57}{16} g_2^{4} m^2

-  \frac{54}{5} M_\Delta^2 g_1^{4}

- 60 M_\Delta^2 g_2^{4}

-  \frac{153}{5} g_1^{2} \gamma^{2}

- 105 g_2^{2} \gamma^{2}

+ \frac{36}{5} \lambda_3 g_1^{2} m^2

+ 36 \lambda_3 g_2^{2} m^2

-  \frac{288}{5} \lambda_4 M_\Delta^2 g_1^{2}

- 192 \lambda_4 M_\Delta^2 g_2^{2}

+ 60 \lambda_3 \gamma^{2}

+ 36 \lambda_4 \gamma^{2}

- 48 \lambda_5 \gamma^{2}

- 15 \lambda_3^{2} m^2

- 3 \lambda_4^{2} m^2

- 6 \lambda_5^{2} m^2

+ 24 \lambda_4^{2} M_\Delta^2

+ 48 \lambda_5^{2} M_\Delta^2

+ \frac{17}{4} g_1^{2} m^2 \tr\left(Y_u Y_u^{\dagger} \right)

+ \frac{5}{4} g_1^{2} m^2 \tr\left(Y_d Y_d^{\dagger} \right)

+ \frac{15}{4} g_1^{2} m^2 \tr\left(Y_e Y_e^{\dagger} \right)

+ \frac{3}{4} g_1^{2} m^2 \tr\left(Y_\Phi Y_\Phi^{\dagger} \right)

+ \frac{45}{4} g_2^{2} m^2 \tr\left(Y_u Y_u^{\dagger} \right)

+ \frac{45}{4} g_2^{2} m^2 \tr\left(Y_d Y_d^{\dagger} \right)

+ \frac{15}{4} g_2^{2} m^2 \tr\left(Y_e Y_e^{\dagger} \right)

+ \frac{15}{4} g_2^{2} m^2 \tr\left(Y_\Phi Y_\Phi^{\dagger} \right)

+ 40 g_3^{2} m^2 \tr\left(Y_u Y_u^{\dagger} \right)

+ 40 g_3^{2} m^2 \tr\left(Y_d Y_d^{\dagger} \right)

+ 36 \gamma^{2} \tr\left(Y_u Y_u^{\dagger} \right)

+ 36 \gamma^{2} \tr\left(Y_d Y_d^{\dagger} \right)

+ 12 \gamma^{2} \tr\left(Y_e Y_e^{\dagger} \right)

+ 12 \gamma^{2} \tr\left(Y_\Phi Y_\Phi^{\dagger} \right)

+ 24 \gamma^{2} \tr\left(Y_\Delta Y_\Delta^{*} \right)

- 36 \lambda_3 m^2 \tr\left(Y_u Y_u^{\dagger} \right)

- 36 \lambda_3 m^2 \tr\left(Y_d Y_d^{\dagger} \right)

- 12 \lambda_3 m^2 \tr\left(Y_e Y_e^{\dagger} \right)

- 12 \lambda_3 m^2 \tr\left(Y_\Phi Y_\Phi^{\dagger} \right)

+ 48 \lambda_4 M_\Delta^2 \tr\left(Y_\Delta Y_\Delta^{*} \right)

-  \frac{27}{2} m^2 \tr\left(Y_u Y_u^{\dagger} Y_u Y_u^{\dagger} \right)

- 21 m^2 \tr\left(Y_u Y_u^{\dagger} Y_d Y_d^{\dagger} \right)

-  \frac{27}{2} m^2 \tr\left(Y_d Y_d^{\dagger} Y_d Y_d^{\dagger} \right)

-  \frac{9}{2} m^2 \tr\left(Y_e Y_e^{\dagger} Y_e Y_e^{\dagger} \right)

- 7 m^2 \tr\left(Y_e Y_e^{\dagger} Y_\Phi^{*} Y_\Phi^{\trans} \right)

- 9 m^2 \tr\left(Y_e Y_e^{\dagger} Y_\Delta^{*} Y_\Delta \right)

-  \frac{9}{2} m^2 \tr\left(Y_\Phi Y_\Phi^{\dagger} Y_\Phi Y_\Phi^{\dagger} \right)

- 9 m^2 \tr\left(Y_\Phi Y_\Phi^{\dagger} Y_\Delta Y_\Delta^{*} \right)

+ 2 \tr\left(Y_e Y_e^{\dagger} Y_\Phi^{*} M_\nu M_\nu^{\dagger} Y_\Phi^{\trans} \right)

+ 2 \tr\left(Y_\Phi M_\nu^{\dagger} M_\nu Y_\Phi^{\dagger} Y_e^{*} Y_e^{\trans} \right)

- 16 \tr\left(Y_\Phi M_\nu^{\dagger} Y_\Phi^{\trans} Y_\Phi^{*} M_\nu Y_\Phi^{\dagger} \right)

- 14 \tr\left(Y_\Phi Y_\Phi^{\dagger} Y_\Phi M_\nu^{\dagger} M_\nu Y_\Phi^{\dagger} \right)

- 14 \tr\left(Y_\Phi Y_\Phi^{\dagger} Y_\Phi M_\nu^{*} M_\nu^{\trans} Y_\Phi^{\dagger} \right)

- 12 \tr\left(Y_\Phi M_\nu^{\dagger} M_\nu Y_\Phi^{\dagger} Y_\Delta Y_\Delta^{*} \right)

- 12 \tr\left(Y_\Delta Y_\Phi^{*} M_\nu M_\nu^{\dagger} Y_\Phi^{\trans} Y_\Delta^{*} \right)
\end{autobreak}
\end{align*}
\begin{align*}
\begin{autobreak}
\beta^{(1)}(M_\Delta^2) =

-  \frac{18}{5} M_\Delta^2 g_1^{2}

- 12 M_\Delta^2 g_2^{2}

+ 2 \gamma^{2}

- 2 \lambda_4 m^2

+ 8 \lambda_1 M_\Delta^2

+ 2 \lambda_2 M_\Delta^2

+ 4 M_\Delta^2 \tr\left(Y_\Delta Y_\Delta^{*} \right)
\end{autobreak}
\end{align*}
\begin{align*}
\begin{autobreak}
\beta^{(2)}(M_\Delta^2) =

-  \frac{9}{5} g_1^{4} m^2

- 10 g_2^{4} m^2

+ \frac{2001}{50} M_\Delta^2 g_1^{4}

+ 12 M_\Delta^2 g_1^{2} g_2^{2}

+ 63 M_\Delta^2 g_2^{4}

-  \frac{6}{5} g_1^{2} \gamma^{2}

-  \frac{12}{5} \lambda_4 g_1^{2} m^2

- 12 \lambda_4 g_2^{2} m^2

+ \frac{192}{5} \lambda_1 M_\Delta^2 g_1^{2}

+ \frac{48}{5} \lambda_2 M_\Delta^2 g_1^{2}

+ 128 \lambda_1 M_\Delta^2 g_2^{2}

+ 32 \lambda_2 M_\Delta^2 g_2^{2}

- 8 \lambda_1 \gamma^{2}

- 2 \lambda_2 \gamma^{2}

- 20 \lambda_4 \gamma^{2}

+ 16 \lambda_5 \gamma^{2}

+ 4 \lambda_4^{2} m^2

+ 8 \lambda_5^{2} m^2

- 20 \lambda_1^{2} M_\Delta^2

- 10 \lambda_1 \lambda_2 M_\Delta^2

-  \frac{15}{2} \lambda_2^{2} M_\Delta^2

- 2 \lambda_4^{2} M_\Delta^2

- 4 \lambda_5^{2} M_\Delta^2

+ 3 M_\Delta^2 g_1^{2} \tr\left(Y_\Delta Y_\Delta^{*} \right)

+ 15 M_\Delta^2 g_2^{2} \tr\left(Y_\Delta Y_\Delta^{*} \right)

- 12 \gamma^{2} \tr\left(Y_u Y_u^{\dagger} \right)

- 12 \gamma^{2} \tr\left(Y_d Y_d^{\dagger} \right)

- 4 \gamma^{2} \tr\left(Y_e Y_e^{\dagger} \right)

- 4 \gamma^{2} \tr\left(Y_\Phi Y_\Phi^{\dagger} \right)

+ 12 \lambda_4 m^2 \tr\left(Y_u Y_u^{\dagger} \right)

+ 12 \lambda_4 m^2 \tr\left(Y_d Y_d^{\dagger} \right)

+ 4 \lambda_4 m^2 \tr\left(Y_e Y_e^{\dagger} \right)

+ 4 \lambda_4 m^2 \tr\left(Y_\Phi Y_\Phi^{\dagger} \right)

- 32 \lambda_1 M_\Delta^2 \tr\left(Y_\Delta Y_\Delta^{*} \right)

- 8 \lambda_2 M_\Delta^2 \tr\left(Y_\Delta Y_\Delta^{*} \right)

+ 4 \sqrt{2} \gamma \tr\left(Y_\Phi M_\nu^{\dagger} Y_\Phi^{\trans} Y_\Delta^{*} \right)

+ 4 \sqrt{2} \gamma \tr\left(Y_\Delta Y_\Phi^{*} M_\nu Y_\Phi^{\dagger} \right)

- 6 M_\Delta^2 \tr\left(Y_e Y_e^{\dagger} Y_\Delta^{*} Y_\Delta \right)

- 6 M_\Delta^2 \tr\left(Y_\Phi Y_\Phi^{\dagger} Y_\Delta Y_\Delta^{*} \right)

- 36 M_\Delta^2 \tr\left(Y_\Delta Y_\Delta^{*} Y_\Delta Y_\Delta^{*} \right)

+ 8 \tr\left(Y_\Phi M_\nu^{\dagger} M_\nu Y_\Phi^{\dagger} Y_\Delta Y_\Delta^{*} \right)

+ 8 \tr\left(Y_\Delta Y_\Phi^{*} M_\nu M_\nu^{\dagger} Y_\Phi^{\trans} Y_\Delta^{*} \right)
\end{autobreak}
\end{align*}
}

\subsection{Fermion mass couplings}
{\allowdisplaybreaks

\begin{align*}
\begin{autobreak}
\beta^{(1)}(M_\nu) =

M_\nu^{\trans} Y_\Phi^{\dagger} Y_\Phi

+ Y_\Phi^{\trans} Y_\Phi^{*} M_\nu
\end{autobreak}
\end{align*}
\begin{align*}
\begin{autobreak}
\beta^{(2)}(M_\nu) =

\frac{51}{40} g_1^{2} M_\nu Y_\Phi^{\dagger} Y_\Phi

+ \frac{51}{40} g_1^{2} Y_\Phi^{\trans} Y_\Phi^{*} M_\nu^{\trans}

+ \frac{51}{8} g_2^{2} M_\nu Y_\Phi^{\dagger} Y_\Phi

+ \frac{51}{8} g_2^{2} Y_\Phi^{\trans} Y_\Phi^{*} M_\nu^{\trans}

-  \frac{1}{4} M_\nu^{\trans} Y_\Phi^{\dagger} Y_e^{*} Y_e^{\trans} Y_\Phi

-  \frac{1}{4} Y_\Phi^{\trans} Y_e Y_e^{\dagger} Y_\Phi^{*} M_\nu

+ 2 Y_\Phi^{\trans} Y_\Phi^{*} M_\nu Y_\Phi^{\dagger} Y_\Phi

+ 2 Y_\Phi^{\trans} Y_\Phi^{*} M_\nu^{\trans} Y_\Phi^{\dagger} Y_\Phi

-  \frac{1}{4} M_\nu^{\trans} Y_\Phi^{\dagger} Y_\Phi Y_\Phi^{\dagger} Y_\Phi

-  \frac{1}{4} Y_\Phi^{\trans} Y_\Phi^{*} Y_\Phi^{\trans} Y_\Phi^{*} M_\nu

-  \frac{3}{2} M_\nu^{\trans} Y_\Phi^{\dagger} Y_\Delta Y_\Delta^{*} Y_\Phi

-  \frac{3}{2} Y_\Phi^{\trans} Y_\Delta^{*} Y_\Delta Y_\Phi^{*} M_\nu

-  \frac{9}{2} \tr\left(Y_u Y_u^{\dagger} \right) M_\nu Y_\Phi^{\dagger} Y_\Phi

-  \frac{9}{2} \tr\left(Y_u Y_u^{\dagger} \right) Y_\Phi^{\trans} Y_\Phi^{*} M_\nu^{\trans}

-  \frac{9}{2} \tr\left(Y_d Y_d^{\dagger} \right) M_\nu Y_\Phi^{\dagger} Y_\Phi

-  \frac{9}{2} \tr\left(Y_d Y_d^{\dagger} \right) Y_\Phi^{\trans} Y_\Phi^{*} M_\nu^{\trans}

-  \frac{3}{2} \tr\left(Y_e Y_e^{\dagger} \right) M_\nu Y_\Phi^{\dagger} Y_\Phi

-  \frac{3}{2} \tr\left(Y_e Y_e^{\dagger} \right) Y_\Phi^{\trans} Y_\Phi^{*} M_\nu^{\trans}

-  \frac{3}{2} \tr\left(Y_\Phi Y_\Phi^{\dagger} \right) M_\nu Y_\Phi^{\dagger} Y_\Phi

-  \frac{3}{2} \tr\left(Y_\Phi Y_\Phi^{\dagger} \right) Y_\Phi^{\trans} Y_\Phi^{*} M_\nu^{\trans}

-  12 \sqrt{2} \gamma Y_\Phi^{\trans} Y_\Delta^{*} Y_\Phi
\end{autobreak}
\end{align*}
}
\bibliography{neutrino.bib}

\end{document}